\documentclass[traditabstract]{aa}
\usepackage{graphics}
\usepackage{amssymb, amsmath}
\usepackage{natbib}
\usepackage{lineno}
\bibpunct{(}{)}{;}{a}{}{,}

\DeclareRobustCommand{\ion}[2]{%
\relax\ifmmode
\ifx\testbx\f@series
{\mathbf{#1\,\mathsc{#2}}}\else
{\mathrm{#1\,\mathsc{#2}}}\fi
\else\textup{#1\,{\mdseries\textsc{#2}}}%
\fi}

\def\apjl{ApJ}%
\def\apjs{ApJS}%
\def\ao{Appl.~Opt.}%
\def\aap{A\&A}%
\begin{document}

\title{Variation in the dust emissivity index across M33 with Herschel and Spitzer (HerM33es)}
\titlerunning{Dusty ISM of M33 }

\author{ F.~S. Tabatabaei\inst{1}, J. Braine\inst{2}, E.\,M. Xilouris\inst{3},  C. Kramer\inst{4}, M. Boquien\inst{5}, F. Combes\inst{6}, C. Henkel\inst{7,8}, M.~Relano\inst{9}, S. Verley\inst{9}, P. Gratier\inst{10},  F. Israel\inst{11}, M.\,C. Wiedner\inst{6}, M.~R\"ollig\inst{12}, K.\,F.~Schuster\inst{10}, P.~\,van\,der\,Werf\inst{11}}
\authorrunning{Tabatabaei et al.}

\institute{Max-Planck-Institut f\"ur Astronomie, K\"onigstuhl 17, 69117 Heidelberg, Germany
\and Laboratoire d'Astrophysique de Bordeaux, Universit\'e de Bordeaux and CNRS UMR 5804, 33271 Floirac, France 
\and Institute for Astronomy, Astrophysics, Space Applications and Remote 
Sensing, National Observatory of Athens, GR-15236 Athens, Greece.
\and Instituto Radioastronomia Milimetrica, 18012 Granada, Spain
\and Aix Marseille Universit\'e, CNRS, LAM (Laboratoire d'Astrophysique de Marseille) UMR 7326, 13388, Marseille, France
\and Observatoire de Paris, LERMA, CNRS, 61 Av. de l'Observatoire, 75014 Paris, France
\and Max-Planck Institut f\"ur Astronomie, Auf dem H\"ugel 69, 53121 Bonn, Germany 
\and Astron. Dept., King Abdulaziz University, P.O. Box 80203, Jeddah, Saudi Arabia
\and Department F\'isica Te\'orica y del Cosmos, Universidad de Granada, Spain
\and IRAM, 300 rue de la Piscine, 38406 St. Martin d'H\'eres, France
\and Sterrewacht Leiden, Leiden University, PO Box 9513, 2300 RA, Leiden, The Netherlands
\and I. Physikalisches Institut, Universit\"at zu K\"oln, Z\"ulpicher Str. 77, D-50937 K\"oln, Germany
%
%\and Leiden Observatory, Leiden University, P.O. Box 9513, NL-2300 RA Leiden, The Netherlands 
%
%\and SUPA, Institute of Astronomy, University of Edinburgh, Royal Observatory, Blackford Hill, Edinburgh EH9 3HJ, UK
}
\offprints{F.\,S. Tabatabaei \\ taba@mpia.de}
%   \date{Received September 15, 2004; accepted December 16, 2004}

\abstract{We study the wavelength dependence of the dust emission as a function of
position and environment across the disk of M33 at a linear resolution of 160\,pc
using {\it Spitzer} and {\it Herschel} photometric data.   M33 is a Local Group
spiral with a slightly subsolar metallicity, making it an ideal stepping-stone to
less regular and lower metallicity objects such as {  dwarf} galaxies and, probably,
young universe objects.
Expressing the emissivity of the dust as a power law, the power-law exponent
($\beta$) is estimated from 
two independent approaches designed to {  properly treat} the degeneracy between $\beta$ and the
dust temperature.  
Both $\beta$ and the dust temperature are higher in the inner disk than in the outer
disk, contrary to reported $\beta - T$ anti-correlations {  found in other sources}. 
In the cold $+$ warm dust model, the warm component and the ionized gas (H$\alpha$)
have a very similar distribution across the galaxy, demonstrating that the model
separates the components in an appropriate fashion.
The flocculent spiral arms and the dust lanes are evident in the map of
the cold component. 
Both cold and warm dust column densities are high in star forming regions and reach
their maxima toward the giant star forming complexes NGC604 and NGC595.  $\beta$ declines from
close to 2 in the center to about 1.3 in the outer disk.   $\beta$ is positively
correlated with star formation and with molecular gas column, as traced by H$\alpha$
and CO emission. 
The lower dust emissivity index in the outer parts of M33 is likely related to the
reduced metallicity (different grain composition) and possibly different size 
distribution.  
It is not due to the decrease in stellar radiation field or temperature in a 
simple way because the FIR-bright regions
in the outer disk {  also} have a low $\beta$.  Like most spirals, M33 has a (decreasing) radial gradient in star formation and molecular-to-atomic gas ratio such that the
regions bright in H$\alpha$ or CO tend to trace the inner disk, making it difficult
to distinguish between their effects on the dust.  
{  The assumption of a constant emissivity index $\beta$ is obviously
not appropriate. } } 
\maketitle
%(30-80\,K)

%________________________________________________________________

\section{Introduction}
Dust and gas are thoroughly mixed in the interstellar medium (ISM), with approximately half of the metals (atomic number Z$>$2) in each.  Unlike molecules, which radiate at specific frequencies, dust emission covers all frequencies and carries much more energy.  Dust emission thus provides an alternative to spectral lines to study the mass distribution in the ISM.  Dust has the advantage over the commonly-used CO molecule of not being photo-dissociated by UV photons.  However, dust emission depends strongly on the grain temperature, such that it is not straightforward to measure the dust mass.  In order to estimate the grain temperature or range in temperatures, it is necessary to know how grain emission varies with wavelength.

For dust grains in local thermal equilibrium, the temperature is usually obtained using a modified black body (MBB) emission given by 
\begin{eqnarray}
S_{\nu} = \, B_{\nu}(T)\, [1 - exp(- \tau_{\nu})], 
\end{eqnarray}
where $S_{\nu}$ is the intensity, $B_{\nu}$ represents the Planck function,
and $\tau_{\nu}$ denotes the optical depth of the dust, depending on
frequency $\nu$. In the optically thin limit, the above equation converts to
\begin{eqnarray}
S_{\nu} = \, B_{\nu}(T)\, \tau_{\nu} = \, B_{\nu}(T) \, \kappa_{\nu} \, \Sigma_{\rm dust},
\end{eqnarray}
where $\Sigma_{\rm dust}$ is the dust mass surface density and $\kappa_{\nu} = \kappa_0 \, (\nu/\nu_0)^{\beta}$ is the dust emissivity with index $\beta$ ($\kappa_0$ is the grain cross-section per gram at frequency $\nu_0$).

There is some evidence supporting variations of the dust emission spectrum with environmental conditions from both Galactic \citep[e.g.][]{Paradis} and extra-galactic studies \citep[e.g.][]{Lisenfeld_0}.  The dust emissivity index $\beta$  could change depending on grain properties such as structure, size distribution, or chemical composition and is still a matter of debate. 
These properties may be affected by different physical/environmental processes like shattering, sputtering, grain - grain collisions \citep[mainly due to shocks, see e.g.][and references therein]{Jones_04}, condensation of molecules onto grains, and coagulation \citep[e.g.][]{Draine_06}. {  The
extra-galactic observations presented here sample a beam of 160\,pc with
a depth of few 100\,pc (Combes et al. 2012). Along each
line-of-sight, we therefore expect to sample regions of widely
different physical properties, e.g. temperatures and densities. The
observed dust temperatures and $\beta$ indices are therefore weighted
averages of the local conditions along the lines of sight. This is
further discussed in Section 6.}
Neglecting the variation in the dust emissivity {  index} could be misleading when determining the 
dust mass \citep[e.g.][]{Malinen}.  For instance, from the long-wavelength COBE data \cite{Reach_95},  assuming a constant $\beta$, found a very massive cold dust component in the Milky Way (MW) but this was not confirmed by \cite{Lagache}. 

Nearby galaxies provide ideal astrophysical laboratories 
(1) with different chemical abundances and (2) where - for face on galaxies - the galactic disk is only crossed once by our line of sight, so it is straightforward to distinguish {  spiral arms, associate gas/dust clouds with star forming regions, and inter-arm regions.}

The Herschel Space Observatory \citep{Pilbratt} provides far-infrared (FIR) and
sub-millimeter (sub-mm) data at high {  angular} resolution and sensitivity over the wavelengths
required to investigate the physical properties of dust in galaxies.  Recent studies have suggested that the emission of dust grains could differ from the standard MBB model (see Eq.~1) which assumes a $\beta$=2 emissivity law. % or the use of graphite grains to model carbon dust. 
This can lead to unphysical gas-to-dust mass ratios compared
to those expected from the metallicity of the galaxies \citep{Meixner,Galametz,Galliano,Israel_11}.
The emissivity indices reported from observations range from $\beta = 1$ to $\beta =
2.5$ \citep{Chapin,Casey_11,Boselli_12}. Theoretically, $\beta<1$ is excluded at the FIR and submm wavelengths, otherwise the Kramers-Kronig relation does not hold.  Mie theory implies that $\beta=2$ for spherical grains of idealized dielectrics and metals, while $\beta=1$ if the complex optical constant does not depend on frequency \citep[][ chapters~2, 3, and 8]{krugel}.

{\bf There is a systematic degeneracy between $\beta$ and the dust temperature $T$ given by Eq.(2), with an amplitude which depends on the signal-to-noise ratio, S/N,  and the spectral coverage. Therefore,}  $\beta$ and $T$ cannot be reliably determined for a single position by fitting the FIR-submm dust emission, even for a homogeneous and single-temperature cloud \citep[e.g. ][]{Kelly}.  This is because equally good fits \citep[i.e. well within observational noise even for high S/N data, see e.g. Fig.~3 in][]{Plank_3}
%to the 160-500 micron emission 
can be obtained for different ($\beta$, $T$) pairs. %ApJ 696, 676  
The degeneracy is such that $\beta$ increases as $T$ decreases and vice-versa.  The scatter in the fits can produce an apparent anti-correlation between $\beta$ and $T$  \citep[][]{Shetty_09b,Shetty_09}.

\begin{figure*}
\begin{center}
\resizebox{12cm}{!}{
\includegraphics*{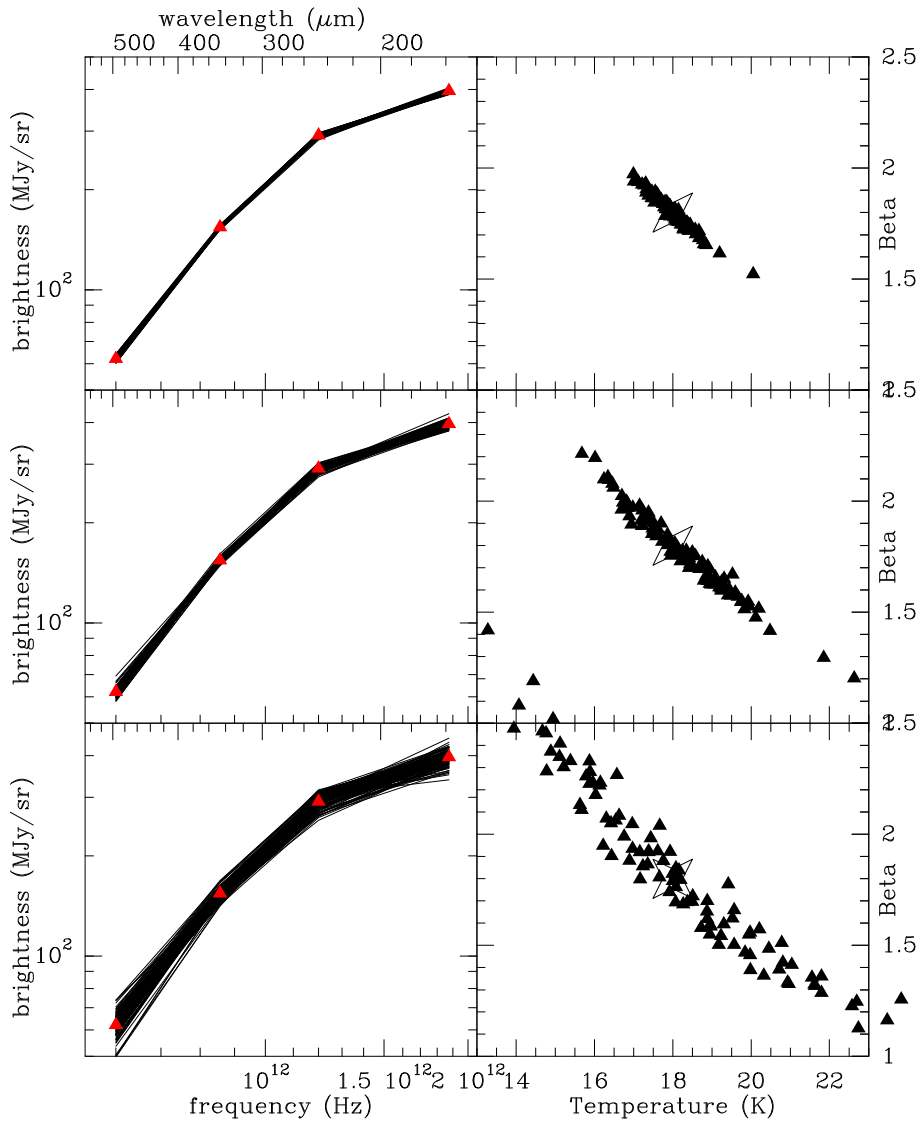}}
\caption[]{Illustration of the $\beta-T$ degeneracy:{  the panels on the left hand side} show fits to the 4 data
points to which noise has been added and the right column shows the values of
$\beta$ and T.  The large triangles {  on the left}  show the original data points
(before adding noise) which follow a modified black-body law with $\beta = 1.8$ and
$T=18$, rather typical values for normal spiral galaxies.  The large open stars show
this position in  the right-hand column.  From bottom to top, the noise added
follows a Gaussian distribution with standard deviation of 5\% (bottom panels), 2\%
(middle panels) and 1\% (top panels).  Despite this very low noise level, the
scatter in $\beta$ and T is significant and correlated. {  An enlargement of the spectral coverage, if no other processes like emission from small or spinning dust grains play a role within the extended wavelength range, would help reduce the degeneracy.}}
\label{fig:degeneracy}
\end{center}
\end{figure*}

{  The power of this degeneracy is illustrated in Fig.~\ref{fig:degeneracy}.  For a modified black body of temperature 18~K and index $\beta$ = 1.8, we add noise with a standard deviation of 1, 2, and 5\% (left {  pannels} from top to bottom) of the signal, corresponding to S/N ratios of 100, 50, and 20.  The mock data points are fit and the fits are plotted; this is done 100 times.  The red triangles are the original no-noise data points at 160, 250, 350, and 500 microns.
The right column shows the derived $\beta$ and T values and the large open star indicates the (T$=18$K, $\beta = 1.8$) point.   Despite the very high S/N ratios, the scatter is broad and follows a banana-shaped anti-correlation pattern.  {  The anti-correlation is generated by the noise despite the unrealistically high signal to noise levels used in
these plots.  For the S/N ratios of 20, 50, and 100 used in Fig.~\ref{fig:degeneracy}, 
the uncertainties (standard deviation) are 3~K, 1.1~K, and 0.5~K in temperature
and 0.4, 0.15, and 0.08 in $\beta$, respectively.}  For T$=18$K and $\beta = 1.8$, the data can be fit by a variety of values varying such that $\beta \approx 1.8 - 0.155 (T - 18)$. Further below, we will
present a more detailed statistical (Monte Carlo) analysis showing how
we constrain the dust temperature and $\beta$. }

A number of authors have found an anti-correlation between $\beta$ and T
\citep[e.g.][]{Dupac}. 
While {  many} authors \citep[e.g.][]{Paradis,Ysard} believe the anti-correlation is not
entirely due to the $\beta - T$ degeneracy, it remains difficult to isolate and
identify the physical and degenerate parts of this relation
\citep[e.g.][]{Galametz_12, Shetty_09b}. 

Laboratory experiments have also found an anti-correlation between $\beta$ and $T$.  However, in space, the anti-correlation is typically for dust temperatures of 10-20\,K (the dominant temperatures of dust emitting in the submm) whereas laboratory experiments find the difference between 300\,K dust (room temperature) and 10-20\,K dust.  The difference in dust spectra between 10\,K and 20\,K  is very difficult to identify in laboratory experiments \citep[e.g.][]{coupeaud}.  It is currently not clear whether laboratory measurements support a physical $\beta - T$ anti-correlation for dust in the 10-20\,K range.

\begin{table}
\begin{center}
\caption{Positional data on M33.}
\begin{tabular}{ l l } 
\hline
\hline
Nucleus position\,(J2000)$^{1}$    & RA\,=\,$1^{h}33^{m}51.0^{s}$      \\
    &  DEC\,=\,$30^{\circ}39\arcmin37.0\arcsec$\\
Position angle of major axis$^{2}$   &23$^{\circ}$ \\
Inclination$^{3}$    & 56$^{\circ}$ \\
Distance$^{4}$\,(1$\arcsec$=\,4\,pc)   & 840\,kpc\\
\hline
\noalign {\medskip}
\multicolumn{2}{l}{$^{1}$ \cite{devaucouleurs_81}}\\
\multicolumn{2}{l}{$^{2}$ \cite{Deul}}\\
\multicolumn{2}{l}{$^{3}$ \cite{Regan_etal_94}}\\
\multicolumn{2}{l}{$^{4}$ \cite{Freedman_etal_91}}\\
\end{tabular}
\end{center}
\end{table}
The nearest Scd galaxy, M~33\,(NGC\,598), at a distance of 840 kpc \citep[1$\arcsec \simeq$ 4\,pc, ][]{Freedman_etal_91} has been extensively studied at IR wavelengths. Its mild inclination  \citep[$i=56^{\circ}$,][Table~1]{Regan_etal_94} makes it a suitable target for mapping a large variety of astrophysical properties.
Herschel observations of M~33 (HerM33es\footnote{Herschel M33 extended survey open time key project, http://www.iram.es/IRAMES/hermesWiki}, Kramer et al. 2010) enable us to study the dust SED  at a resolution where complexes of giant molecular clouds and star forming regions are resolved.  
{  Using the Herschel PACS and SPIRE data together with the Multiband Imaging Photometer Spitzer (MIPS) fluxes, \cite{Kramer10} studied the dust SED in radial intervals of 2\,kpc width (out to 8\,kpc galactocentric radius) as well as the integrated SED of M\,33. Assuming a constant $\beta$, they found that a two-component MBB fits the SEDs better than the isothermal models. They also showed that the global SED fits better with $\beta$=1.5 than with $\beta$=2. 
In a pixel-by-pixel approach, \cite{Xilouris} derived the  temperature and luminosity density distribution of a two-component MBB for the best-fitted $\beta$ of 1.5, which was fixed across the galaxy. 
This paper investigates possible variations of $\beta$ radially and in the disk of M33, using different single- and two-component MBB approaches in which the degeneracy between $\beta$ and $T$ is treated properly. Studying both single- and two-component MBBs are important to assess the temperature mixing along the line of sight, which could in principle lead to a lower $\beta$ value if a single MBB model is assumed \citep[][]{Malinen}.}
In order to obtain as much information as possible about the warm dust and to avoid  as much as possible contamination by non-thermalized grains  (e.g. very small grains), we use the Herschel SPIRE/PACS and Spitzer MIPS maps at wavelengths $\lambda \geq$\,70\,$\mu$m, probing primarily the dust emission from the big grains which are believed to be in thermal equilibrium with the interstellar radiation field \citep[see e.g.][]{Carey}. 
In the  diffuse Galactic  interstellar medium (ISM), these grains emit as a MBB with an equilibrium temperature of $17-18$\,K  \citep[e.g.][]{Boulanger}.

{  The paper is organized as follows. The data sets used in this study are described in Sect.~2. We present our methods and approaches as well as their resulting physical parameters  in Sect.~3. Then, we  investigate the systematics of the two-component MBB approaches by performing a Monte-Carlo simulation (Sect.~4). Uncertainties of the physical parameters are presented in Sect.~5. We discuss and summarize the results in Sects.~6 and 7. }  

\begin{table*}
\begin{center}
\caption{Images of M\,33 used in this study. }
\begin{tabular}{ l l l l} 
\hline

Wavelength & Resolution  & rms noise& Telescope \\

\hline
500\,$\mu$m  &  $37\arcsec$& 8\,mJy/beam &Herschel-SPIRE$^{1}$\\
350\,$\mu$m  &  $25\arcsec$& 9.2\,mJy/beam & Herschel-SPIRE$^{1}$\\
250\,$\mu$m  &  $18\arcsec$&14.1\,mJy/beam  &Herschel-SPIRE$^{1}$\\
160\,$\mu$m  &  $11\arcsec$& 6.9\,mJy\,pix$^{-2}$ &Herschel-PACS$^{2}$\\
100\,$\mu$m  &  $7\arcsec$ & 2.6\,mJy\,pix$^{-2}$ &Herschel-PACS$^{2}$\\
70\,$\mu$m  &  $18\arcsec$ & 10\,$\mu$Jy\,arcsec$^{-2}$&Spitzer-MIPS$^{3}$\\
6570\AA{}\,(H$\alpha$)& $2\arcsec$& 0.3\,cm$^{-6}$\,pc & KPNO$^{4}$ \\
HI-21\,cm   &  $17\arcsec$ &50\,K\,km\,s$^{-1}$ &  VLA$^{5}$\\
CO(2-1)     &  $12\arcsec$ & 0.06\,K\,km\,s$^{-1}$& IRAM-30m$^{5}$\\
\hline
\noalign {\medskip}
\multicolumn{3}{l}{$^{1}$ \cite{Kramer10,Xilouris}}\\
\multicolumn{3}{l}{$^{2}$ \cite{Kramer10,Boquien_11}}\\
\multicolumn{3}{l}{$^{3}$ \cite{Tabatabaei_1_07}}\\
\multicolumn{3}{l}{$^{4}$ \cite{Hoopes_et_al_97H}}\\
\multicolumn{3}{l}{$^{5}$ \cite{Gratier}}\\%\cite{Deul_87}}\\
%\multicolumn{3}{l}{$^{6}$ \cite{Gratier}}\\
\end{tabular}
\end{center}
\end{table*}
\begin{figure*}
\begin{center}
\resizebox{\hsize}{!}{\includegraphics*{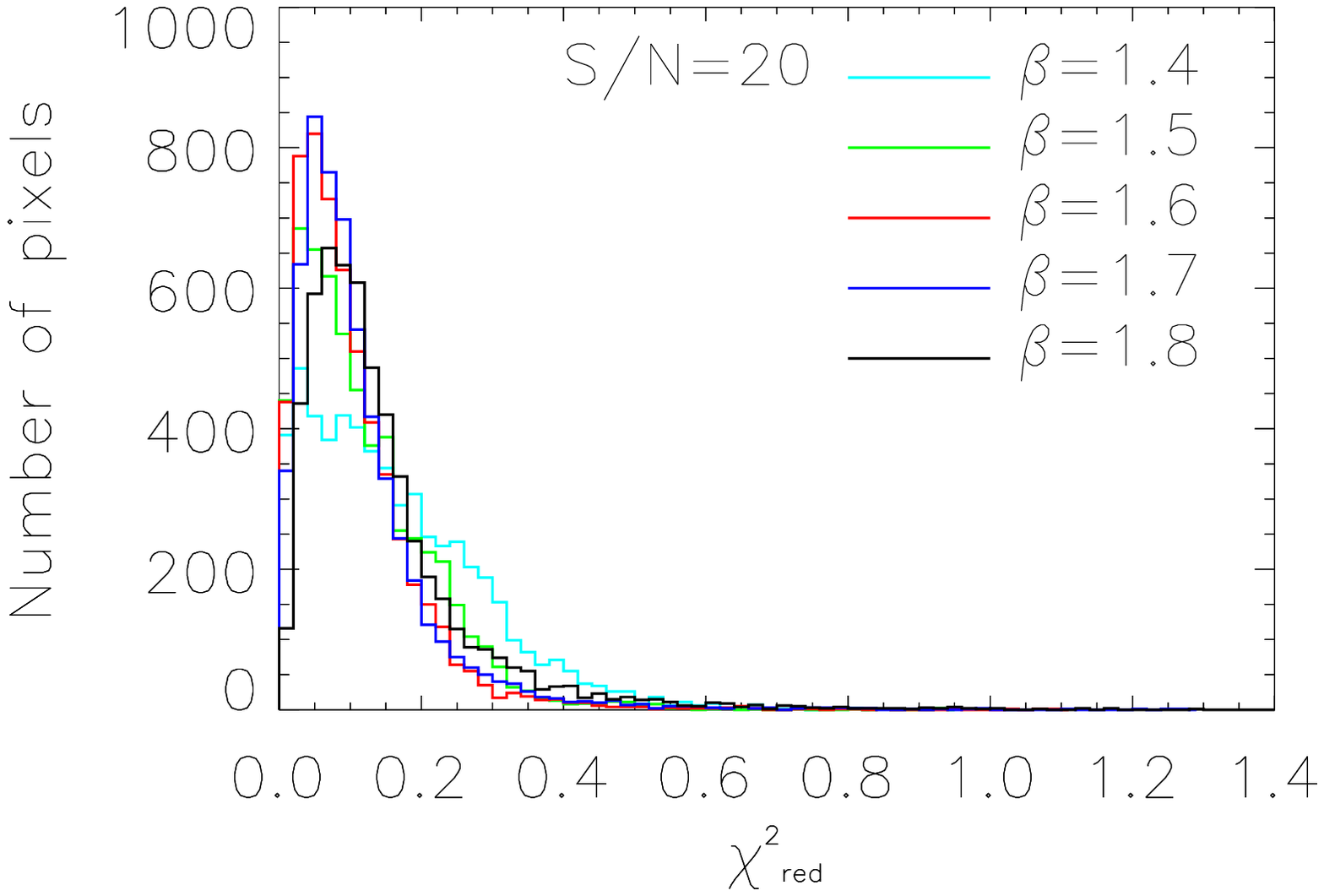}\includegraphics*{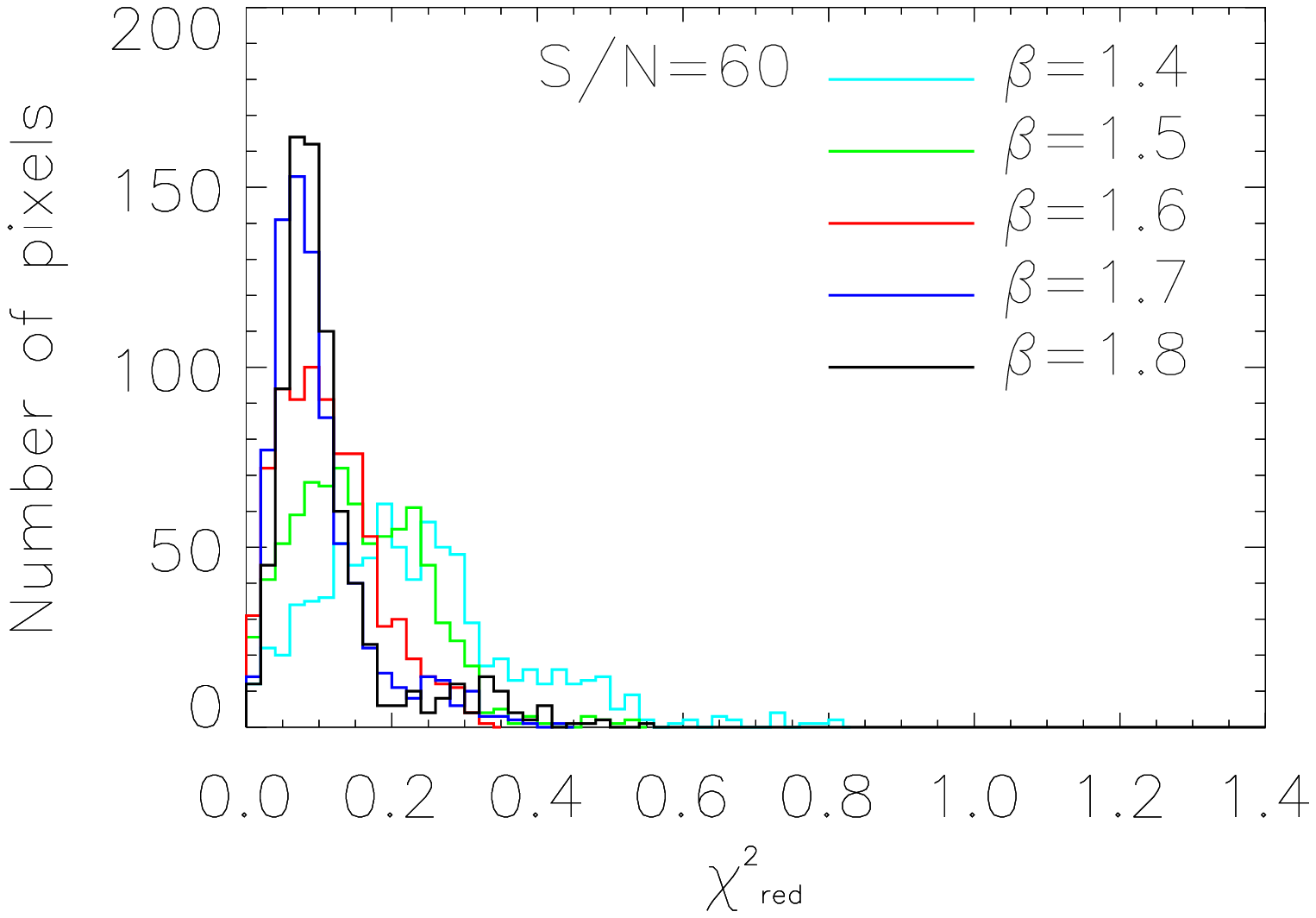}}
\caption[]{ Histogram of the reduced $\chi^2$ in the single MBB fitting approach for selected $\beta$ values between 1.4 and 1.8. Results are shown for fits to the high (20$\sigma$, {\it left}) and very high (60$\sigma$, {\it right}) signal to noise data. }
\label{fig:x20sig}
\end{center}
\end{figure*}
\section{Data}
Table~2 summarizes the data used in this study.
M33 was observed at 250\,$\mu$m, 350\,$\mu$m, and 500\,$\mu$m  using the SPIRE photometer \citep{Griffin} onboard the Herschel space telescope. The data were reduced using the Herschel data reduction software HIPE 7.0 as detailed in \cite{Xilouris}. We also used the 160\,$\mu$m  data taken with the Herschel PACS instrument \citep{Poglitsch}.   The PACS {  data were} reduced using the scanamorphos algorithm \citep{Roussel_12} as discussed in detail in \cite{Boquien_11}. {  At shorter wavelengths, we used the 70\,$\mu$m Spitzer MIPS data in the two-component MBB approach\footnote{  The PACS data at 100\,$\mu$m were not used due to {  their} low sensitivity compared to the MIPS data \citep[see e.g.][]{Aniano_12,Hinz_12}. In the single-component MBB approach, we used the PACS data at 100\,$\mu$m to reject SED fits having an excess of 100\,$\mu$m emission compared to the observations.}\citep{Tabatabaei_1_07}}.

{  The maximum calibration uncertainties (i.e., for diffuse emission) is 15\% for the MIPS \citep{Carey}, 20\% for the PACS \citep{Poglitsch}, and 15\% for the SPIRE bands \citep{Griffin}.}

We also used  H$\alpha$, HI, and CO(2-1) line emission data to investigate the connection between the dust spectrum and tracers of the ionized and neutral gas phases. The CO(2-1) line emission was  observed with the IRAM-30m telescope and detailed in \cite{Gratier}. The HI-21\,cm line was mapped with the VLA \citep{Gratier}.   The H$\alpha$ data is from the Kitt Peak National Observatory (KPNO) \citep{Hoopes_et_al_97H}.
% The Gratier et al 2010 HI data are MUCH better than Deul (sensitivity, resolution, and total flux).
The maps used in our analysis were all convolved to the resolution of the 500\,$\mu$m SPIRE image ($\sim 40\arcsec$) by using the dedicated convolution kernels provided by \cite{Aniano_11} and projected to the same grid of $10\arcsec$ pixel and center position. 

\begin{figure}
\begin{center}
\resizebox{7.2cm}{!}{\includegraphics*{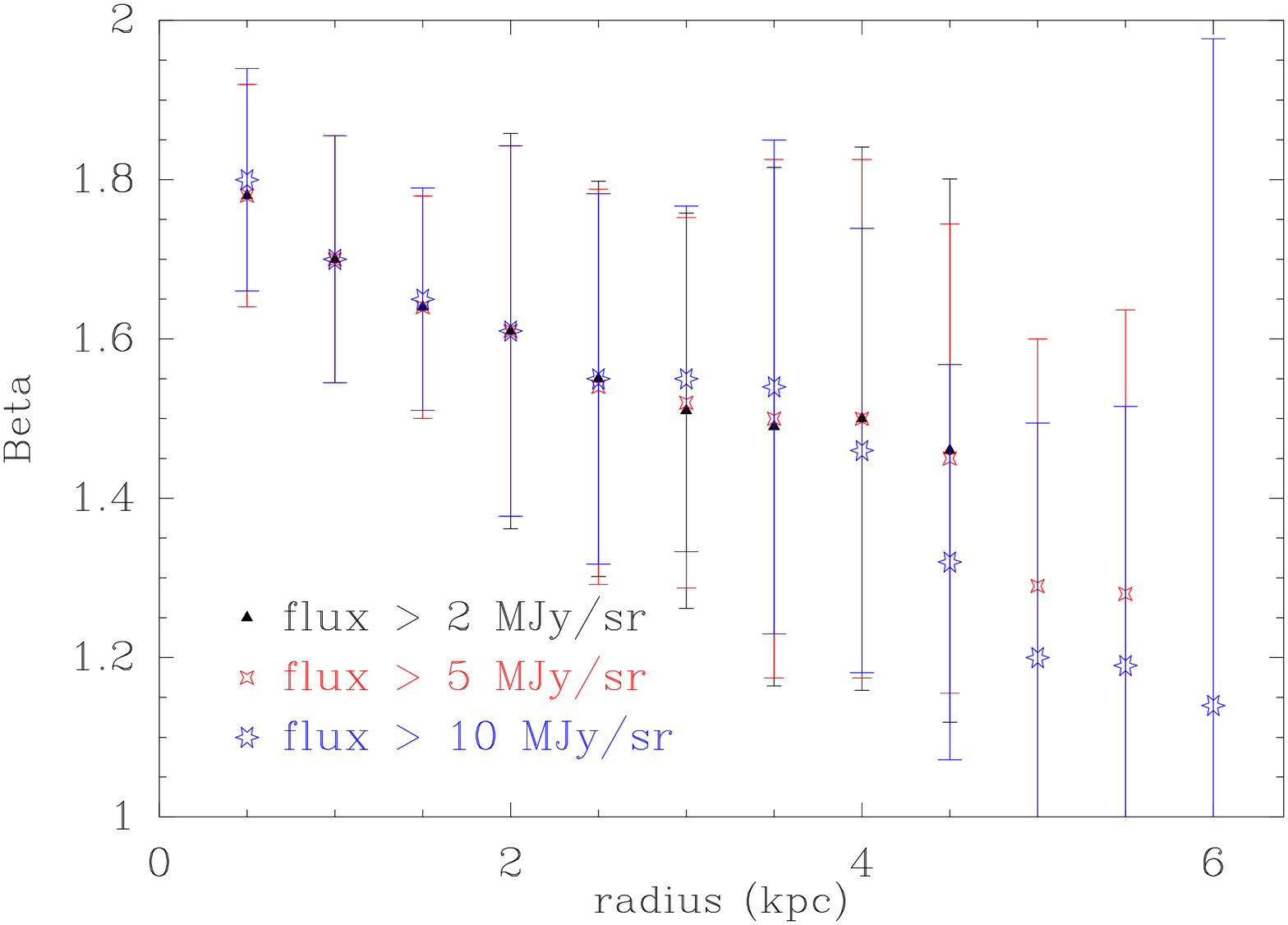}}
\resizebox{7.2cm}{!}{\includegraphics*{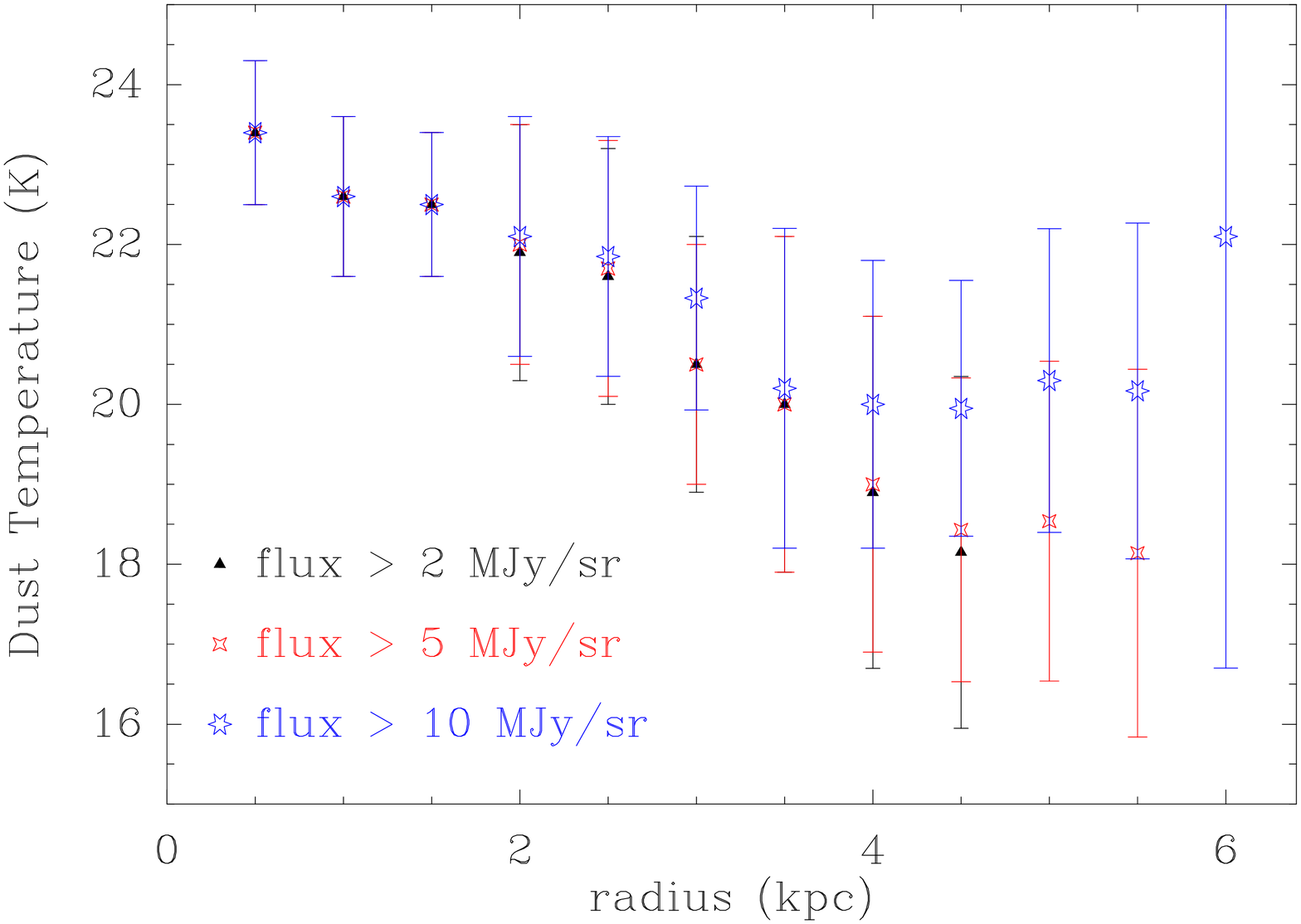}}
\caption[]{ Radial variation of the `best-$\beta$' calculated by minimizing the reduced $\chi^2$ for each of the annuli  in the single modified black body (MBB) fitting approach ({\it top}).  The different symbols show a coherent decrease in $\beta$ with radius and represent values obtained for different flux cutoffs as indicated.  ({\it Bottom}): dust temperature calculated for each flux cutoff at each radius. The error bars show the 1 sigma dispersion. }
\label{fig:radial}
\end{center}
\end{figure}

\section{Methodology and Results}
We {  constrain independently} $\beta$ and $T$ by using ($a$) 
single component MBB fits where it is assumed that the grains are described by a fixed $\beta$ but a varying temperature and all physically reasonable values of $\beta$ are tested to see which $\beta$ results in the lowest residuals and ($b$)  2-component MBB models where an iterative method is applied to solve systems of independent equations  using the Newton-Raphson method.  Approach ($a$) aims to derive $\beta$ and $T$ as a function of the galactocentric radius and to look for trends with FIR luminosity.  We derive $\beta$, $T$, and  mass surface densities pixel-by-pixel across the galaxy in approach ($b$).  

\begin{figure}
\begin{center}
\resizebox{6.3cm}{!}{\includegraphics*{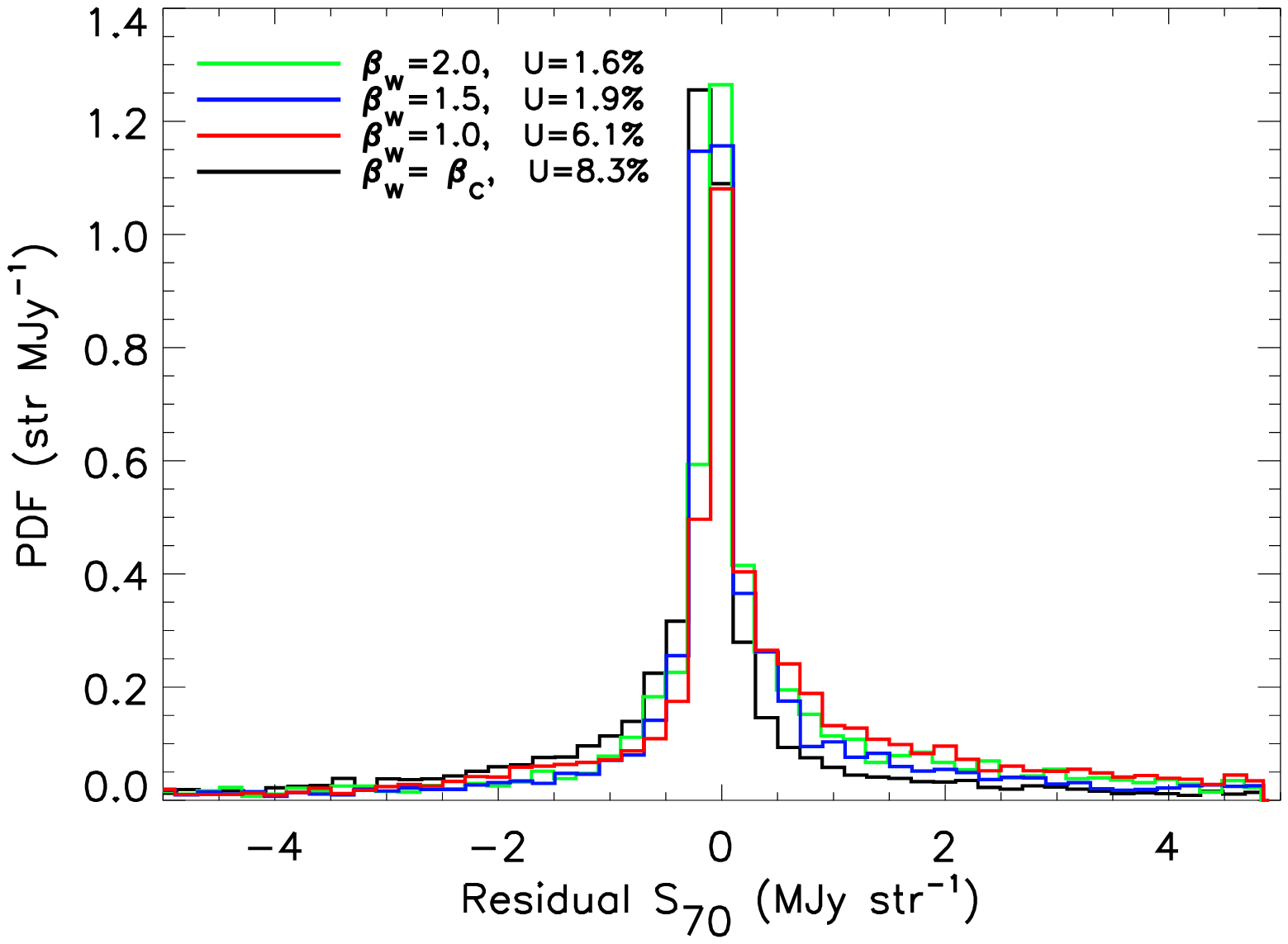}}
\resizebox{6.3cm}{!}{\includegraphics*{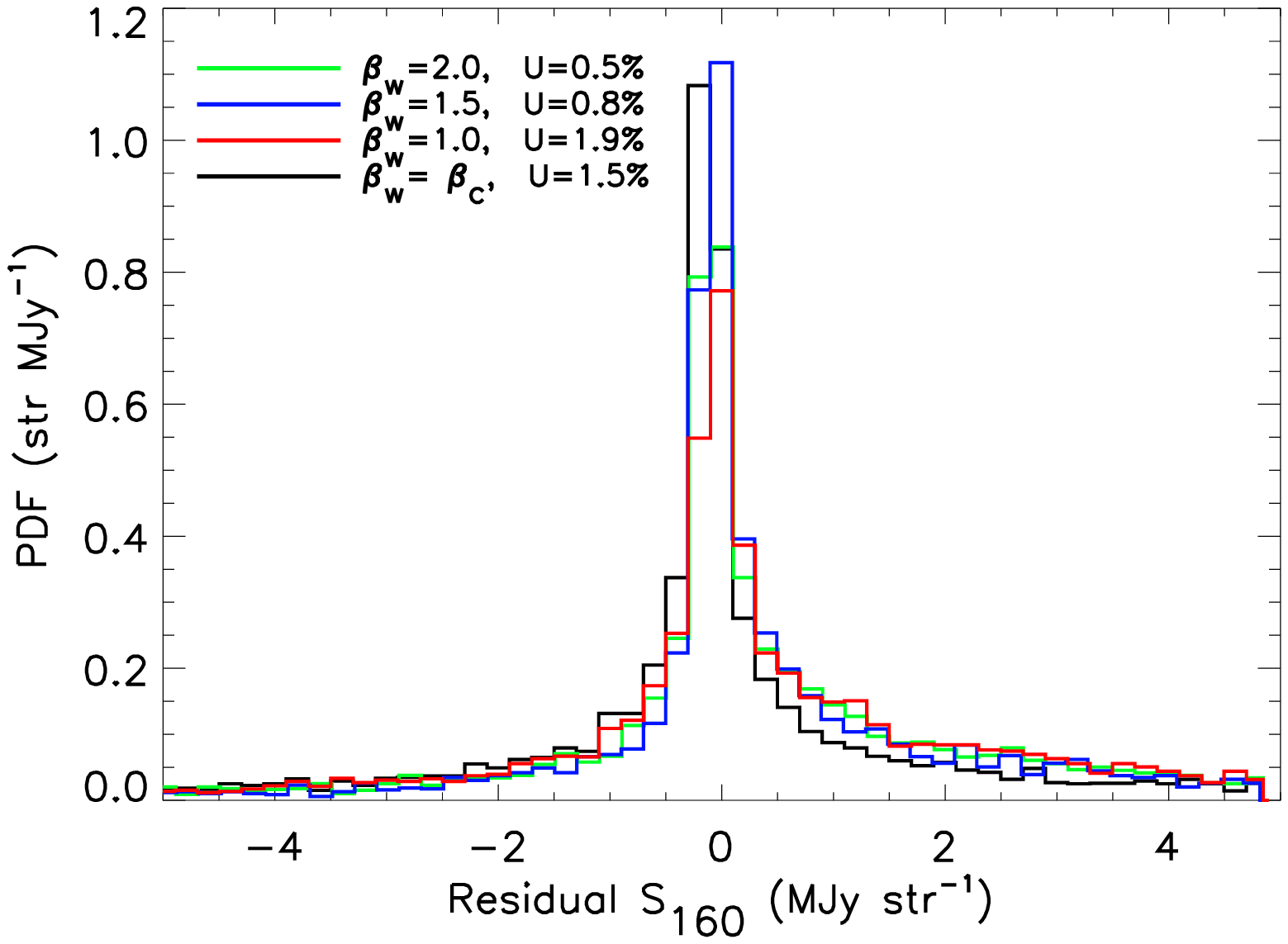}}
\resizebox{6.3cm}{!}{\includegraphics*{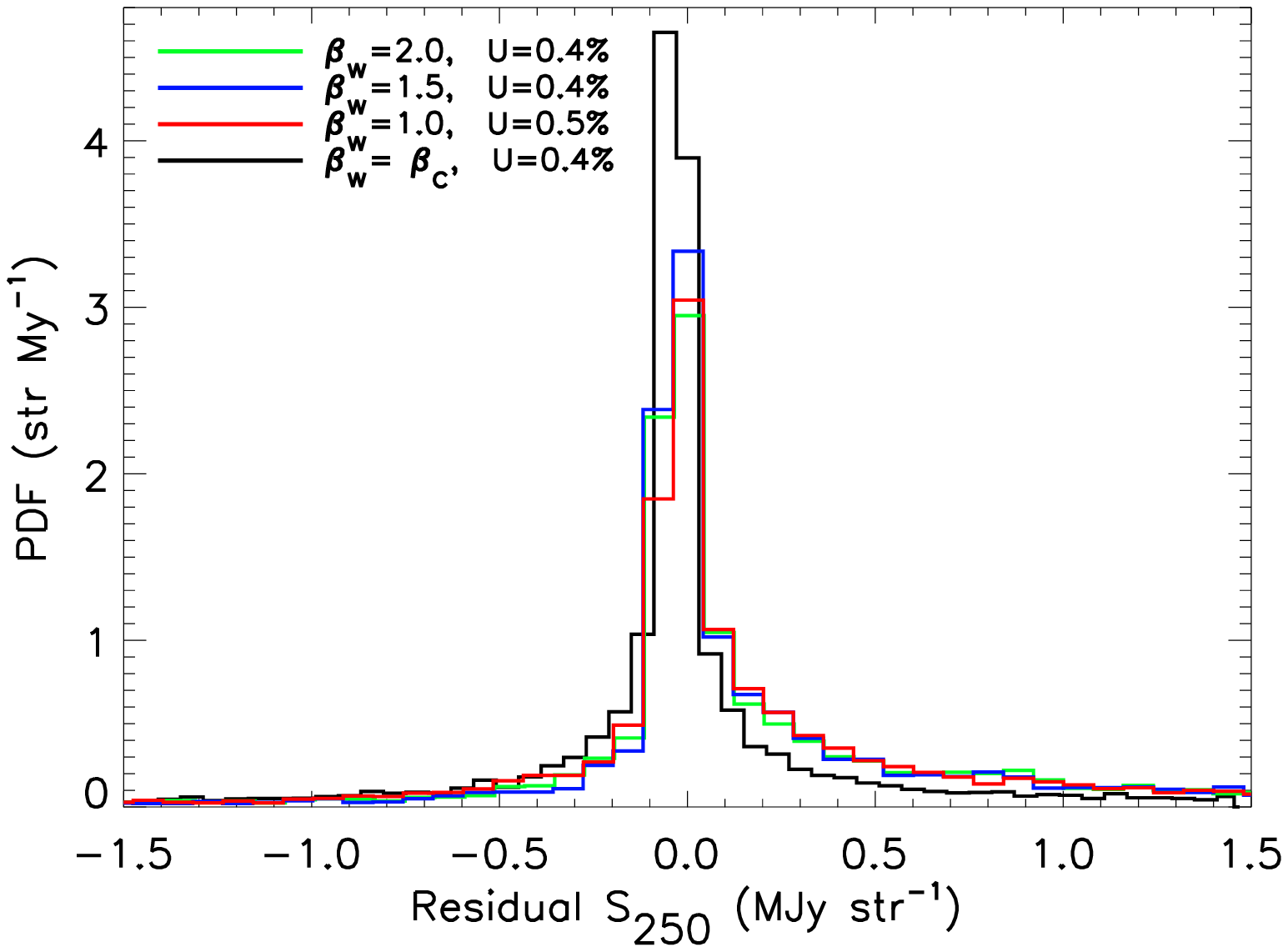}}
\resizebox{6.3cm}{!}{\includegraphics*{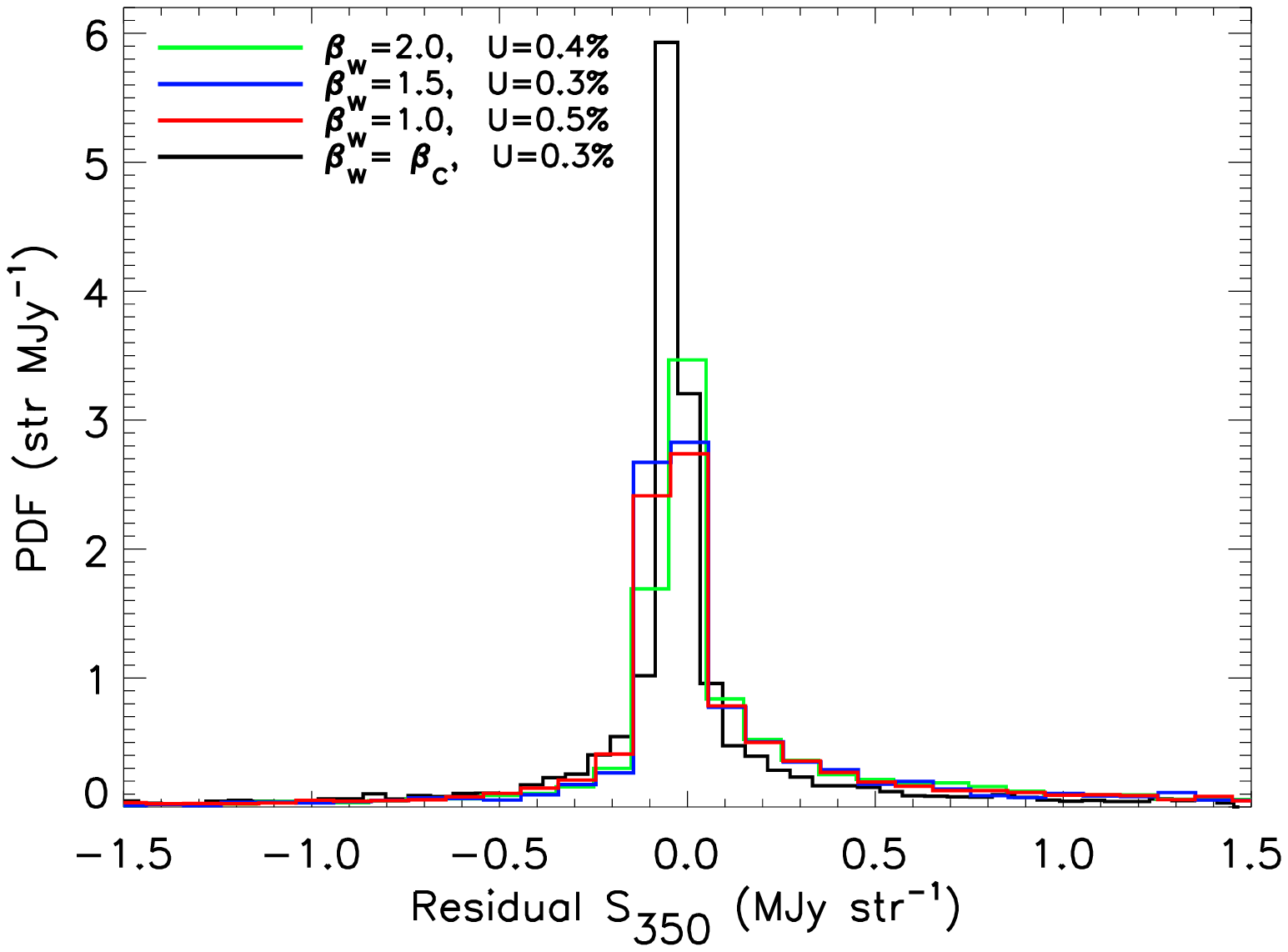}}
\resizebox{6.3cm}{!}{\includegraphics*{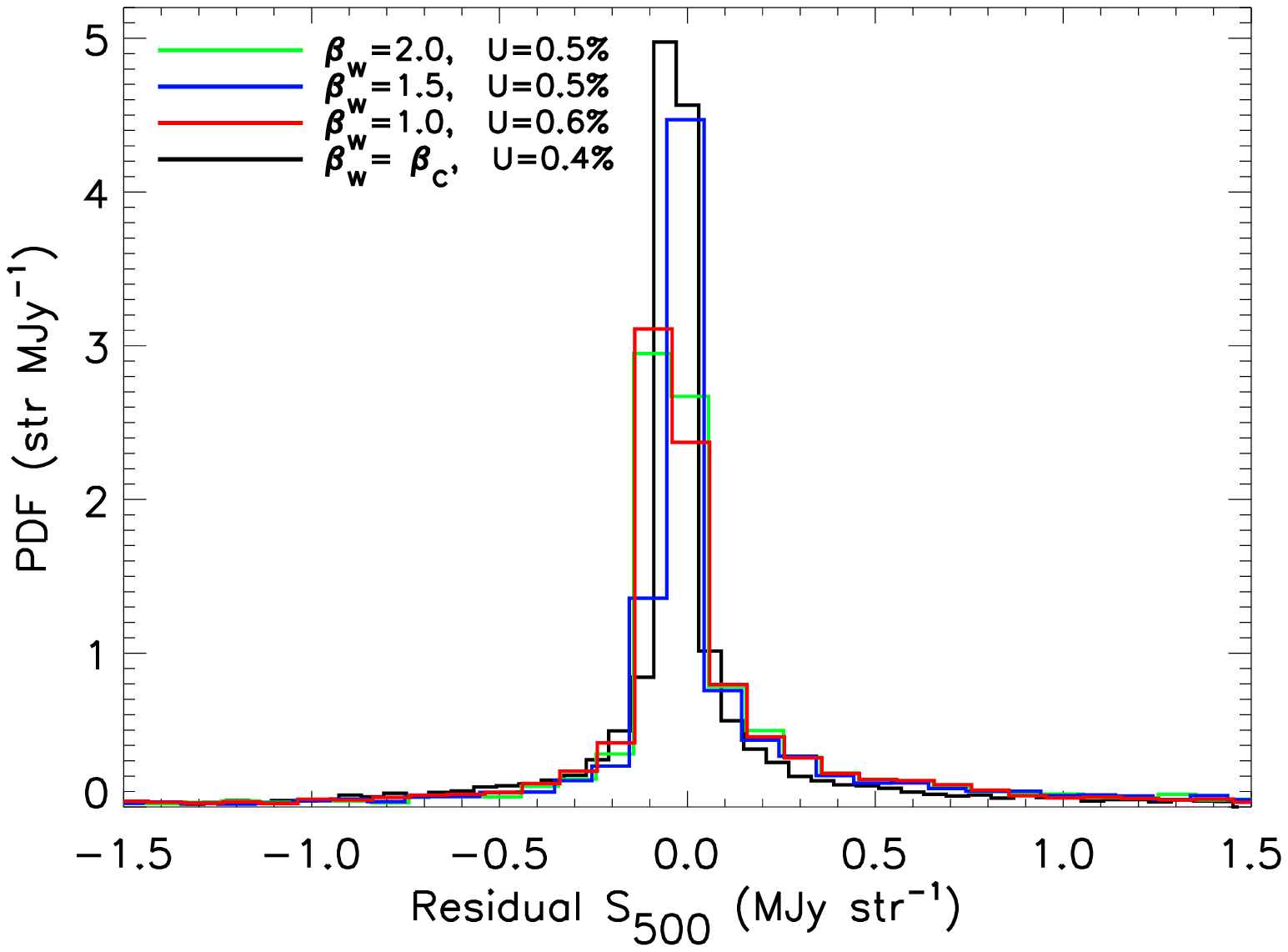}}
\caption[]{Probability density function (PDF) of residuals between observations and model for the two-component MBB approaches at ({\it from top to bottom}) 70, 160, 250, 350, and 500\,$\mu$m. The uncertainty  of each approach (U) is also given for comparison.}
\label{fig:U1}
\end{center}
\end{figure}

\subsection{Single-component approach}
In the single-component MBB model (Eq.~1),
it is assumed that  the grain properties do not change over the region or structure studied so for each pixel we fit only the temperature (and the optical depth, which acts as a general scaling).  The temperature is fit for a fixed $\beta$ but we repeat the fitting process for many different values of $\beta$.  For a given $\beta$, the overall quality of the fit is determined by minimizing the reduced $\chi^2$ as in \cite{Xilouris} over the whole set of pixels\footnote{We minimize $\Sigma (S_{\rm obs} -S_{\rm fit})^2/S^2_{\rm fit}$ over all bands; for 4 bands fitting two parameters (temperature and optical depth), this is equivalent to a reduced $\chi^2$.}.  This process, while making the hypothesis that $\beta$, representing the grain properties, is unlikely to undergo real variations {  at the scale of} the regions studied, avoids the $\beta-T$ degeneracy while allowing us to determine the `best $\beta$'. 

For each pixel, fits are first made for the 160-500\,$\mu$m  bands where cold dust dominates \citep{Compiegne}.  {  At 100\,$\mu$m, the emission comes from a mixture of {  cold and warm dust}.  The observed 100\,$\mu$m emission thus represents an upper limit to the emission from the cool component which we fit from 160-500\,$\mu$m.}  If the 100\,$\mu$m flux derived from the fit is higher than the observed flux, then we refit using the 100\,$\mu$m flux as well, resulting in a more {  realistic} dust temperature.

Exploring the best $\beta$ for a selected set of $\beta$=1, 1.3, 1.5,
1.7 and 2, \cite{Xilouris} showed that the data from all wavelengths above the $3\sigma$ rms noise level is best fitted by $\beta$=1.5.  Similarly, we explore how the best $\beta$ value varies when selecting progressively higher S/N data.  While choosing only high-quality data is desirable as long as enough pixels are available (which is the case), it introduces a selection of regions with more and more star formation and probably also a higher fraction of molecular gas as compared to the lower S/N regions.  Comparing to the S/N=3 case \citep[][see Fig.~3 in]{Xilouris}, where $\beta=1.5$ yields the lowest  residuals, we find that the more actively star-forming high S/N regions have steeper $\beta$, e.g. $\beta=1.6$ for S/N=20 and $\beta=1.7$ for S/N=60 (see Fig.~\ref{fig:x20sig}).  

In order to better investigate the change in $\beta$, we divided M33 into radial bins  of 0.5\,kpc and made the same calculations for each $\beta$ between 1 and 2.4 with steps of 0.1, looking for the lowest residuals.  {  Figure~\ref{fig:radial}} shows that the best value for $\beta$ decreases with radius from 1.8 to about 1.2. Though differently, the dust temperature also decreases with radius.   
The higher flux cutoff (10\,MJy/sr) has a higher temperature for two reasons.  First, at higher fluxes, dust tends to be warmer.  Secondly, since the $\beta$ calculated at large radii are lower for the 10 MJy/sr cutoff, the temperatures obtained are higher.  
These results are very similar to those of Fig.~1 in \cite{Taba_12} but using entirely different numerical techniques.

\subsection{Two-component approaches}
Here we introduce a second dust component  heated mainly by young  massive  stars to a higher temperature  (warm dust) in addition to the dust heated  by the general {  interstellar radiation field (ISRF, cold dust)}. The aim of this approach is to differentiate the dust properties in arm/interarm and inner/outer disk regions, including diffuse emission.  Hence, we perform this analysis pixel-by-pixel using the most sensitive Herschel and Spitzer maps available in the wavelength range of 70$\mu$m to 500$\mu$m. 
The specific intensity of a two-component MBB dust emission is given by  
\begin{eqnarray}
S_{\nu} &=& \, B_{\nu}(T_{c}) [1\,-\,exp (-\tau_{\nu,{ c}} )] \\
& &  + \, B_{\nu}(T_{ w}) [1\,-\,exp (-\tau_{\nu,{ w} })] \nonumber,
\end{eqnarray}
where the indices $c$ and $w$ stand for the cold and warm dust, respectively. 
%We derived the optical depths of the cold and warm dust components as $\tau_{\nu, {\rm c}}= \kappa_{\nu,c} \Sigma_{\rm dust,c}$ and $\tau_{\nu, {\rm w}} = \kappa_{\nu,w}\, \Sigma_{\rm dust,w}$. 
The Newton-Raphson iteration method from Numerical Recipes was used to solve the 5 equations corresponding to the five wavelengths at 70, 160, 250, 350, and 500\,$\mu$m in order to derive $T_c$, $T_w$, $\Sigma_{{\rm dust},c}$, $\Sigma_{{\rm dust},w}$, and $\beta$. 
\begin{figure}
\begin{center}
\resizebox{7cm}{!}{\includegraphics*{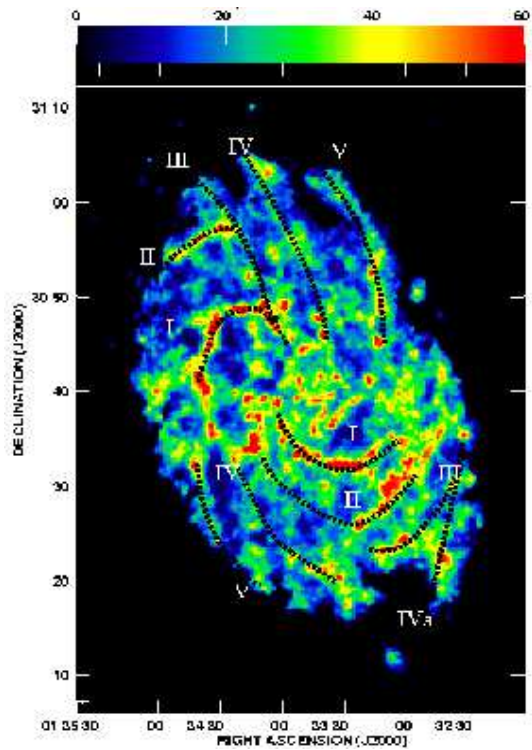}}
\caption[]{  Distribution of the cold dust mass surface density in M33. A sketch of the optical spiral arms \citep{Sandage} is overlaid on top of the cold dust map. The {  wedge} gives the surface density units in $\mu$g\,cm$^{-2}$.  }
\label{fig:column}
\end{center}
\end{figure}

\begin{figure*}
\begin{center}
\resizebox{7cm}{!}{\includegraphics*{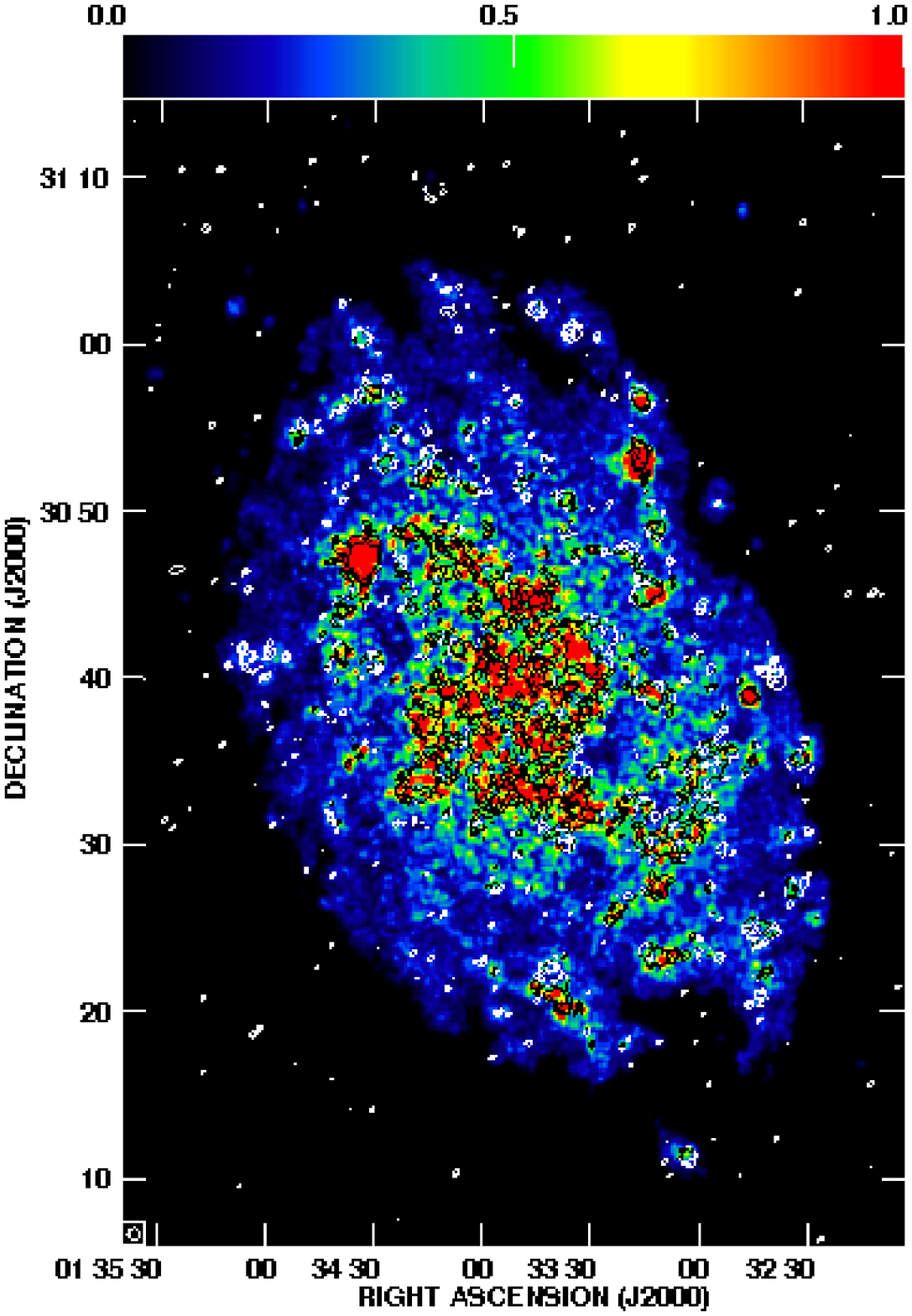}}
\resizebox{7.3cm}{!}{\includegraphics*{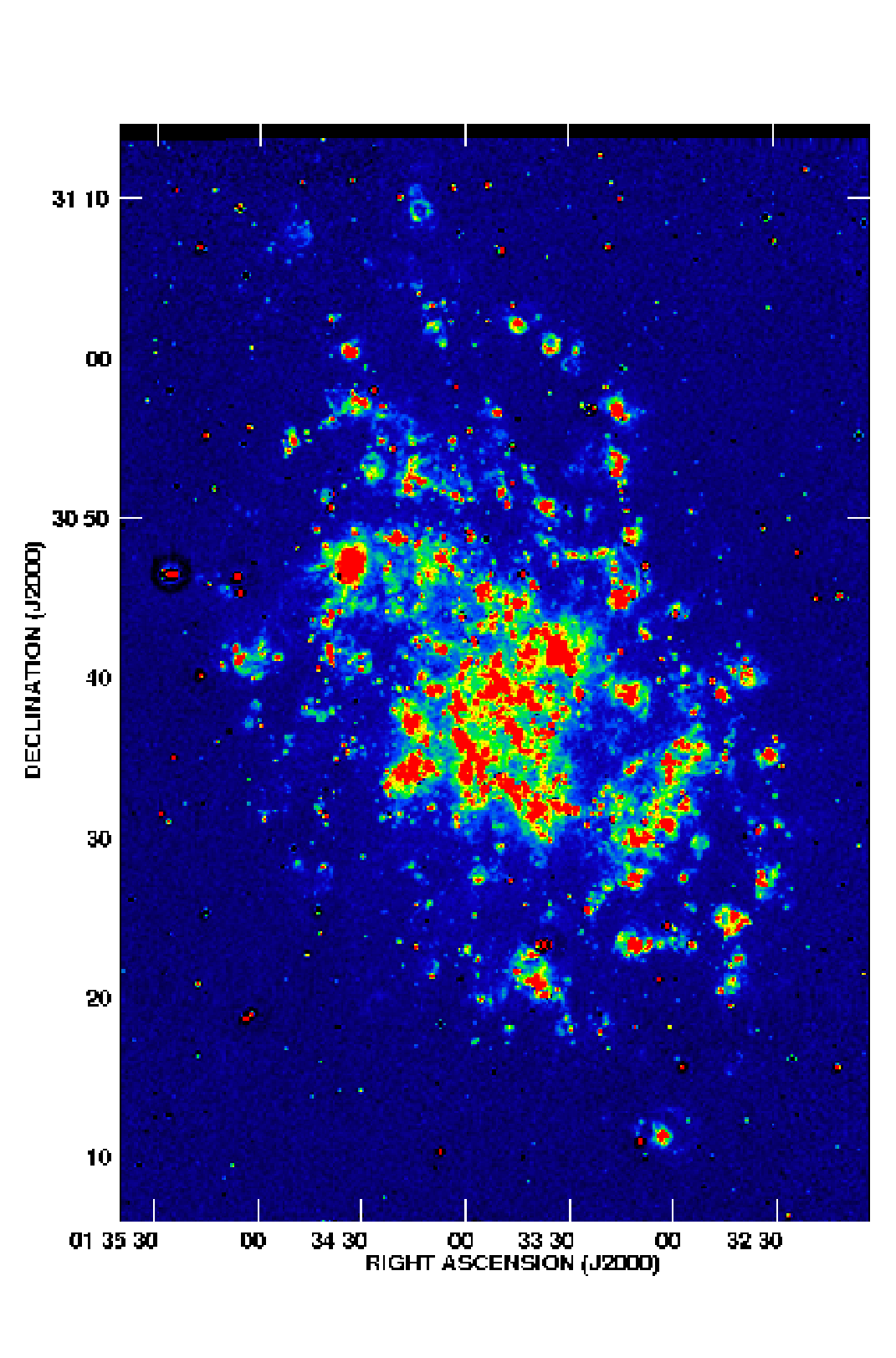}}
\caption[]{  {\it Left}: Distribution of the warm  dust mass surface density in M33. Overlaid are contours of the H$\alpha$ emission, shown in black or white for a better contrast against background.  The {  wedge} gives the surface density units in $\mu$g\,cm$^{-2}$. {\it Right}:  The H$\alpha$ map of M33.}
\label{fig:column2}
\end{center}
\end{figure*}

In order to decrease the probability of falling into a local minimum of the merit function, the initial guess for the physical parameters is determined through synthesizing the observed intensities.
We first generate uniform samples of 5000 random sets of variables corresponding to each physical parameter: cold and warm dust temperatures, mass surface densities, and $\beta$. Each random set of physical parameters is then used to synthesize intensities given by {  Eq. (3)} at each wavelength. To define the initial values for the Newton-Raphson method, 20 parameter sets leading to the synthesized intensities closest to the observed values are selected  by maximizing the likelihood.  Thus, the Newton-Raphson method is applied 20 times for each pixel of the maps resulting in 20 sets of solutions out of which the set minimizing the merit function ($\equiv \sum \frac{(S_{\rm model} -S_{\rm obs})^2}{\sigma^2_{\rm obs}}$, {  where $\sigma_{\rm obs}$ is the corresponding calibration uncertainty}) was taken as the final set of solutions. {  The initialization scheme is completely independent for every two pixels. Hence, there is no correlated error due to this procedure. }
\begin{figure}
\begin{center}
\resizebox{7cm}{!}{\includegraphics*{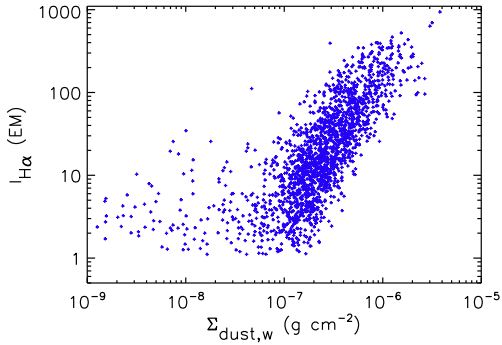}}
\caption[]{  Linear correlation between the warm dust surface density $\Sigma_{{\rm dust},w}$ and the H$\alpha$ emission measure. The scatter in the low surface density tail coincides with the systematic of the approach for this range of surface density (see Sect.~4).}
\label{fig:ha_warm}
\end{center}
\end{figure}

Considering the cold and warm dust, different treatments for $\beta$ are used: \\
I) Both components of dust are assumed to have the same variable emissivity index ($\beta_{c}=\beta_{w}$=$\beta$).  \\
II) As most of the dust is in the cool phase, the variable $\beta$ is attributed to this major component ($\beta_{c}=\beta$), while the warm dust is assumed to emit with a fixed emissivity index across the disk. Although we are left with a fixed value of $\beta_{w}$ (note that the number of unknowns must be equal to the number of equations), we repeat the analysis for different possibilities of $\beta_w$=1, 1.5, and 2.\\

The above calculations are all performed for every pixel of the maps with intensities higher than the 3$\sigma$ noise level at each wavelength.   The pixel-by-pixel analysis further helps to reduce dust temperature mixing as the range of temperatures is limited in a given pixel. Using a Monte-Carlo simulation,  the two-component MBB approaches are then synthesized to determine the systematics and reliability of each {  method}. 
\begin{figure}
\begin{center}
\resizebox{7cm}{!}{\includegraphics*{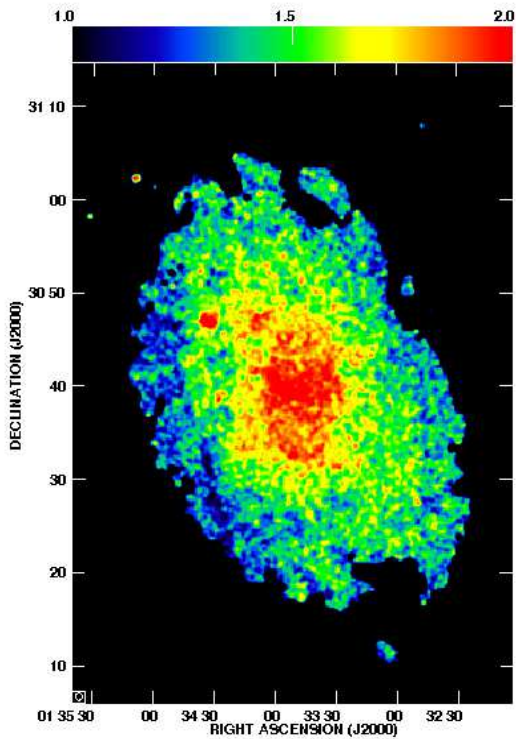}}
\caption[]{Distribution of the dust emissivity index $\beta$ in the disk of M\,33 obtained using the two-component approach with $\beta_w$=2.}
\label{fig:beta}
\end{center}
\end{figure}
The temperature and surface density of the cold dust do not depend on the choice of $\beta_w$ but the physical parameters of the warm dust  are best reproduced using the $\beta_w$=2 model (see Sect.~4). This agrees with \cite{Casey}  who derived a mid-infrared power law slope of $\simeq$2 indicative of optically-thin dust with a shallow radial density profile from  clumpy, hot  regions. 
Hence, we focus on the results of the two-component MBB with $\beta_w$=2, although the general distributions of the physical parameters  across the galaxy  are similar to those of other two-component MBB models. The PDF (probability density function) of residuals between the observed and modeled intensities (Fig.~\ref{fig:U1}) further shows the preference for the $\beta_w$=2 case for the warm dust component as it results in a smaller uncertainty U\footnote{U is defined as the mean of relative residuals  $\mid~\frac{S_{\rm obs}-S_{\rm model}}{S_{\rm obs}}~\mid$ over all pixels.} at 70$\mu$m than the other 2-component MBB models (see the U values in {  Fig.~\ref{fig:U1}}, upper panel). 

\subsubsection{Distribution of cold and warm dust}
Figures~\ref{fig:column} and \ref{fig:column2}\footnote{  The maps of the physical parameters shown in Figures~\ref{fig:column}, \ref{fig:column2}, \ref{fig:beta}, and \ref{fig:temper} are slightly smoothed to improve legibility.} show the resulting distributions of the mass surface densities of the cold and warm dust for the $\beta_w$=2 case. The cold dust has been found not only in the main spiral arms (arm I in the north and the south), but also in the weaker arms, flocculent structures and dust lanes. Dense patches of cold dust with surface densities of $\Sigma_{{\rm dust},c} >$\,45\,$\mu$g\,cm$^{-2}$ occur along  arm I and some parts of arms II and  III (in both the north and the south).
% \citep[a sketch of the spiral arms has been shown in Fig.~3 of][]{Tabatabaei_2_07}. 
Moreover, filaments of  dense dust have been detected in the central 2.5\,kpc.
The densest region corresponds to the high concentration of dust in the giant HII/star forming complex NGC~604. 
%This agrees with the strong CO(2-1) emission from this object \citep{Gratier}.
   
The warm dust is mostly concentrated in the central 2.5\,kpc and along the main spiral arm I showing  clumpy structures. A good match between  clumps of warm dust and star forming/HII regions is found by a comparison with the H$\alpha$ {  emission  (Fig.~\ref{fig:column2}, right)}. Apart from the central region and the main arms, the rest of the galaxy is covered by many spotty warm dust features, most of them can be found in the H$\alpha$ map as well.   Therefore, not only the bright HII regions (like NGC\,604, IC\,133, NGC\,595, NGC\,588, and B690) but also the weaker ones can be traced in the warm dust map. 

{  In addition to the tight spatial correlation between H$\alpha$ emission and the warm dust in Fig.~\ref{fig:column2},  Fig.~\ref{fig:ha_warm} shows that the warm dust column and the H$\alpha$ emission are proportional, with $I_{\rm H\alpha} \propto (0.92\pm\,0.02) \Sigma_{{\rm dust},w}$.  The correlation is best (largest Pearson correlation coefficient, see Sect.~6) when $\beta_w=2$ and this is one of the reasons for preferring the $\beta_w=2$ model over $\beta_w=1$ and 1.5.  The Monte-Carlo simulation presented in Sect.~4 provides further support for $\beta_w=2$.}

The average mass surface density of the warm dust in star forming complexes in the central 4\,kpc is  $\Sigma_{{\rm dust},w}= 0.8\pm 0.3\,\mu$g\,cm$^{-2}$. {  The cold-to-warm dust mass ratio has a gaussian distribution (in log) which peaks at $\sim$100, typical of spiral galaxies \citep{Misiriotis_06,Taba_cold_13}, and a standard deviation of one {\bf order of magnitude}. We stress that this result is derived without any pre-assumption about the filling factor of the cold or warm dust.  }

\subsubsection{Distribution of the dust emissivity index $\beta$}
Figure~\ref{fig:beta} shows that $\beta$ varies between  1.2 and 2 across the galaxy, in agreement
with Fig.~\ref{fig:radial}. 
The mean value of the dust emissivity index over the disk is $\beta=1.5\pm\,0.2$. This is in agreement with $\beta$ derived globally by \cite{Kramer10} using the integrated SED and \cite{Xilouris} by exploring the best $\beta$ described in Sect.~3.1.  As in the inner disk, a steep dust spectrum with $\beta \simeq2$ is found in NGC604. Out to NGC 604, the main spiral arms are visible in the $\beta$ map having a higher index ($\beta>$1.5) than their adjacent inter-arm regions with smaller $\beta$. This confirms the higher $\beta$  for the high S/N cutoffs derived {  for these radii} in the single-component MBB approach (Sect.~3.1). {  All over the $\beta$ map, there are small-scale fluctuations with variations of $\lesssim$0.2  which could be due to noise or systematics of the approach (Sects.~4 and 5).}  

\subsubsection{Cold and warm dust temperatures}
\begin{figure*}
\begin{center}
\resizebox{13cm}{!}{
\includegraphics*{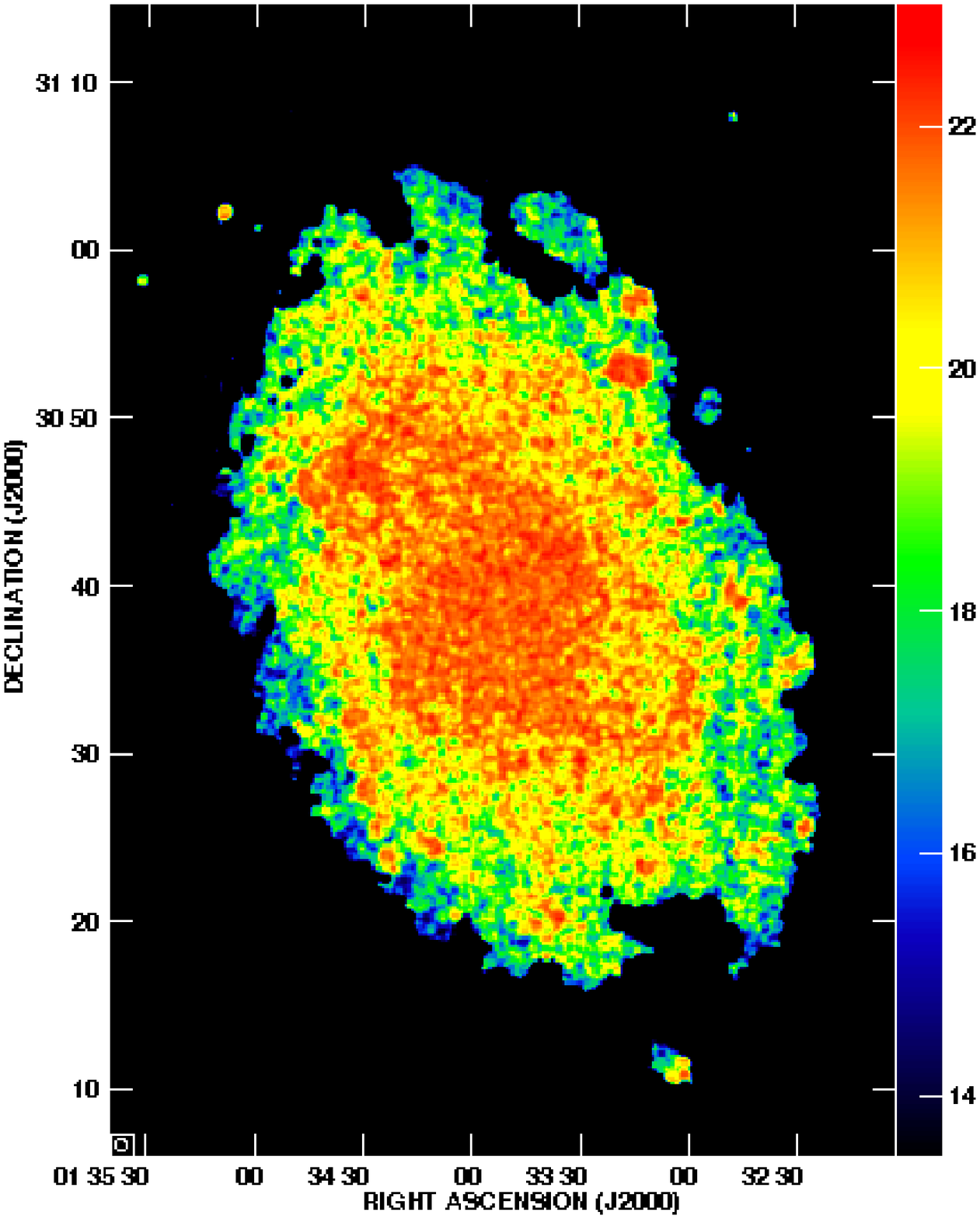}
\includegraphics*{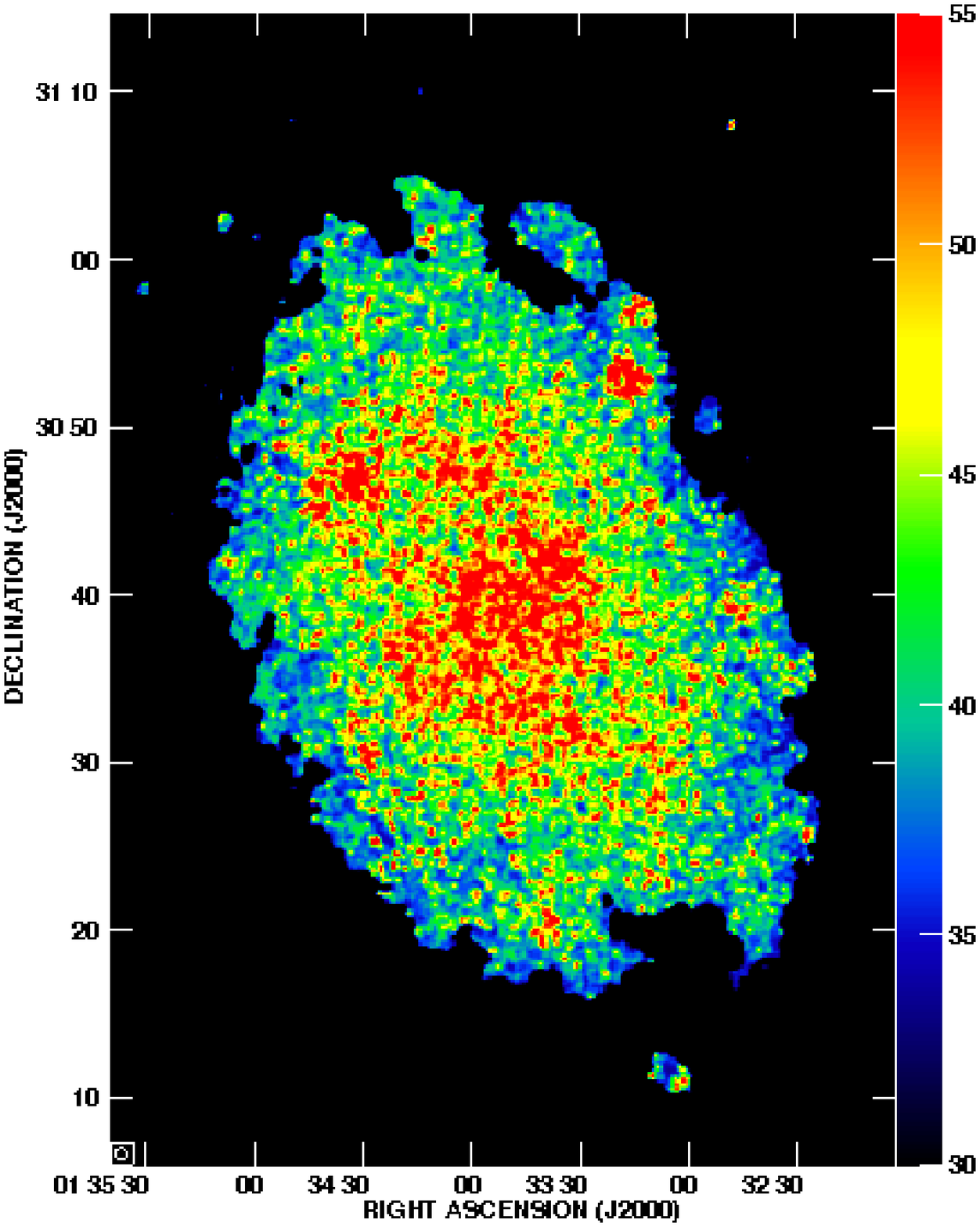}}
\caption[]{Distribution of the {  cold ({\it left}) and warm ({\it right}) dust temperatures} in M33. The {  wedge} shows the temperature values in Kelvin.}
\label{fig:temper}
\end{center}
\end{figure*}

The temperatures of both cold and warm dust components are in general lower in the outer disk than in the inner disk of M33. 
In the central 4\,kpc, the cold dust temperature $T_{c}$ varies mainly between 20\,K and 24\,K with an average of 21$\pm$2\,K (Figs.~\ref{fig:temper} and \ref{fig:radial}). In the outer parts, $T_{c}$ ranges between 14\,K and 19\,K with a mean value of 17$\pm$2\,K. The higher cold dust temperature in the inner disk with respect to the outer disk is consistent with \cite{Kramer10,Braine10,Xilouris}. A similar trend was also found in other disk galaxies based on  Herschel data \citep[e.g.][]{Pohlen,Engelbracht_10,Fritz}.  The corresponding $T_{c}$ values in HII regions agree with those derived by \cite{Relano_13} using \cite{Draine7} models and following the prescription of \cite{Galametz_12}.  
Warm dust temperatures of $T_w\gtrsim$50\,K are found in the spiral arms and star forming regions with a maximum of {  60$\pm$10\,K}  in NGC~604, NGC~595, and IC~133. 
In the outer parts, the warm dust has a temperature of {  $T_{w}=35\pm10$\,K} on the average (the errors {  given are discussed in Sect.~5}).  

\begin{figure}
\begin{center}
\resizebox{7.5cm}{!}{
\includegraphics*{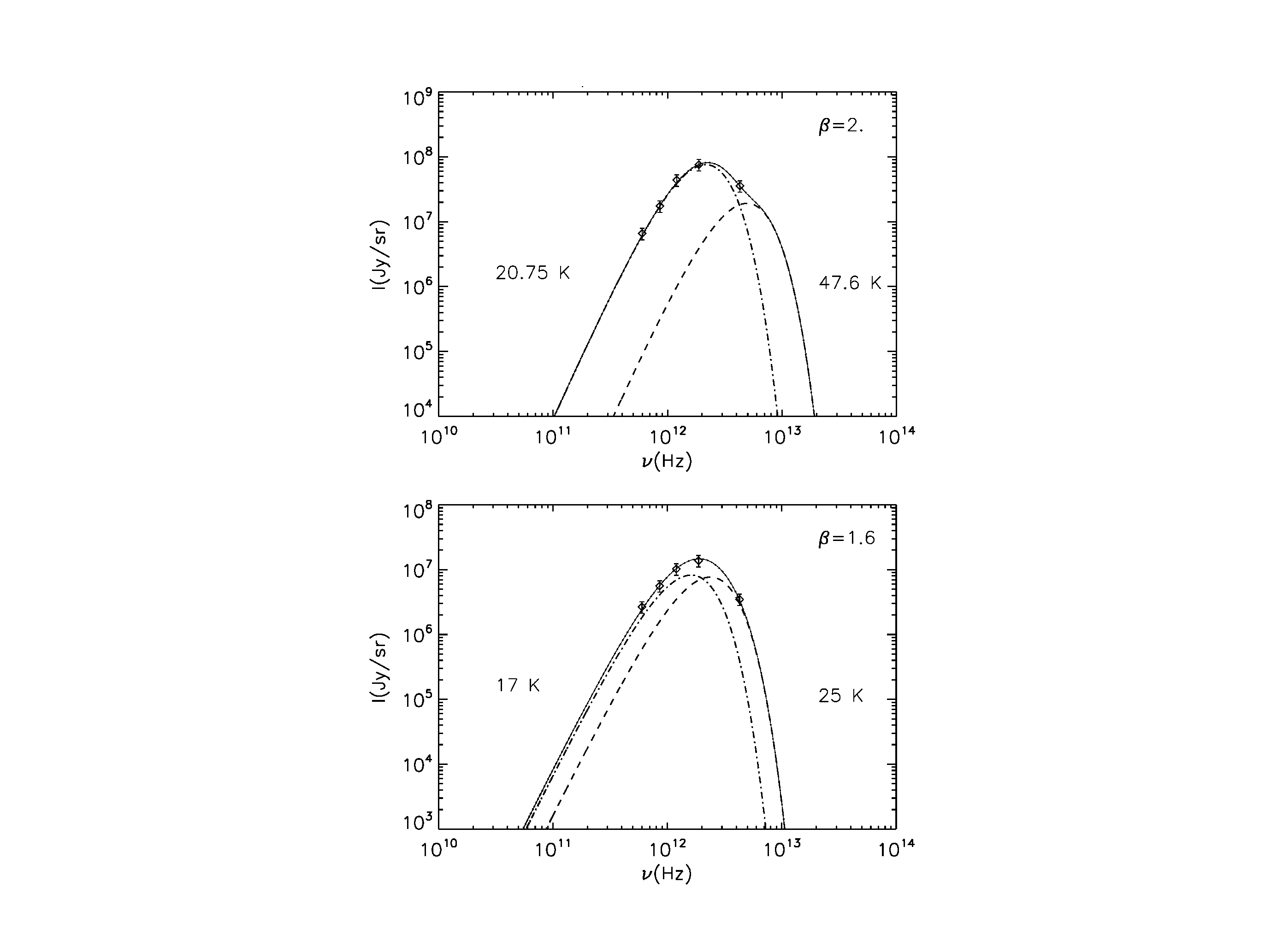}}
\caption[]{Modelled dust SED for the two-component modified black body for 2 different pixels with coordinates RA=1$^{\rm h}$\,33$^{\rm m}$\,31.35$^{\rm s}$, DEC=$30:30:21.18$ ({\it top}) and RA=1$^{\rm h}$\,33$^{\rm m}$\,17.31$^{\rm s}$, DEC=$30:21:47.99$ ({\it bottom}). The dashed, dot-dashed, and solid curves represent the SEDs of the warm, cold, and total dust emission, respectively.  }
\label{fig:sed}
\end{center}
\end{figure}

Based on the physical parameters {  retrieved}, the modelled SEDs are shown for two selected pixels with different $\beta$ in Fig.~\ref{fig:sed}. 

\section{Monte-Carlo simulations with the Newton-Raphson method}
\begin{figure*}
\begin{center}
\resizebox{15cm}{!}{\includegraphics*{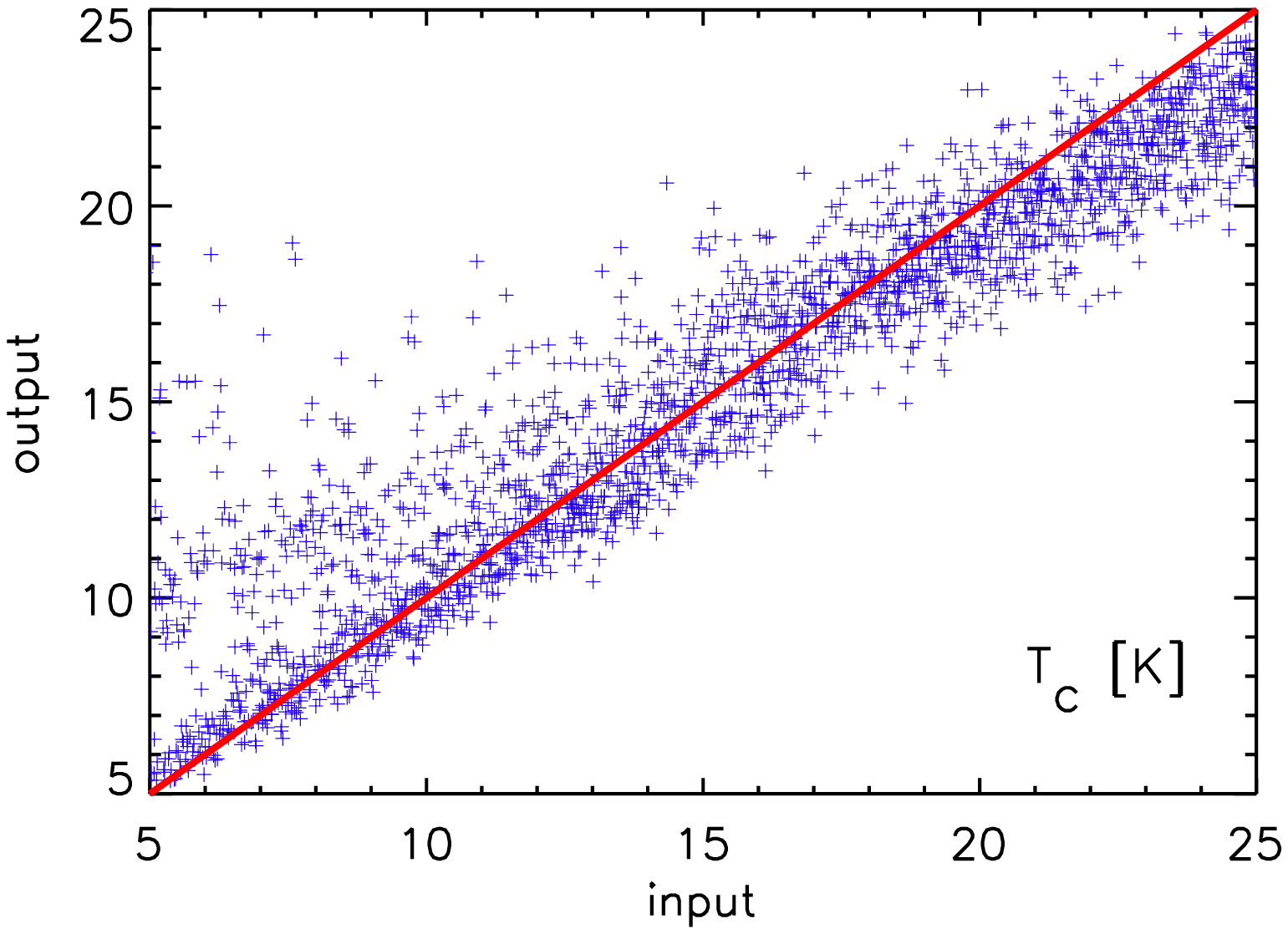}\includegraphics*{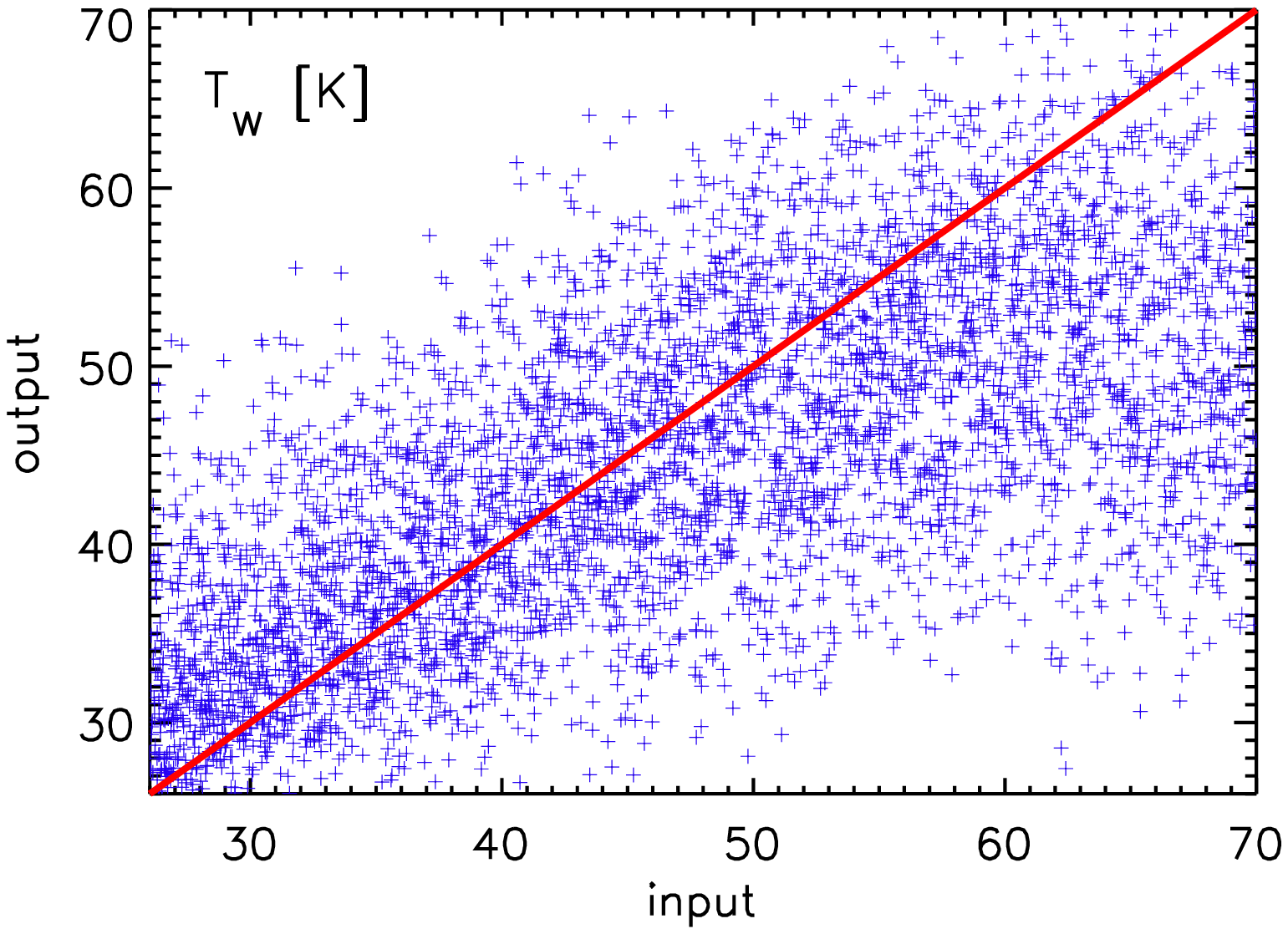}\includegraphics*{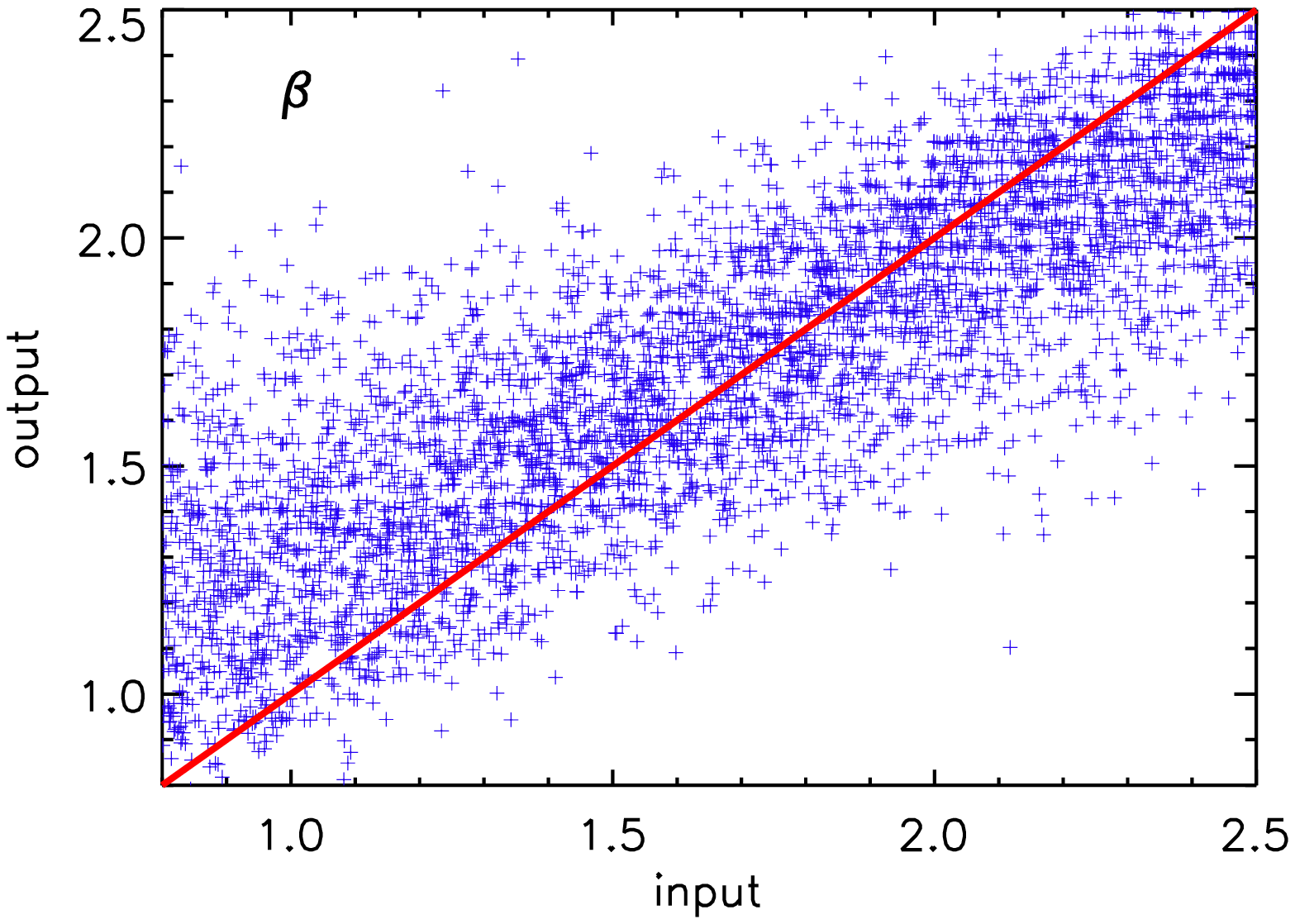}}
\resizebox{15cm}{!}{\includegraphics*{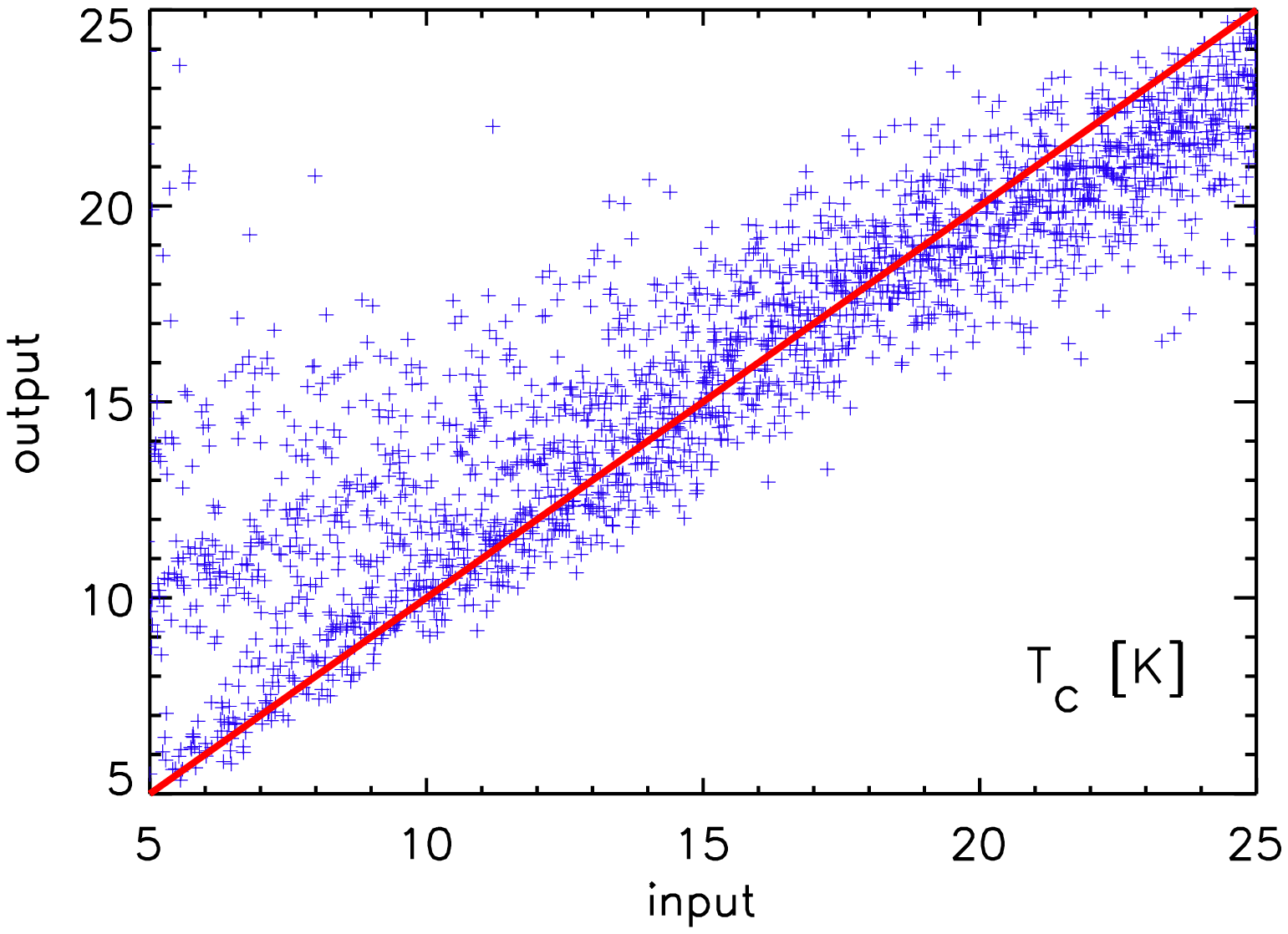}\includegraphics*{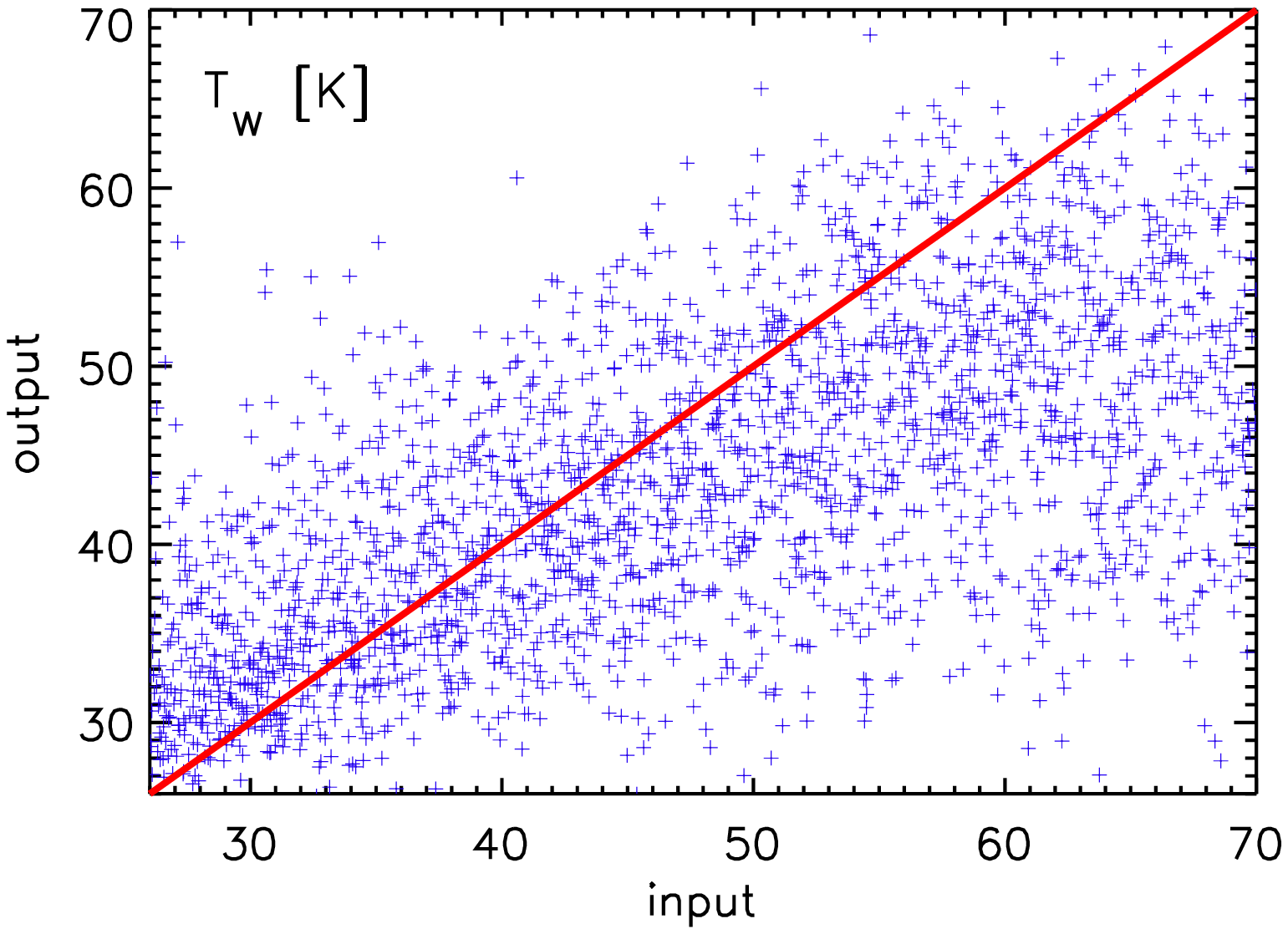}\includegraphics*{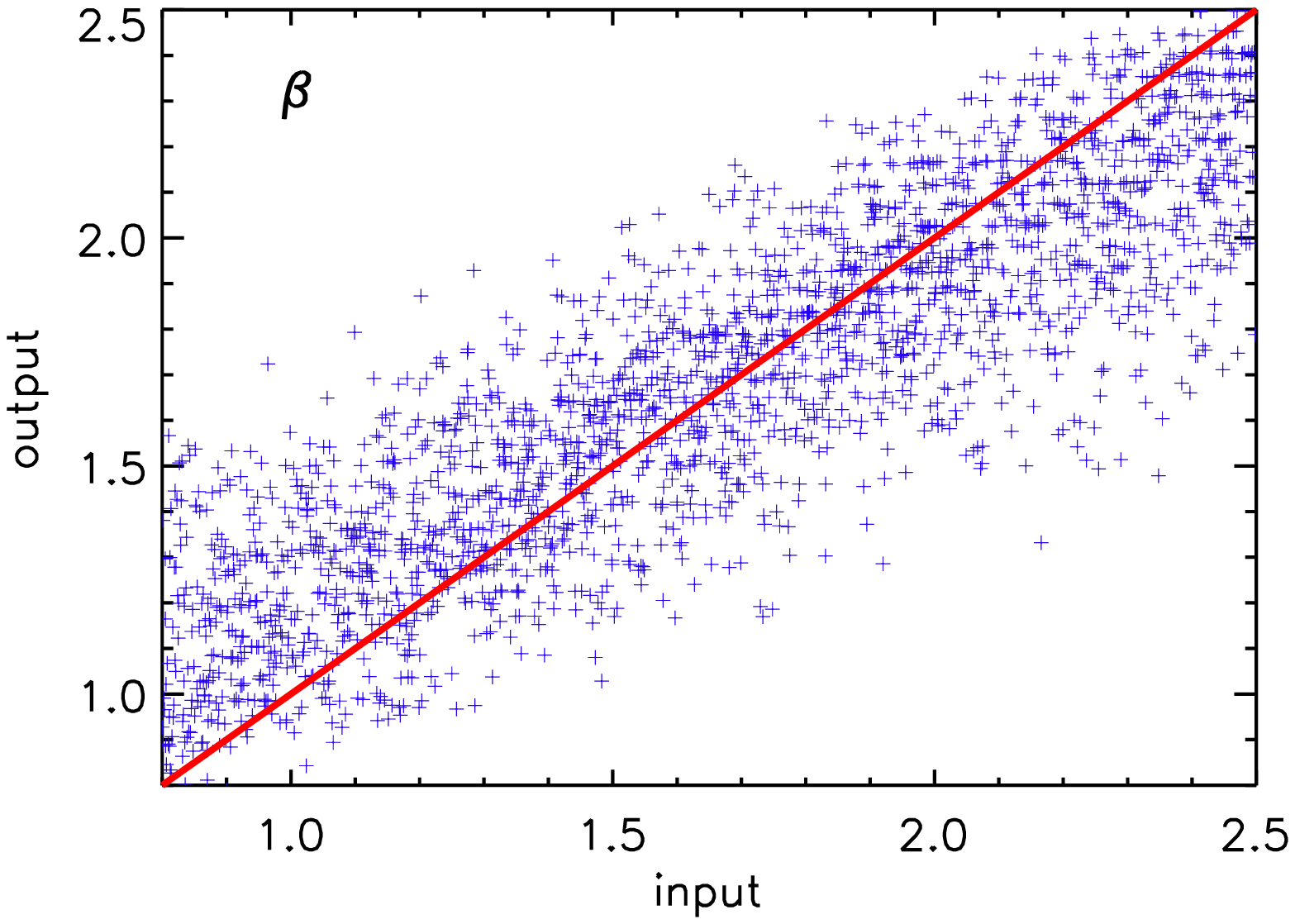}}
\resizebox{15cm}{!}{\includegraphics*{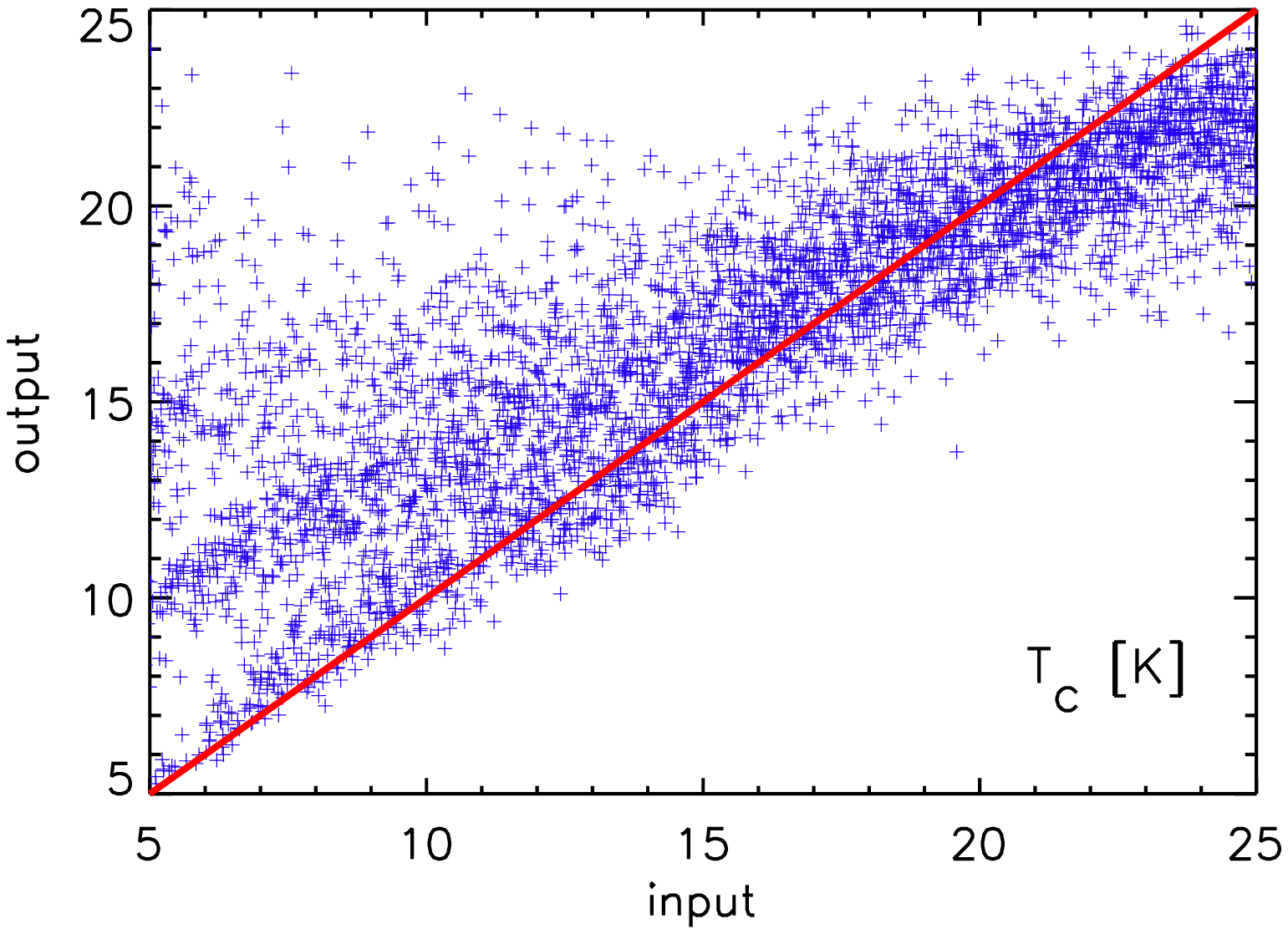}\includegraphics*{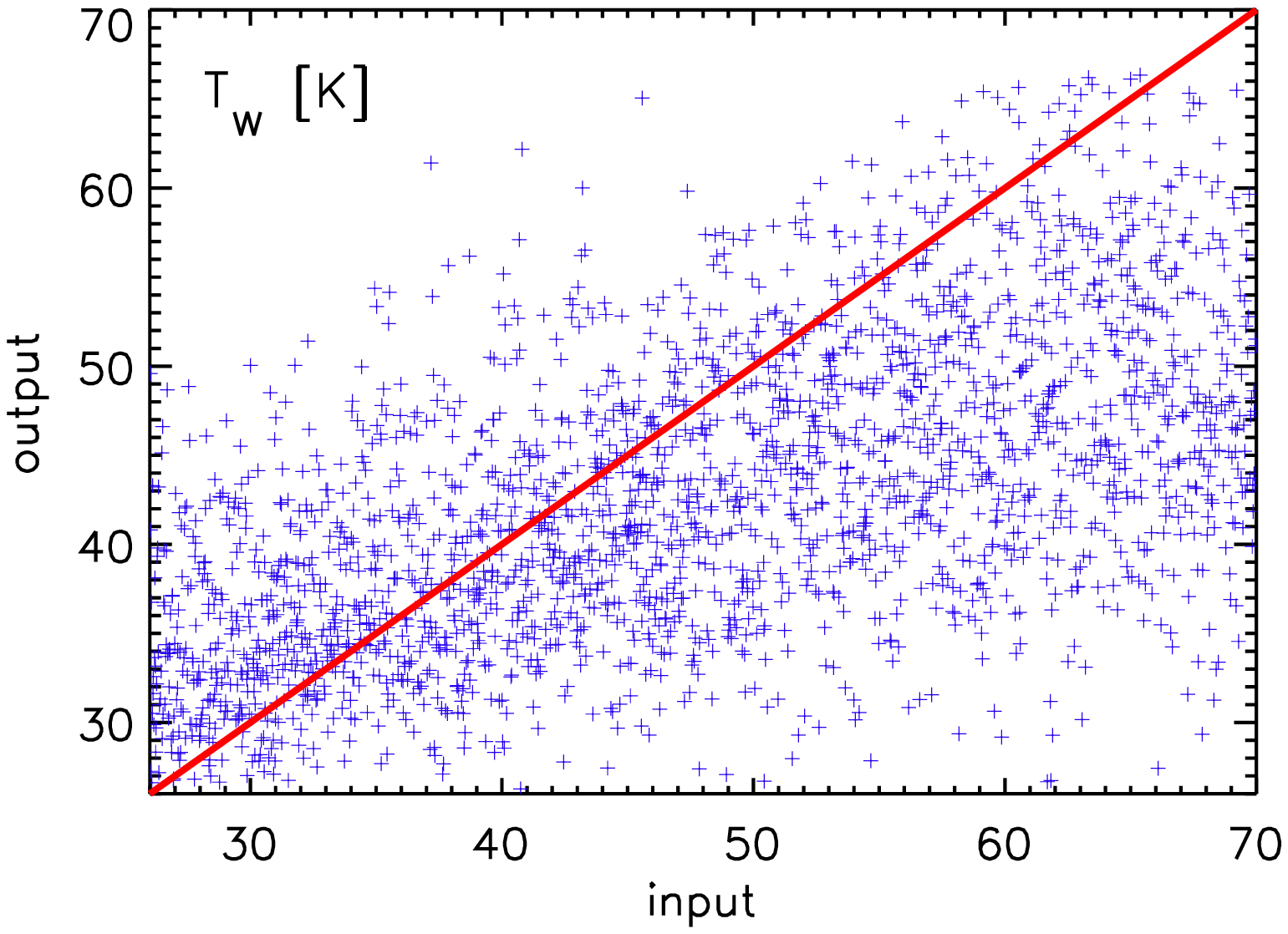}\includegraphics*{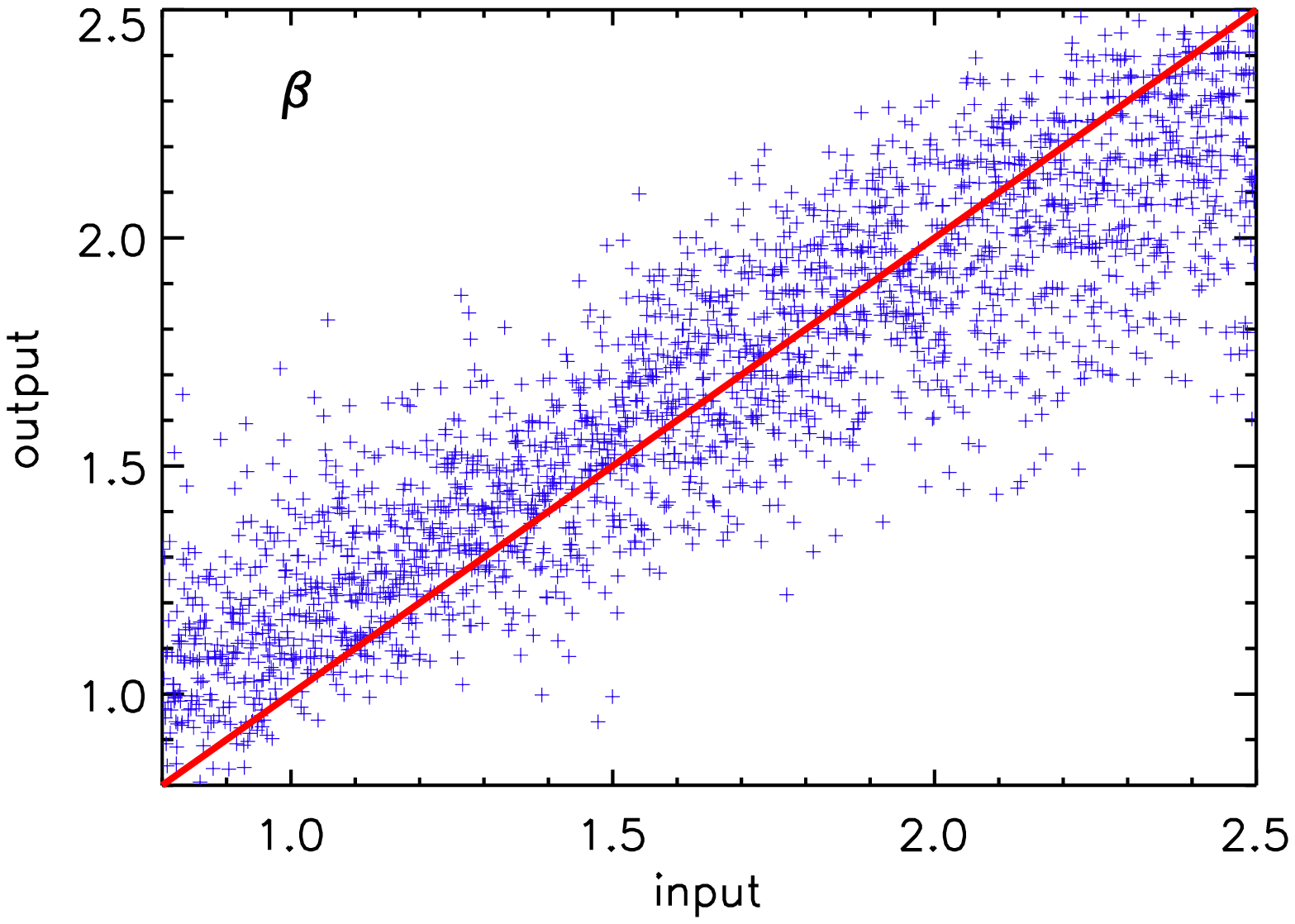}}
\resizebox{15cm}{!}{\includegraphics*{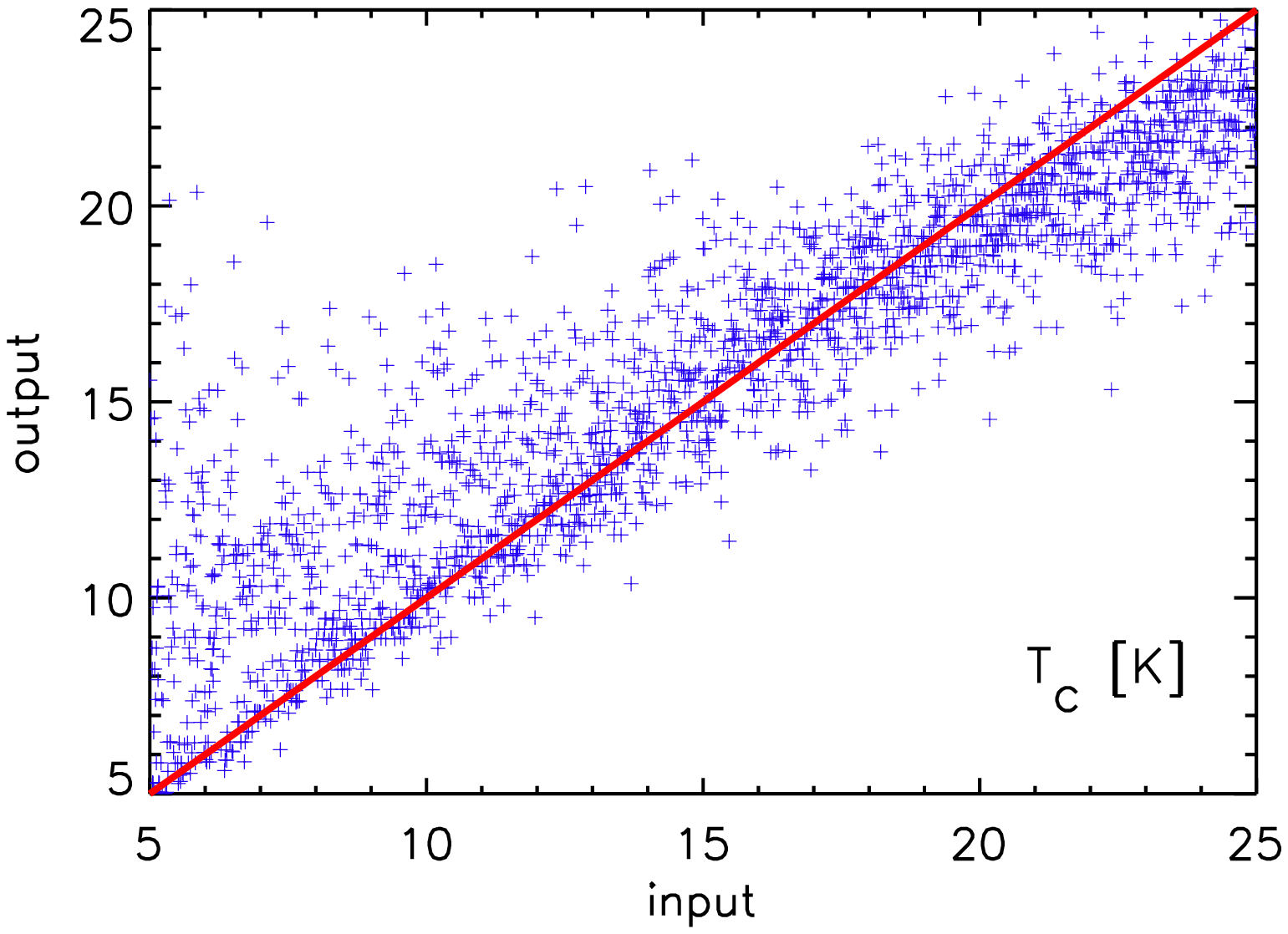}\includegraphics*{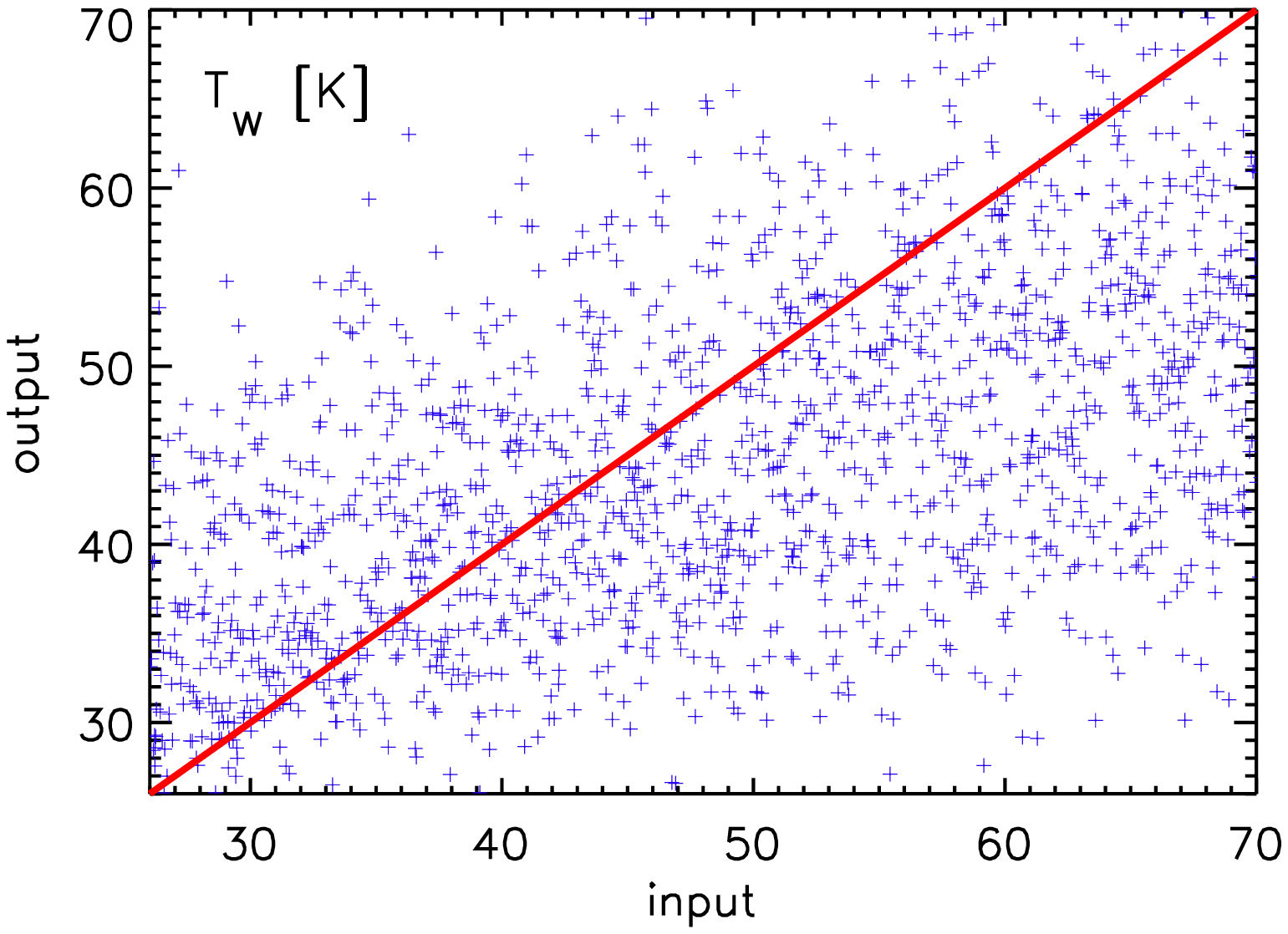}\includegraphics*{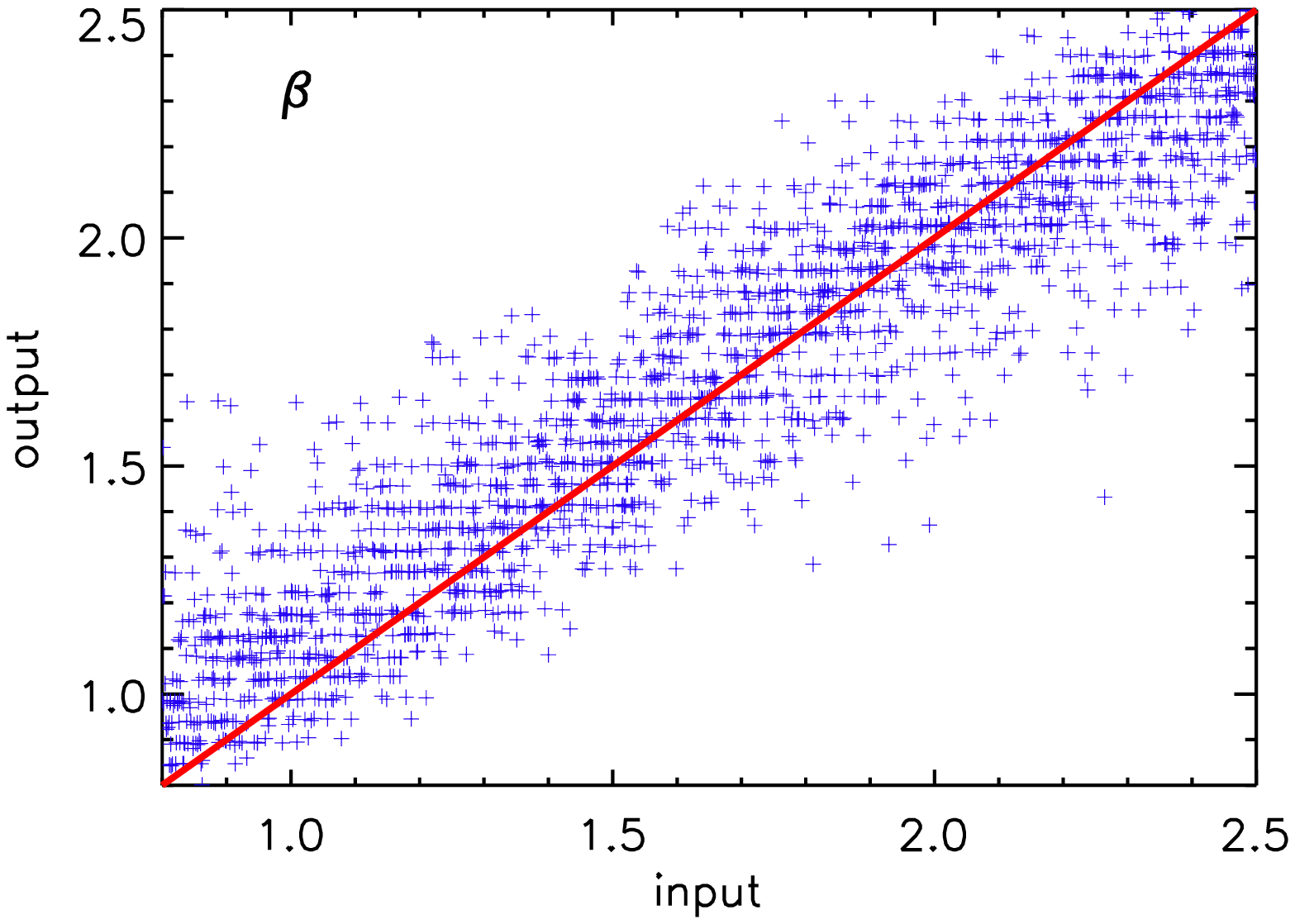}}
\caption[]{  Monte-Carlo simulation of the two-component MBB model.  From left to right:  $T_{\rm c}$, $T_{\rm w}$,  and $\beta_c (=\beta)$. From top to bottom: $\beta_w=2$, $\beta_w=1.5$, $\beta_w=1$, and $\beta_w=\beta_c$.  The red line shows the equality between the output and input parameters.  }
\label{fig:MC1}
\end{center}
\end{figure*}

\begin{figure*}
\begin{center}
\resizebox{12cm}{!}{\includegraphics*{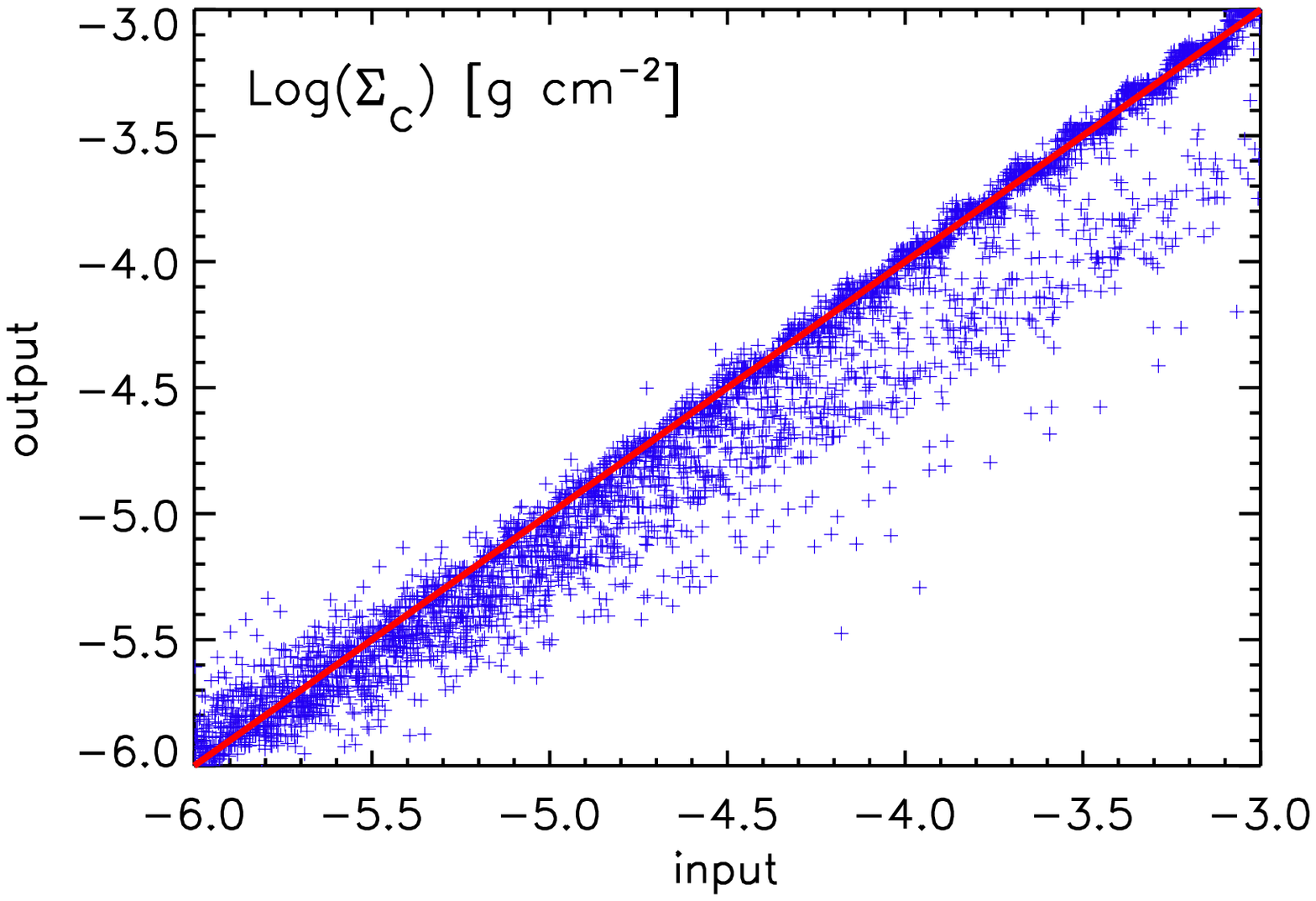}\includegraphics*{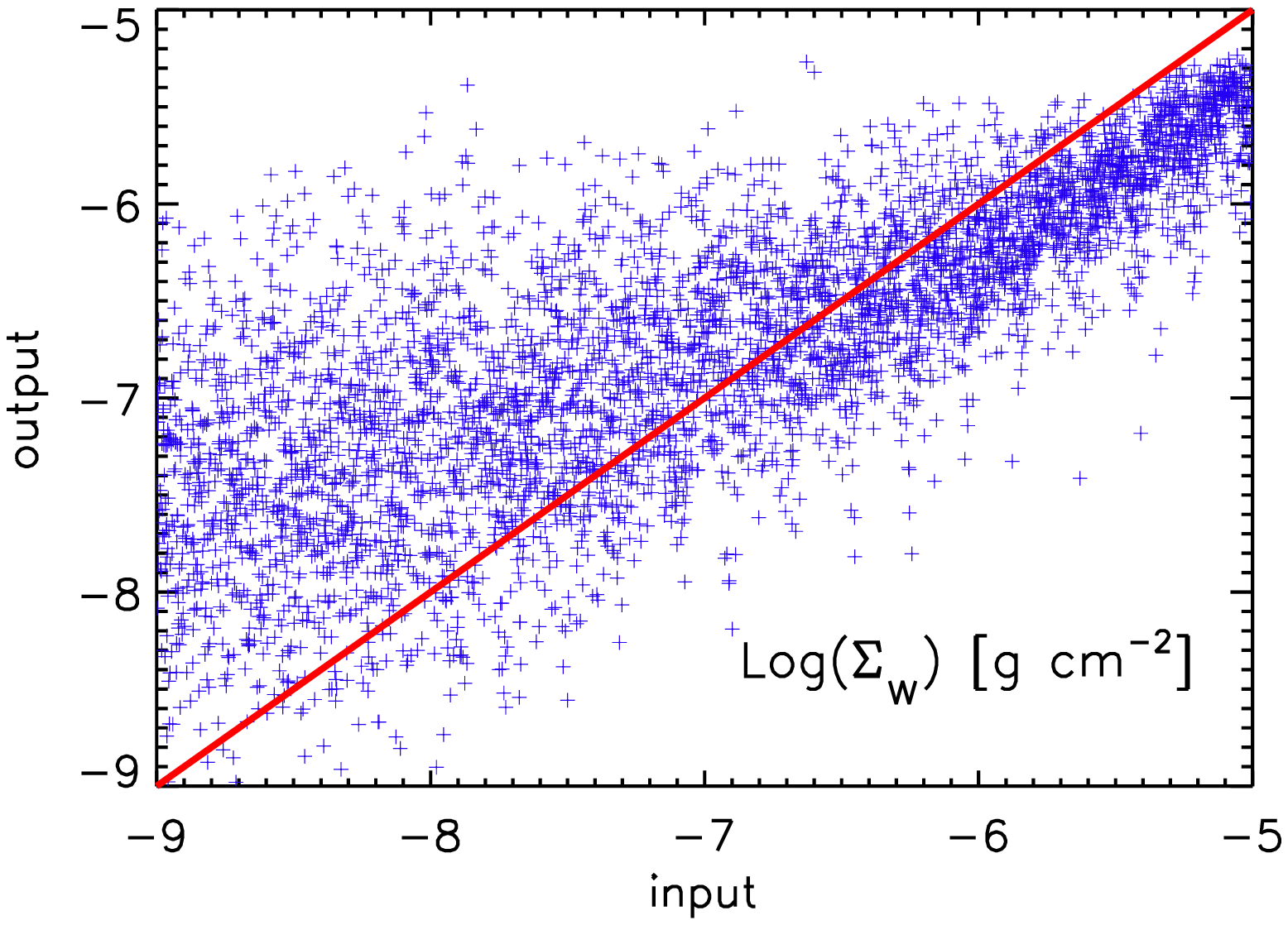}}
\resizebox{12cm}{!}{\includegraphics*{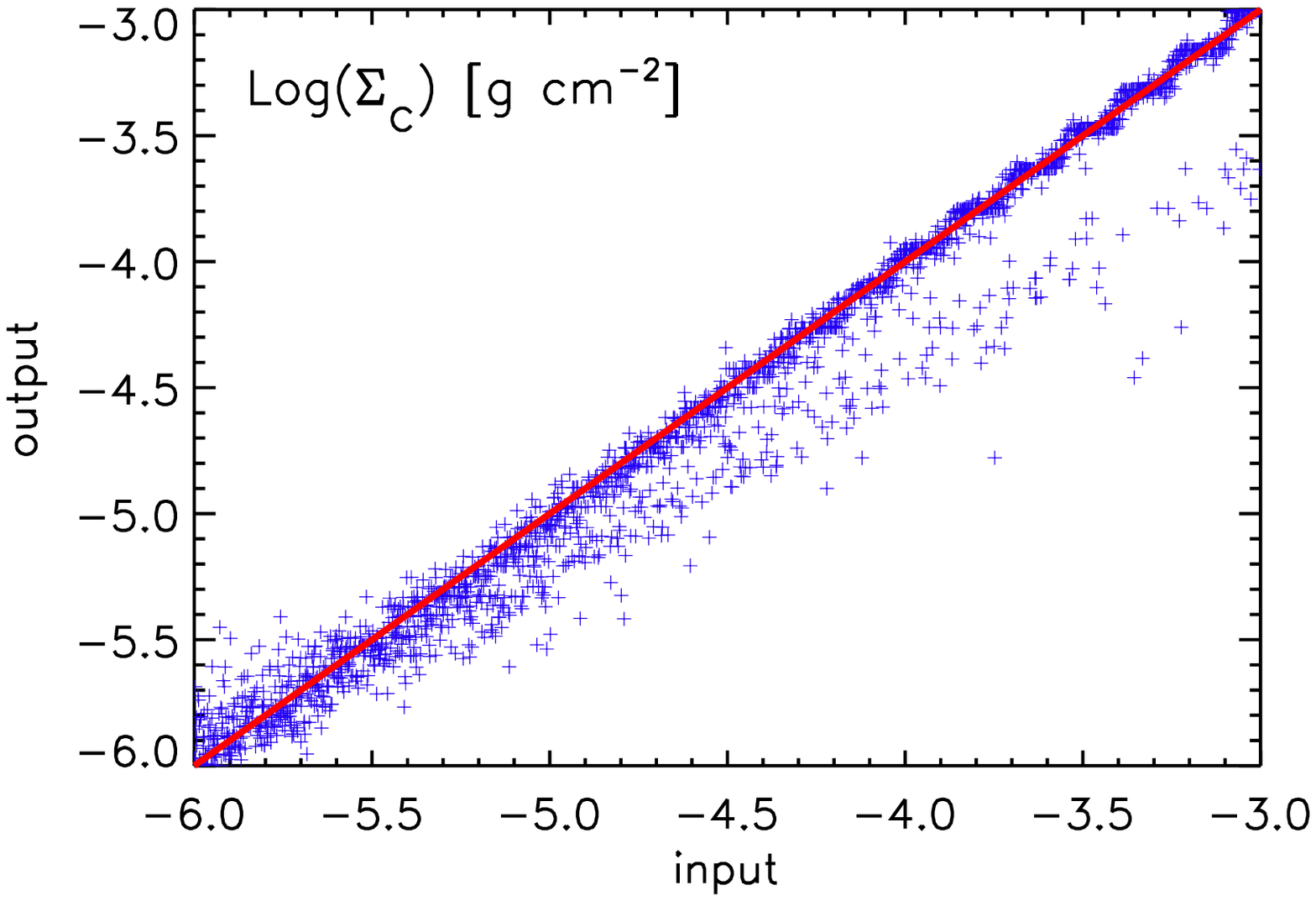}\includegraphics*{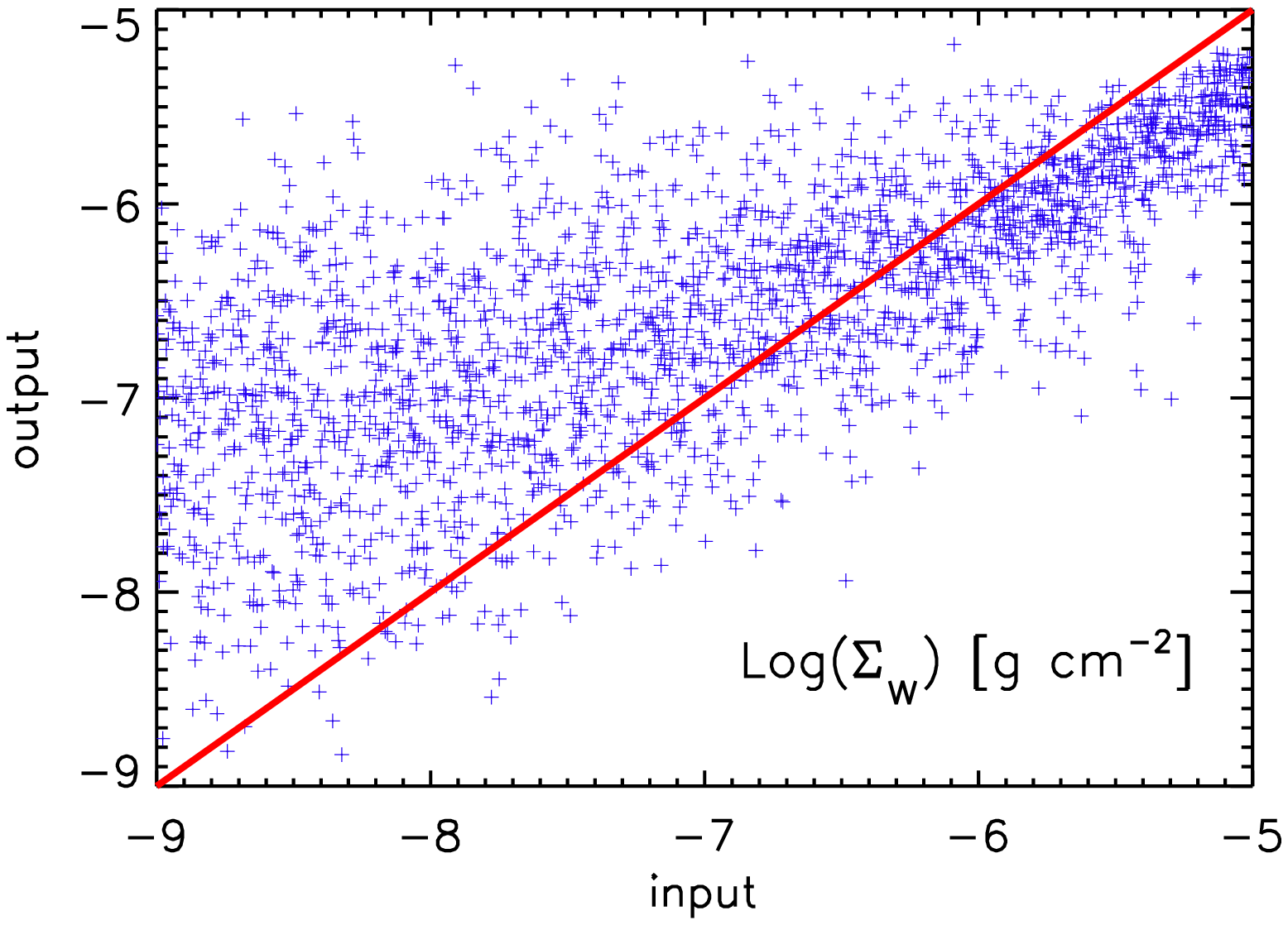}}
\resizebox{12cm}{!}{\includegraphics*{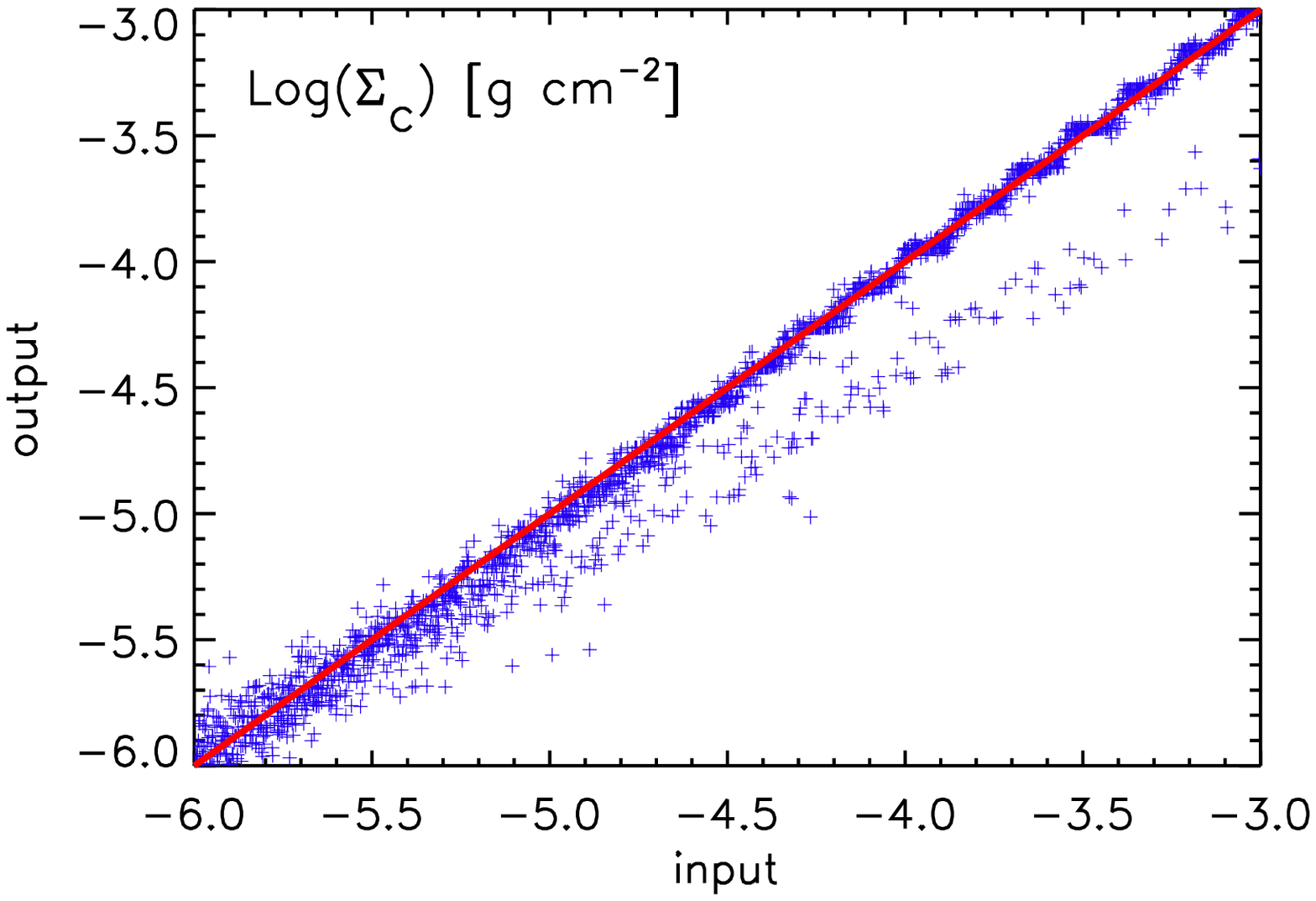}\includegraphics*{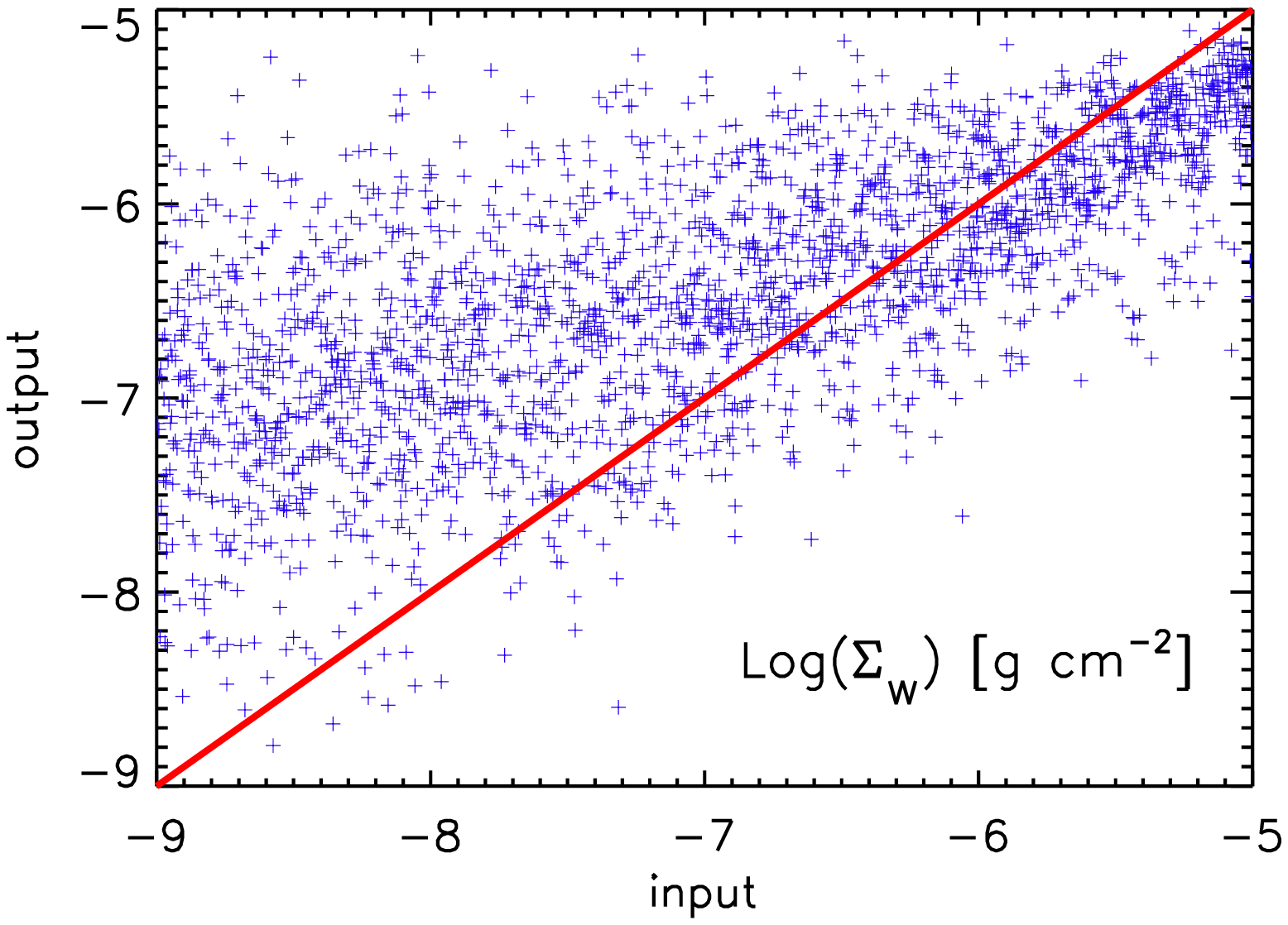}}
\resizebox{12cm}{!}{\includegraphics*{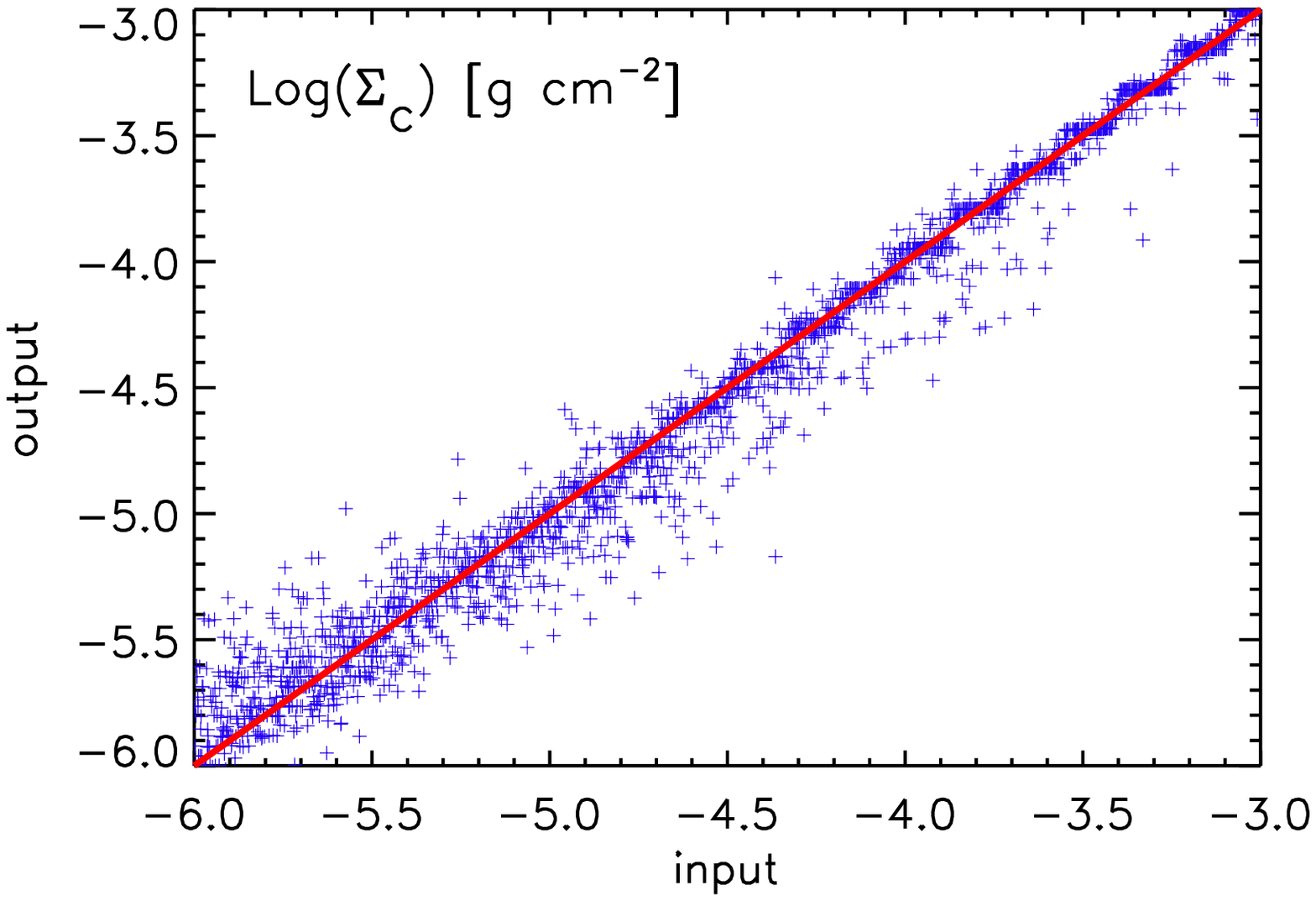}\includegraphics*{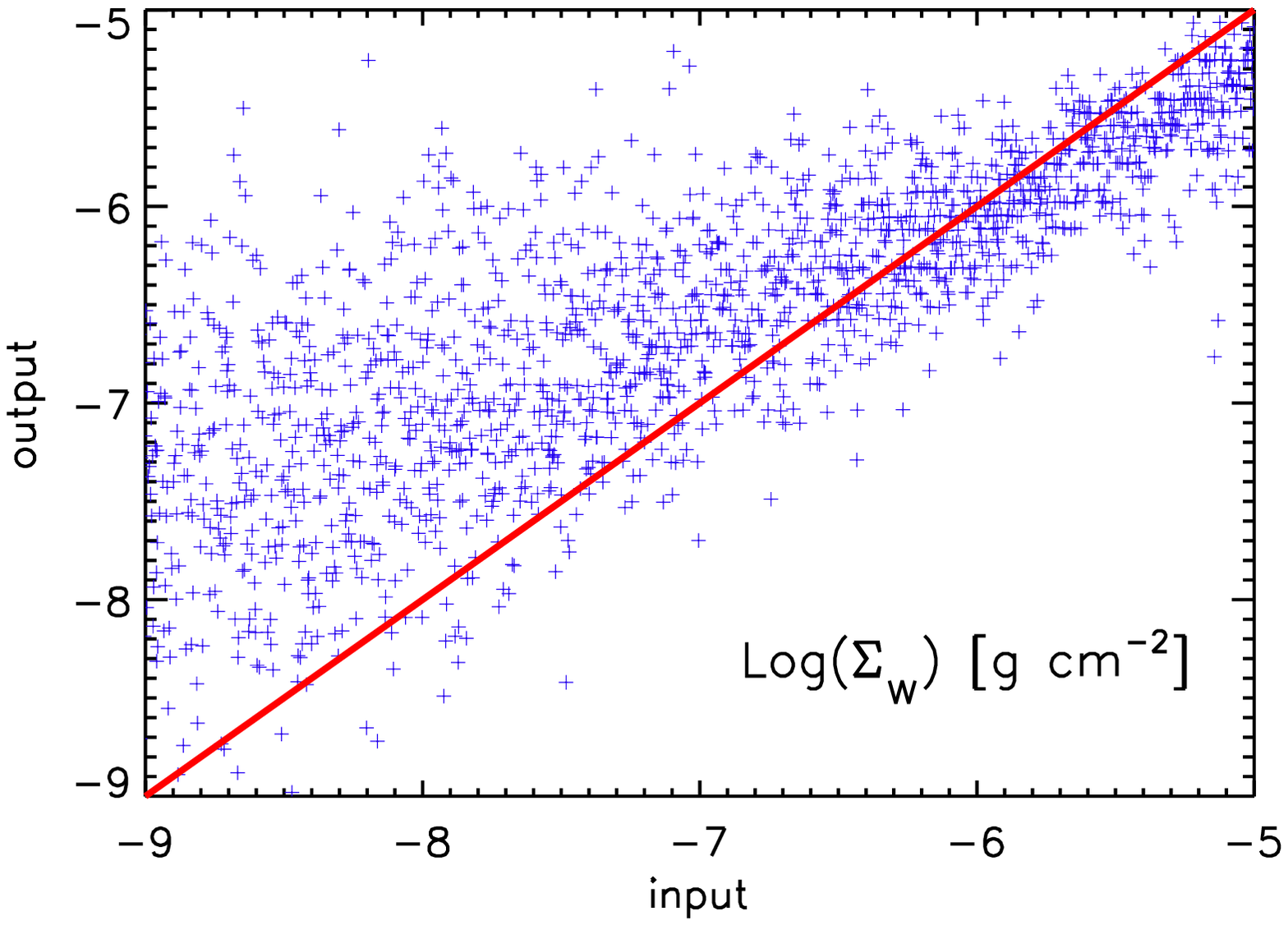}}
\caption[]{  Monte-Carlo simulation of the two-component MBB model. {\it Left:}  $\Sigma_{{\rm dust},c}$ and {\it right:} $\Sigma_{{\rm dust},w}$. From top to bottom: $\beta_w=2$, $\beta_w=1.5$, $\beta_w=1$, and $\beta_w=\beta_c$.  The red line shows the equality between the output and input parameters.  }
\label{fig:MC2}
\end{center}
\end{figure*}

Using a Monte-Carlo simulation,  we {  assess} the systematics and reliability of the two-component approaches. This is particularly useful to check for an artificial correlation between $\beta$ and $T$.
The Monte-Carlo simulations are performed  assuming a uniform probability density function. We first generate random numbers for each of the physical parameters in a  selected  range given in Table~3.  For surface densities, we use a uniform distribution  in logarithmic space. The extreme values in the selected ranges may seem far from reality. Compared to more moderate values, however, this selection ensures that we cover the whole physical parameter space. Each set of physical parameters generated (input) leads to synthesized intensities  using Eq.~(3) as described in Sect. 3.2.  The synthesized intensities are then treated as observed data in the Newton-Raphson method resulting in new sets of physical parameters (output). In the ideal case, i.e. when there are no systematics due to a specific model or approach,  the output parameters are similar to the input parameters.  This case is indicated by a one-to-one (equality) relation between the input and output parameters in {  Figs.~\ref{fig:MC1} and \ref{fig:MC2}}.

In all the two-component MBB models, the simulations show that $T_c$ can be best reproduced between 10\,K and 20\,K  and the scatter around the equality relation is symmetric for the $\beta_w=2$ case (other models tend to overestimate $T_c$). {  In this case, $T_c$ is reproduced with an accuracy better than 3\,K in the range 10-20\,K.} This case also shows a symmetric scatter for $T_w$ for 35\,K~$<T_w<$~50\,K, where it is best reproduced {  (with a dispersion of 11\,K)}, while other models tend to underestimate $T_w$ in this range.  The mass surface density of the cold dust shows generally a small scatter for all models and it is much better constrained than that of the warm dust. {  The accuracy of reproducing the dust mass surface density is about 5.5 times better for the cold than the warm component in logarithmic scale for the $\beta_w=2$ case.  This case}  is again more successful in reproducing $\Sigma_{{\rm dust},w}$ than other cases, especially for  mass surface densities $\Sigma_{{\rm dust},w}< 10^{-6}$\,g\,cm$^{-2}$.  {  Interestingly, we find a coincidence between the low density scatter in the H$\alpha$--$\Sigma_{{\rm dust},w}$ correlation plot (Fig.~\ref{fig:ha_warm}) and the low density deviations in the simulated $\Sigma_{{\rm dust},w}$ {  (Fig.~\ref{fig:MC2})}. } Where  $T_c<$10\,K and  $\Sigma_{{\rm dust},c}>10^{-5}$\,g\,cm$^{-2}$, there are points that follow a shallower slope than the  equality relations. These points are likely local minima in the merit algorithm. 
Synthesizing the observed data, however, reduces the chance of falling into a local minimum, as explained in Sect.~3.2. 

{  Figure~\ref{fig:MC1} shows that $\beta$ is not as well constrained as $T_c$. To give an example, for an input value of  $\beta$=1.5 (with $\beta_w$=2), the output value of  $\beta$ varies between 1.1 and 2 with a  {  dispersion of 0.3}. The method tends to overestimate $\beta$ for  $\beta<$1.5 and underestimate it for   $\beta>$2.}

\begin{figure}
\begin{center}
\resizebox{7cm}{!}{\includegraphics*{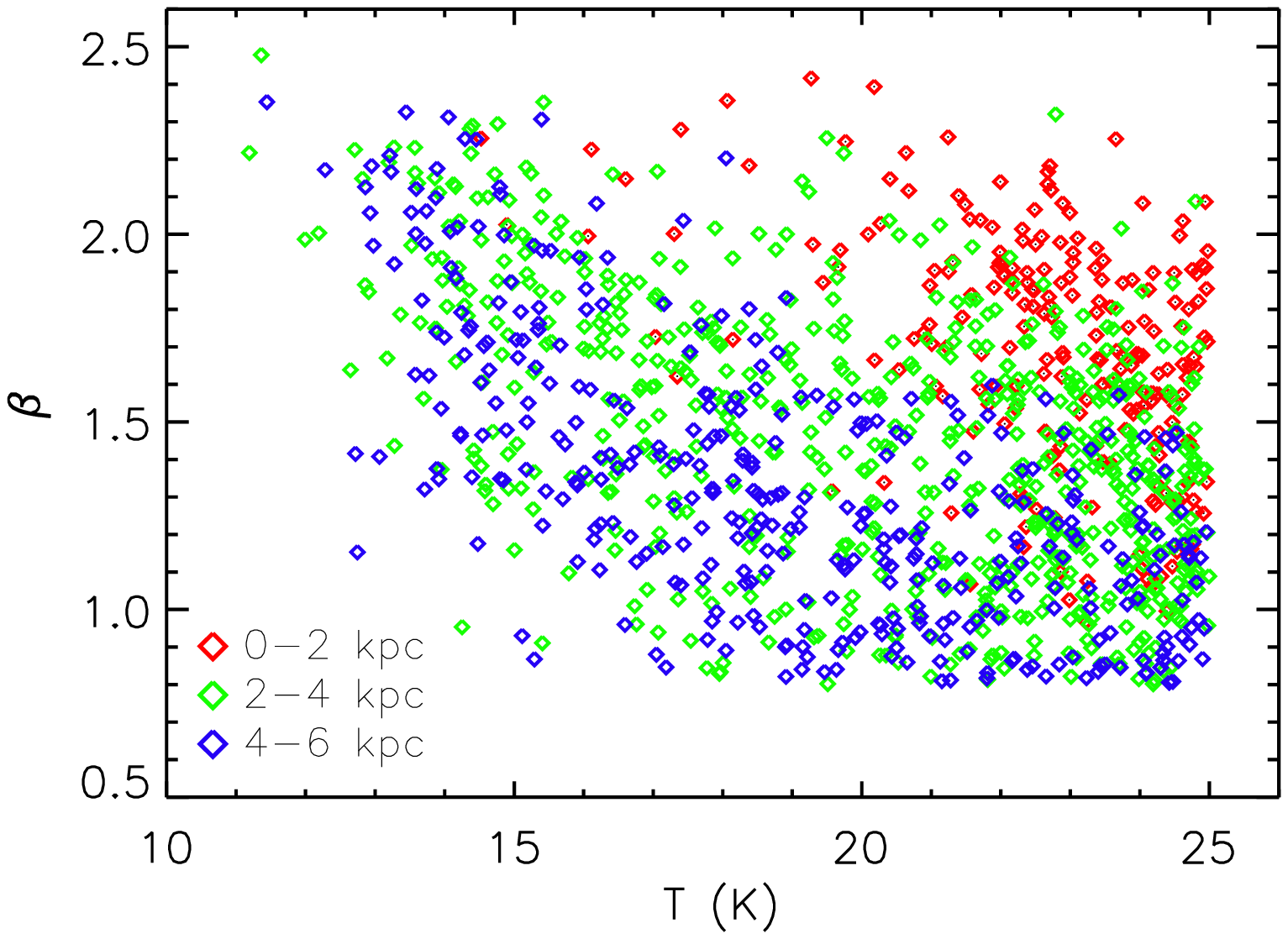}}
\resizebox{7cm}{!}{\includegraphics*{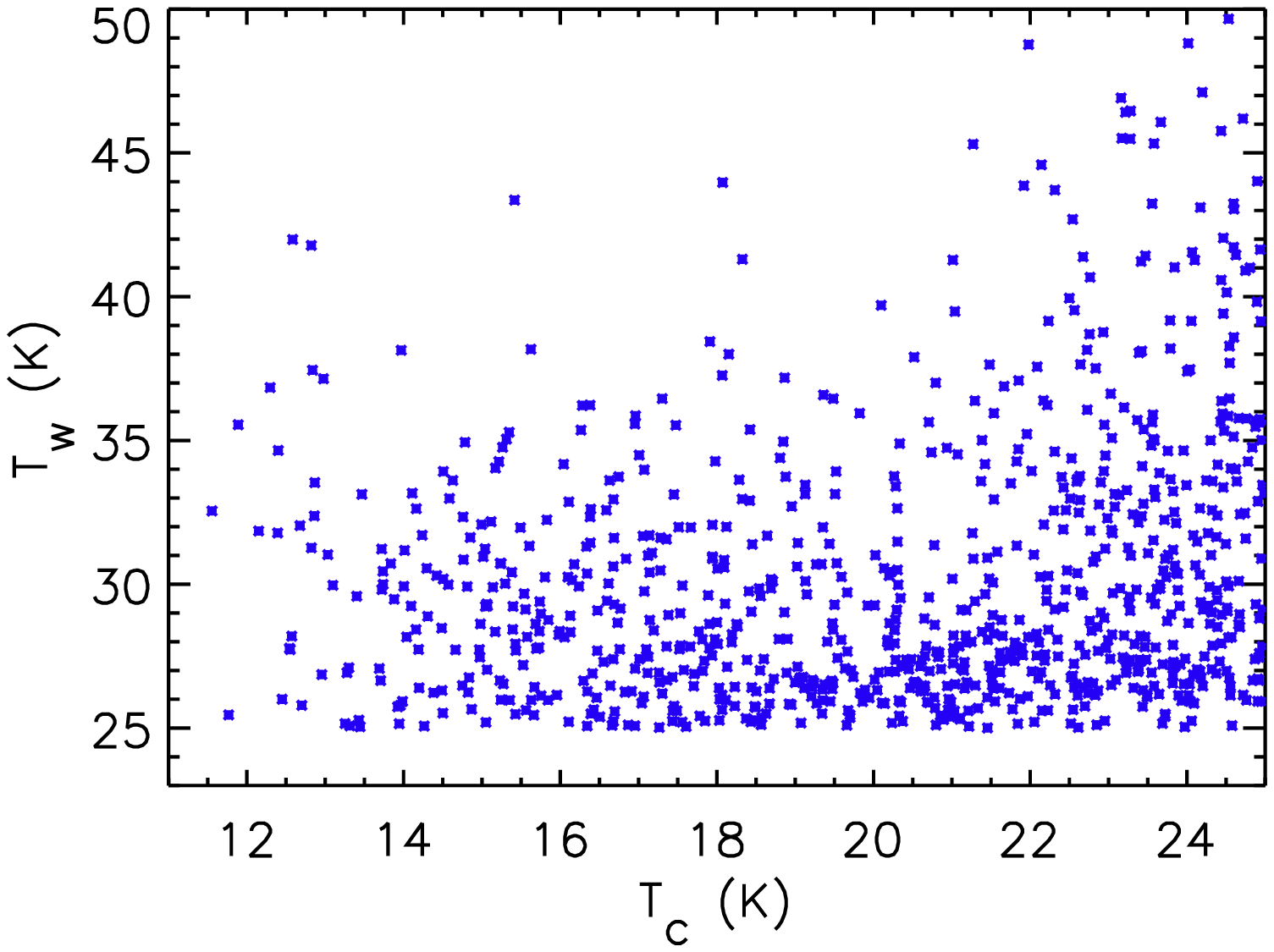}}
\caption[]{{\it Up:} scatter plot of the cold dust emissivity index $\beta$ vs. temperature, {  colour-coded for three radial intervals}. {\it Bottom:} scatter plot of the cold dust temperature vs. the warm dust temperature showing no correlation.}
\label{fig:T-beta}
\end{center}
\end{figure}

{  To study the sensitivity of the different MBB approaches to noise, we add} Gaussian noise with a width of 5$\sigma$ rms of the observed maps to the synthesized intensities at various wavelengths and {  re-run} the simulations\footnote{  Note that Figs.~\ref{fig:MC1} and \ref{fig:MC2} show the simulations before adding the noise.}. Adding the noise increases the dispersion in $\beta$ by $\sim$50\% in the two-component MBB with $\beta_w=2$.  
It is important to note that none of the model approaches lead to an artificial correlation between temperature, mass surface density, and $\beta$ {  for both simulated and real results} (see Fig.~\ref{fig:T-beta} and Sect.~5.1). 

\begin{table*}
\begin{center}
\caption{Setup of the numerical experiment in the two-component MBB approaches. }
\label{tab:lines}\label{tab:var}
\begin{tabular}{c c c c c c c} 
\hline
Approach & T$_{\rm{c}}$ [K] & T$_{\rm{w}}$ [K] & $\log$($\Sigma_{{\rm dust},c}$) [g\,cm$^{-2}$] & $\log$($\Sigma_{{\rm dust},w}$)[g\,cm$^{-2}$] & $\beta$ & Statistics\\\hline
%I\,\,\,\,\,\,\,\,\,\,\, \,\,\,\,\,\,\,\,\,\,& 5$-$100 & $-$   & (-10)$-$(-3) & $-$             & 0.8$-$2.5  & 15\,000\\
$\beta_w$=1\,\, \,& 5$-$25 & 26$-$70 & (-6)$-$(-3) & (-9)$-$(-5)  & 0.8$-$2.5  & 15\,000\\
$\beta_w$=1.5& 5$-$25 & 26$-$70 & (-6)$-$(-3) & (-9)$-$(-5)  & 0.8$-$2.5  & 15\,000\\
$\beta_w$=2\,\,\,\,\,\, & 5$-$25 & 26$-$70 & (-6)$-$(-3) & (-9)$-$(-5)  & 0.8$-$2.5  & 15\,000\\
$\beta_w$= $\beta_c$\, & 5$-$25 & 26$-$70 & (-6)$-$(-3) & (-9)$-$(-5)  & 0.8$-$2.5  & 15\,000\\
\hline
\end{tabular}
\tablefoot{Range of the temperatures T$_{\rm{c}}$ and T$_{\rm{w}}$, mass surface densities $\Sigma_{{\rm dust},c}$ and $\Sigma_{{\rm dust},w}$ (in logarithmic scale), $\beta$,  as well as the statistics of the Monte-Carlo simulations are listed.}
\end{center}
\end{table*} 

\begin{figure*}
\begin{center}
\resizebox{\hsize}{!}{\includegraphics*{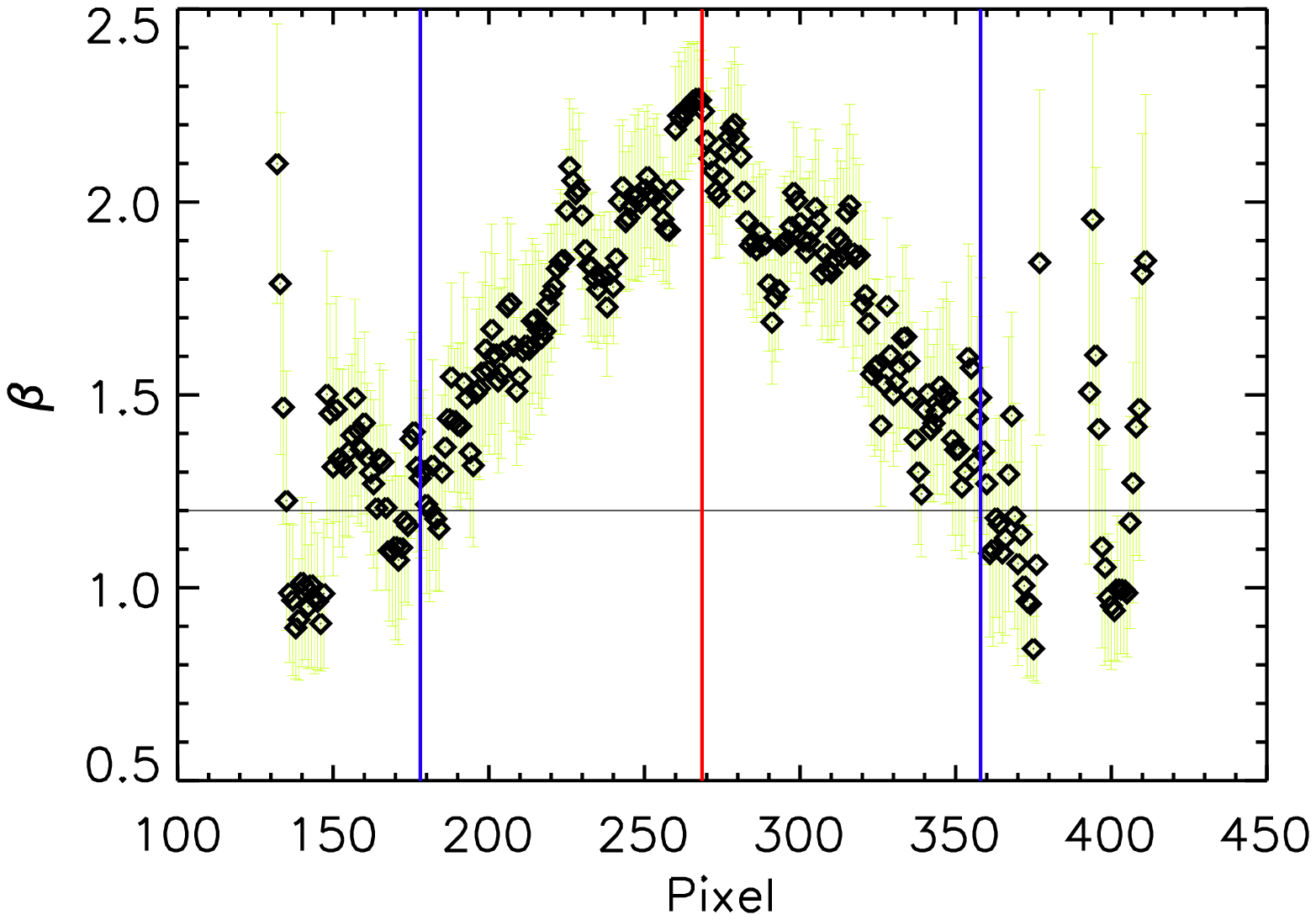}  \includegraphics*{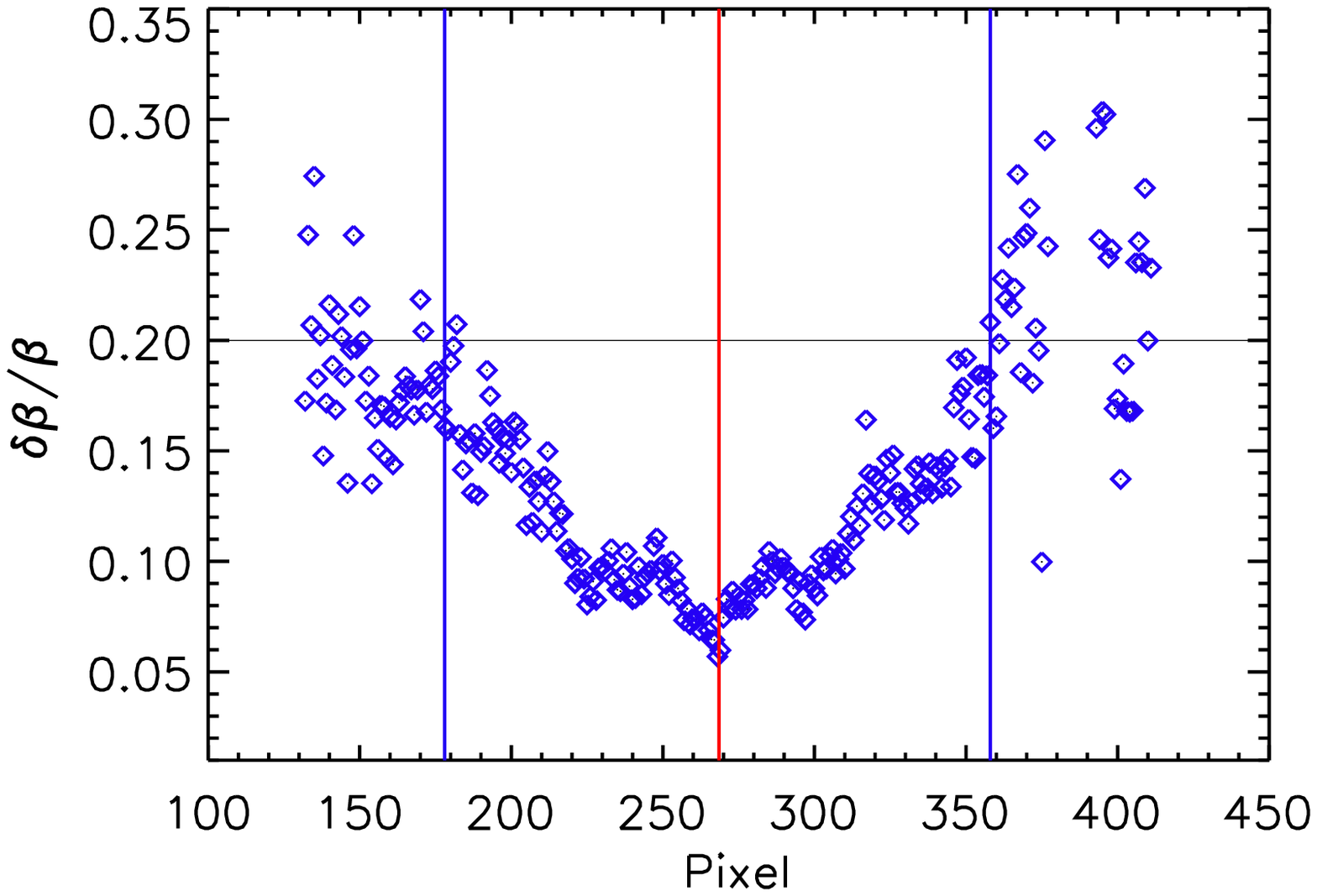}}
\caption[]{  Distribution of the emissivity index $\beta$ ({\it right}) and its relative uncertainty $\delta \beta$ /$\beta$ ({\it left}) along a vertical cut passing through the center. The red and blue lines indicate the center of M\,33 and $R\simeq$\,6\,kpc, respectively.  }
\label{fig:error}
\end{center}
\end{figure*}

{  
\section{Sources of uncertainties}
In addition to the uncertainties due to the method, our results could be affected by statistical and systematic uncertainties on measured fluxes. 
To study the effect of the statistical uncertainties, we first measure the stochastic pixel noise by calculating the dispersion in intensities in background regions. In the reduced images, background dispersion could in principle be due to the propagation of original pixel noise and artifacts into the background pixels as well as possible imperfect foreground subtraction. We then simulated the observed intensities by adding random noise with normal distribution of 3$\sigma$ background noise level measured at each wavelength.  
Using this statistical experiment, we created 1000 mock datasets used as the data points (fluxes) in our two-component approach with $\beta_w=2$ resulting in a set of 1000 values for each physical parameter. The statistical uncertainties ($\delta_{\rm stat}$) were then determined from the standard deviation of the results (Table~4).  
%This test was performed for several pixels and the results were very similar.

\begin{table}
\begin{center}
\caption{  Uncertainties of the obtained physical parameters. }
\label{tab:lines}\label{tab:var}
\begin{tabular}{c c c c} 
\hline
Parameter  & $\delta_{\rm sys}$ & $\delta_{\rm stat}$ & $\delta$\\
\hline
T$_{\rm{c}}$ [K]  & 2.0 & 0.1  & 2.0\\
T$_{\rm{w}}$ [K]& 10.0 & 2.3  & 10.3\\
log($\Sigma_{{\rm dust},c}$) [dex] & 0.08& 0.00  & 0.08\\
log($\Sigma_{{\rm dust},w}$) [dex]& 0.54& 0.34 & 0.64\\
$\beta$ & 0.22 & 0.01 & 0.22\\
\hline
\end{tabular}
\tablefoot{  The values show the corresponding errors in the two-component MBB approach with $\beta_w$=2.}
\end{center}
\end{table} 
%3.5,  11.5, 0.15, 0.82 

The flux uncertainty could also be due to  systematic uncertainties, e.g., calibration errors and artifacts caused by observing cameras  \citep[e.g.][]{Aniano_12},  which do not have the same properties as the statistical errors. In most cases, systematics are addressed via calibration uncertainties, as the systematics due to cameras are not well-defined.  For NGC\,6946 and NGC\,0628, \cite{Aniano_12} investigated the systematics due to the MIPS and PACS cameras by comparing their data at a same wavelength (after convolving them to the same resolution and pixel size). For M33, such a  comparison is possible only at 160$\mu$m, as at other wavelengths the data are taken with only one camera. We consider the relative difference in the 160$\mu$m intensities obtained with MIPS and PACS as an estimate for the systematic error if it is larger than the calibration uncertainty at each pixel. In the region of interest $R<6$\,kpc (inside the optical radius {  $R_{25}\sim 7.5$\,kpc)}, the relative difference between the MIPS and PACS data is $\lesssim$ 20\%, which is comparable to the calibration uncertainty. Hence, the systematics at 160$\mu$m are dominated by the calibration uncertainty. {  We assume a similar situation for the data at other wavelengths. This is supported by using the maximum calibration uncertainties at each wavelength\footnote{The calibration uncertainties used are larger than those of point sources by more than a factor of two. They are also larger than those assumed by \cite{Aniano_12}.} (Sect.~2).
The uncertainties in the physical parameters due to the systematics are then determined by adding the calibration uncertainties (see Sect.~2) to the synthesized intensities retrieved in the Monte-Carlo experiment in Sect.~4. These intensities were then used to derive the physical parameters using the Newton-Raphson method. } This way the systematics due to the method are also taken into account.  %This way the errors are overestimated compared to a case when a uniform distribution of calibration uncertainties are simulated \citep[e.g.][]{veneziani}. 
Table~4  shows the median of the absolute residuals\footnote{The residuals have a Gaussian distribution.} ($\rvert$output -input$\rvert$), $\delta_{\rm sys}$. 

The error on each physical parameter is then given by $\delta = \sqrt{\delta^2_{\rm sta} + \delta^2_{\rm sys}}$.  As expected,  $\delta$ is dominated by the systematics rather than the statistical noise. 

To evaluate the role of the systematics on the decrease of $\beta$ with distance from the center shown in Sect.~3.2.2, yet another experiment is performed.  After adding the systematic uncertainties to the observed intensities, a vertical cut passing through the center of M\,33 was solved 100 times using the Newton-Raphson method. This resulted in 100 sets of solutions for each pixel along the vertical cut\footnote{This cut includes $\sim$ 280 pixels for which $I_{\lambda}>3\sigma$ rms noise.}. The uncertainties in the physical parameters are then determined from the standard deviations of the parameter distributions. Fig.~\ref{fig:error} (left) shows the distribution (profile) of $\beta$ and its uncertainty $\delta \beta$ along the cut. From the center to $R\simeq\,6$\,kpc (indicated by the blue lines), $\delta \beta$ increases from $\simeq$\,0.15 to $\simeq$\,0.3. This is equivalent to a relative change in $\beta$ from $\simeq$\,6\% to $\simeq$\,20\% as shown in Fig.~\ref{fig:error} (right). Hence, the magnitude
of the derived radial decrease of beta is significant, i.e. larger
than three times the uncertainty. }

\section{Discussion}
\subsection{The $\beta$--$T$ correlation}
{\it One of the most important results of this work is that we clearly show that not only the temperature but also $\beta$ decreases with galactocentric distance in M~33.}  This is a real effect, shown by different physical and numerical approaches and because the effect is strong enough to be seen against the well-known numerical $\beta-T$ anti-correlation \citep{Shetty_09}.
%
%An anti-correlation between $\beta$ and $T$ has been already reported by several authors both from observations \citep[e.g.][]{Dupac} and laboratory \citep[e.g.][]{Mennella,Meny}.  However, the temperatures over which the variations are seen are typically not the same in the ISM and in laboratory experiments.  \cite{Shetty_09} have attributed the reported anti-correlation to the effect of noise, or the temperature mixing when the dust is not isothermal along the line of sight. Recent studies based on the Herschel and Planck space observatories show an anti-correlation which, according to the authors, is too strong to be attributed to the noise alone \citep[e.g.][]{Paradis_10,Anderson_10,Planck,Galametz_12}. 

%
The techniques adopted in this work -- solving a system of equations  or deriving the `best-$\beta$' leaving only the temperature as a variable -- are robust despite the  strength of the $\beta-T$ degeneracy.  However, for any given pixel there is a tendency for $\beta$ and $T$ to be anti-correlated simply because dust temperatures derived with lower $\beta$ are higher, whatever the method.  When averaged over radial annuli, it becomes clear that both $\beta$ and $T$ decrease with radius{ , though their decreasing trends are not similar} (e.g. Fig.~\ref{fig:radial}).

%{  To investigate the correlation between $\beta$ and $T_c$, we} calculate Pearson's linear correlation coefficient, $r$,   between the maps of {  the two parameters}. For an ideal correlation, (anti-correlation) $r = 1$ ($r= -1$). The formal error of the correlation
%coefficient depends on the strength of the correlation and the number
%of independent points, $n$, in an image:
%\begin{equation}
%$\Delta r_{c}= \sqrt{1-r^{2}_{c}}/ \sqrt{n-2}.$
%\end{equation}
%
%A set of independent data points,  i.e. a beam area overlap of $<20\%$, was selected by choosing pixels spaced by more than one beamwidth. The resulting {  correlation} coefficient is only $r=-0.26\pm0.03$ for the entire ranges of the two parameters  (Fig.~\ref{fig:T-beta}){ , much weaker than that reported in other galaxies \citep{Smith_12,Galametz_12}}. {  The lack of a positive correlation implies different causes for the decrease of $\beta$ and $T$ with distance from the center.  }

In M31, with similar data and thus similar linear resolution as for M33, \cite{Smith_12} find that $T$ and $\beta$ have opposite behavior as a function of radius (their Fig.~8), such that the temperature increases with radius and $\beta$ decreases.  This may be linked to the $\beta-T$ degeneracy {  because the dust temperature decreases with radius in M31 as shown by \cite{Fritz} using the same data and in other studies \citep{Groves_12,Tabatabaei_10}}.

%Thus,  no significant correlation  (anti-correlation) holds for the entire ranges of the $T_c$ and $\beta$ parameters in M\,33 as found e.g. in M\,31 \citep{Smith}.  The $\beta$--$T$ anti-correlation obtained in M\,31 is considered to be not due to the degeneracy, while a very tight anti-correlation from noise alone was found \citep[see Fig.~15 of][]{Smith}. Moreover, they found an increase of $T$ with radius from 3\,kpc to 10\,kpc radius, unlike $\beta$, which disagrees with earlier works  \citep{Fritz,Groves_12,Tabatabaei_10}.
%This radial increase of $T$ could be possibly influenced by the single-component MBB fitting method with $\beta$ as a variable, as  using the same data, \cite{Fritz} found a `decrease'  of  $T$ with radius instead. 
%
It is already known that the dust temperature decreases with radius in M33 due to the radial decrease in the interstellar radiation field at all wavelengths \citep{Verley_09,Kramer10,Braine10,Boquien_11,Xilouris}. A trend of decreasing $\beta$ could only be expected if $\beta$ is also linked to a factor which varies (on average) with radius. 
%
%We note that the temperature mixing could not explain the radial decrease of $\beta$ as the mixing effect is more severe in the central parts (where many clouds with a broad range of temperature exist) than in the outer parts.  

In addition to the ISRF, many properties of spiral galaxies vary with distance from the center: the metallicity decreases with radius as does the H$_2$/HI ratio, the stellar surface density, and the star formation rate.  Thus, the gradient in $\beta$ we find could be due to any of these factors, among which the metallicity, the H$_2$/HI ratio, and the star formation rate are strongly linked.   
 %The most likely influences are the metallicity, as it has been observed that low-metallicity systems tend to have shallower $\beta$ values and the metallicity could clearly affect dust composition \citep[see also][]{Planck_2}, and the molecular gas fraction, because it is expected that dust grains are more likely to coagulate in the dense environments as well as growe from depletion of molecules onto dust grains. 
This motivates us to look for a correlation between $\beta$ and different environmental conditions imposed by star formation activity and different phases of the ISM. 
\subsection{Connection to star formation and ISM tracers}
Within the inner disk, the spiral arms and NGC\,604 are visible as
high-$\beta$ regions (Fig.~\ref{fig:beta}).  In both Fig.~\ref{fig:radial} and Fig.~\ref{fig:beta}, a clear change in $\beta$ is apparent at 4\,kpc radius, where $\beta$ changes from approximately Galactic \citep[$\beta \simeq 1.8$,][]{Planck_2} to {  lower than that of the LMC \citep[$\beta \simeq 1.5$,][]{Planck}}.  This is also the radius at which \cite{Gardan}  noticed a break in the H$_2$ formation efficiency in M33 (see their Fig.~13). {  Starting at the same radius, the [CII]/FIR ratio is also found to increase
%behave differently vs. radius, showing a break to high ratios 
\citep[][]{Kramer13}}.
 
\begin{figure}
\begin{center}
\resizebox{7cm}{!}{\includegraphics*{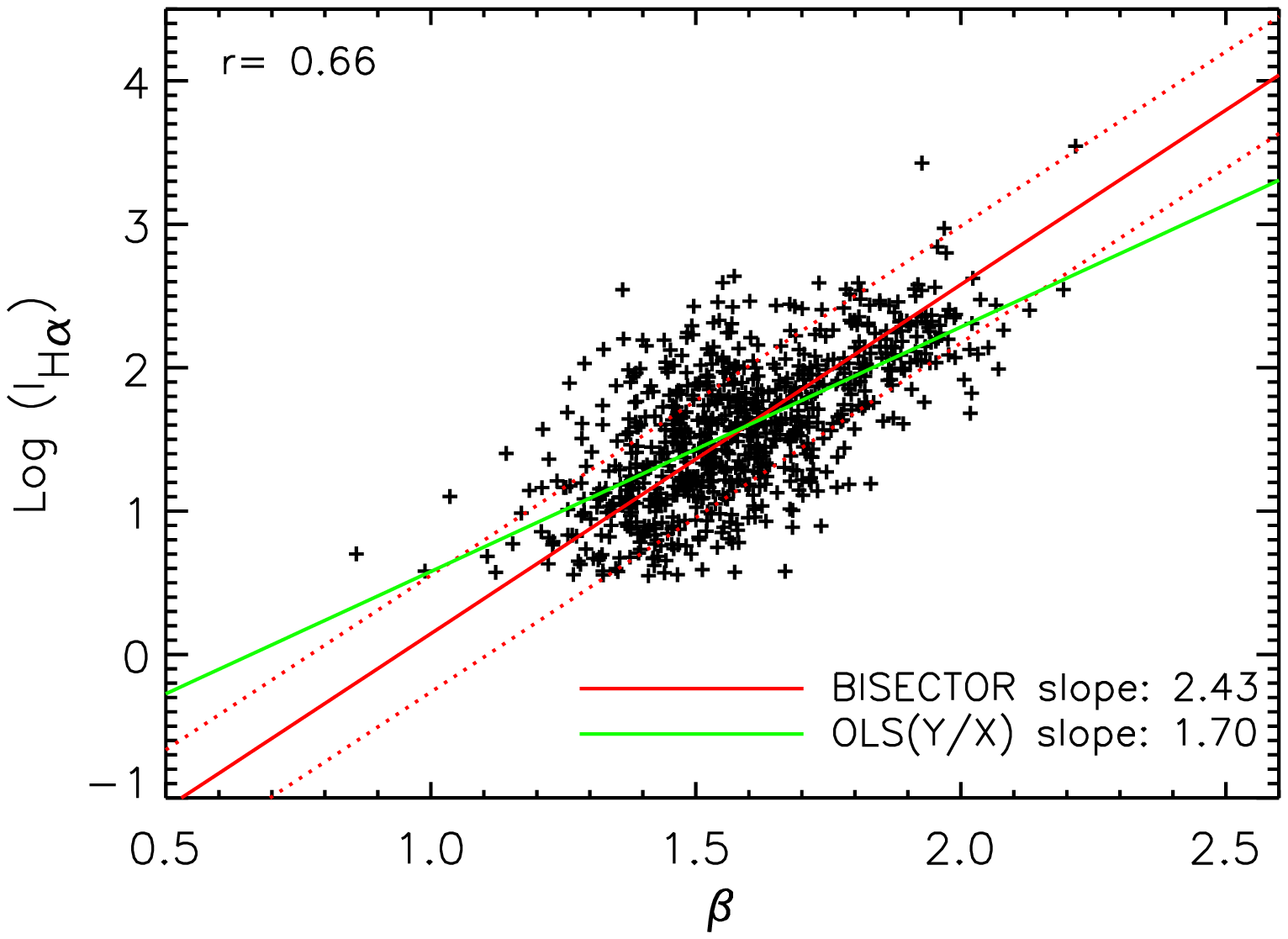}}
\resizebox{7cm}{!}{\includegraphics*{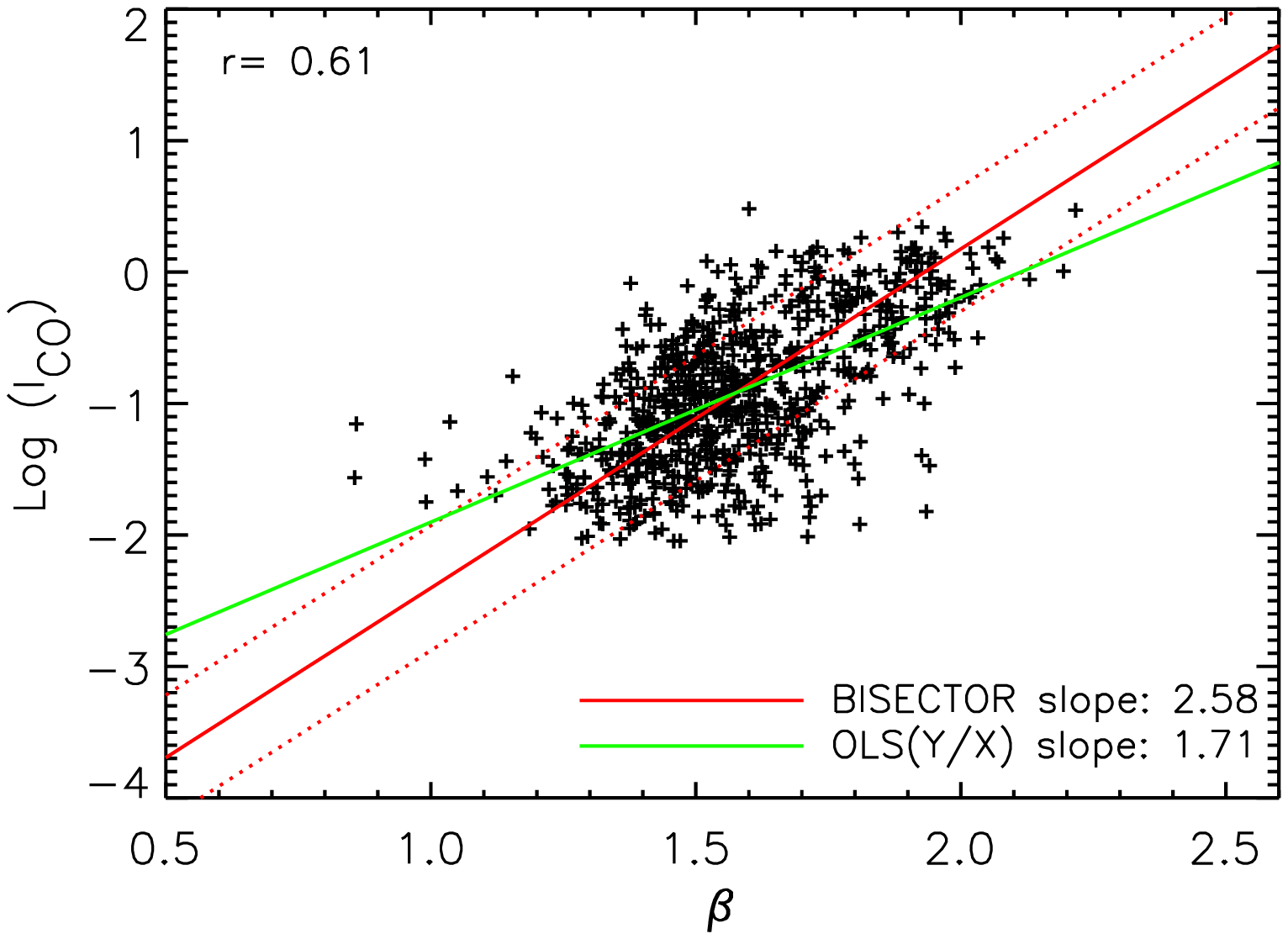}} \resizebox{7cm}{!}{\includegraphics*{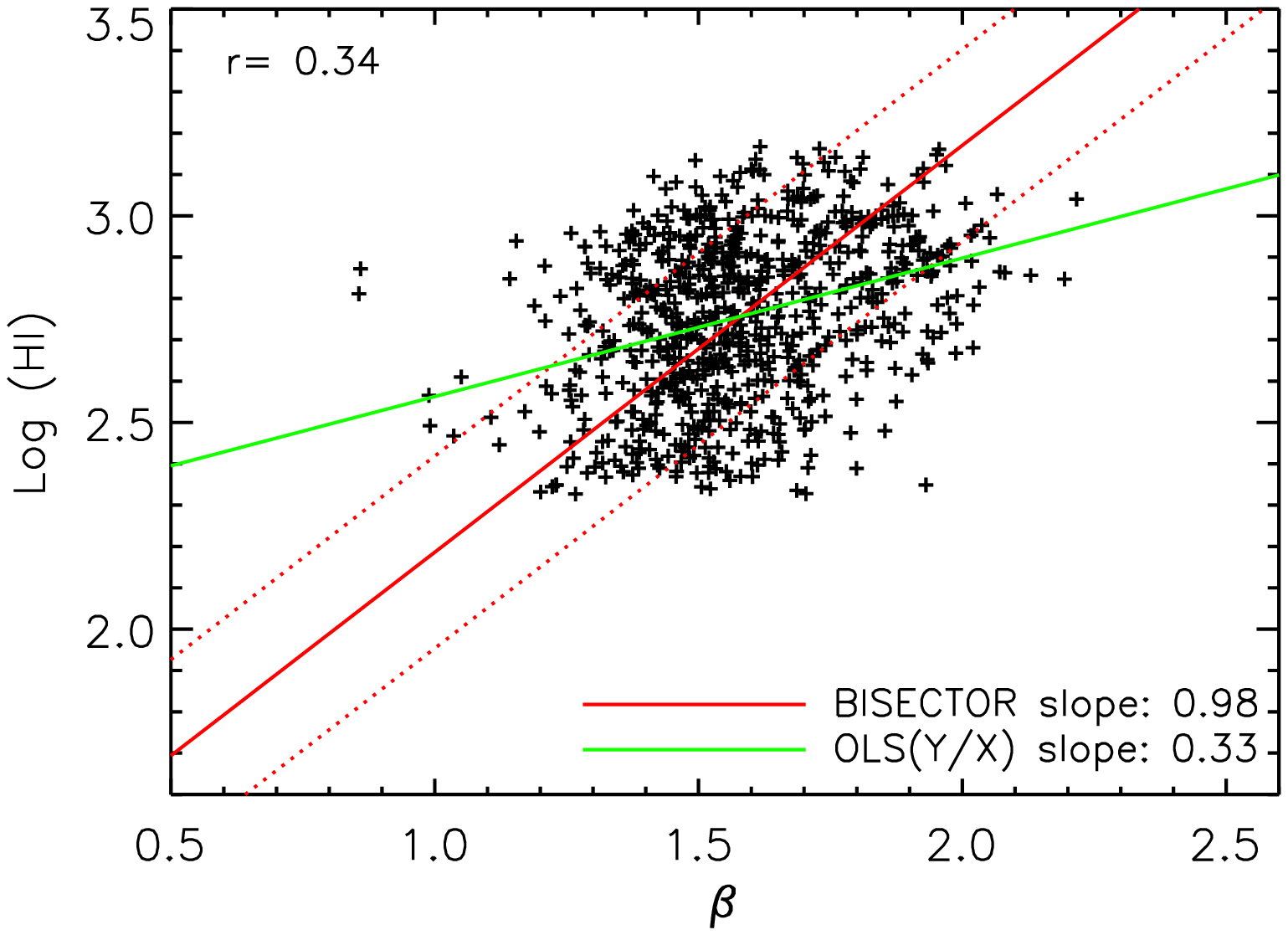}}
\caption[]{Scatter plots of the dust emissivity index $\beta$ versus the H$\alpha$,  CO(1-0), and HI emission from M33. The y-axis presents the logarithm of the intensities in {  units of cm$^{-6}$ pc for the  H$\alpha$ emission and  K\,km\,$s^{-1}$ for the  CO(1-0) and HI emssions}. Also shown are Pearson correlation coefficients (r), {  the ordinary least square (OLS) fits, and the bisector fits.}  }
\label{fig:beta-ISM}
\end{center}
\end{figure}

In Fig.~\ref{fig:beta-ISM}, we investigate  possible connections between the observed $\beta$ and star formation, via the H$\alpha$ brightness, the CO(2-1) and the HI line emission. After excluding pixels with a signal-to-noise ratio smaller than 3, the maps of the  H$\alpha$, CO(2-1), and HI integrated intensities\footnote{  For the sake of consistency, regions not covered by the CO(2-1) observations \citep{Gratier} were masked out in the H$\alpha$ and HI maps before the analysis.} were used to perform cross-correlation with $\beta$ in the same way as in Sect.~5.1. {  Figure~\ref{fig:beta-ISM} shows that  $\beta$ is higher in regions with star formation as
traced by H$\alpha$ emission (top panel).  The link between $\beta$ and the
molecular gas (middle panel), which is the fuel for star formation,  is
slightly weaker and in agreement with the observation that most but not
all molecular clouds in M33 host star formation \citep{Gratier}.  The
atomic gas surface density is only poorly linked to the current star
formation and there is no clear correlation between HI surface density and
$\beta$ (lower panel). The fact that, unlike H$\alpha$ emission and molecular gas, the atomic gas is not  concentrated in  the central disk and along the spiral arms could explain the weak $\beta$--HI correlation obtained. }

%that $\beta$ is indeed correlated with both H$\alpha$ and CO(2-1) as the correlation coefficient is $>$\,0.5. {   Moreover,  H$\alpha$ and CO(2-1) emission are correlated with $\beta$ following a similar relation.}. However, the correlation between $\beta$ and the atomic gas traced by HI is very weak. 
%
Since both H$\alpha$ and molecular gas trace star formation, this indicates that the dust spectrum could differ in starforming and non-starforming regions. Star formation could in principle influence the grain composition  and size distribution (via shattering of dust grains due to strong shocks or dust coagulation), which could modify the dust emissivity index  $\beta$. 
Although such a difference has not been proven, it is possible that the properties of the dust in the atomic and molecular gas clouds differ due to the temperature and density differences.

Star formation replenishes the ISM with metals e.g. through stellar winds of young massive stars and supernovae. Concerning the dust grains, this is likely to increase the abundance ratio of silicates to carbonaceous grains\footnote{The low mass, evolved carbon (AGB) stars are the main source of the carbonaceous grains in the ISM, particularly in low metallicity {  environments} \citep{Lattanzio}.} \citep{Woosley}. According to the standard dust models, $\beta$=2 is consistent with silicate grains, while $\beta$=1 for carbonaceous grains \citep[e.g.][]{Desert}. Observations of the Milky Way show that silicates dominate the ISM \citep[e.g.][crystalline silicates however are not found much in the diffuse ISM]{Molster,Molster_05} which is consistent with the observed dust emissivity index of close to 2  \citep[$\beta \simeq 1.8$][]{Planck_2}. Galaxies with lower metallicity than the Milky Way, however, are found to have flatter SEDs \citep[e.g.][]{Galliano_05,Planck}, {  indicating} the general role of metal abundance on the dust emissivity variation.  Similarly, M33, which has a sub-solar metallicity by about a factor of 2 \citep{Magrini_10}, shows a flatter average value of $\beta$ ($\sim 1.5$),  higher in the central 4\,kpc and lower in the outer parts. The fact that most of the young massive stars are concentrated in the central 4\,kpc of M33, while the evolved carbon stars are smoothly distributed over the disk \citep[e.g.][]{Verley_09}, could provide a distinct difference in the dust composition and hence the emissivity index.
%Late type stars of intermediate mass create a certain amount of carbon. If the metallicity of these stars is low (little oxygen), carbon then overtakes oxygen (C/O $>$ 1) and hence there will be a lot of carbon stars, strongly affecting the interstellar dust with their ejection of carbonaceous grains. If, however, metallicities are high and oxygen is abundant (like in the inner Galaxy), the AGB stars do not produce enough carbon and hence C/O stays $<$\,1, even during the AGB phase. The ejecta from these stars are then less carbonaceous.

The dust grains in the ISM are subject to a variety of destruction and modification processes like vaporization, shattering, and coagulation with the latter two processes being more important in the ISM  \citep[e.g.][]{Tielens_94,Jones_96}. Shattering and fragmentation of dust grains occurs in grain-grain collisions caused by shocks which can shift the peak in the grain size distribution to smaller values \citep[][]{Jones_96} and may change crystalline grains into amorphous \citep{Vollmer_09}, leading to a flatter SED \citep{Seki}. The small grains are more dominant in the diffuse and low-density ISM, considering that they have been accelerated by shock waves traversing the ISM, than in the spiral arms with sites of grain growth in dense molecular clouds. 

{   We note the possibility of systematic deviations of the calculated $\beta$ values from the true $\beta$.}  As shown in Sect.~4, the dynamical range in the true (intrinsic) $\beta$ could be larger than that obtained in the 2-component MBB approach.  The $\beta$ values smaller than $\sim$1.5 could be overestimated and those larger than $\sim$2  underestimated.  
%Hence, mapping  the true $\beta$, a higher contrast is expected. 
The true variation in $\beta$ may be greater than that shown in Fig.~\ref{fig:beta}.  

In our observations, as in likely all astrophysical situations, the dust within the
telescope beam is not at a single temperature.  The emission spectrum of dust at
similar but different temperatures is necessarily broader than single-temperature
dust and, if fit by a single MBB, the resulting $\beta$ will be lower than for the
real grains. {  This however depends strongly on the density \citep{Malinen}.}  

\begin{figure}
\begin{center}
\resizebox{7cm}{!}{\includegraphics*{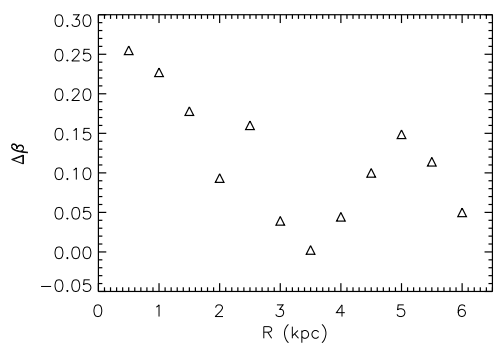}}
\caption[]{  Difference in the dust emissivity index $\beta$ derived using the two-component MBB  and the single-component MBB models vs.  radius. }
\label{fig:beta_dif}
\end{center}
\end{figure}

Could {  the temperature mixing} be particularly {  dominant} for the outer disk of M~33, leading to the radial decrease in $\beta$ that we find? 
{  We have several arguments that suggest this is not the case.\\
- Firstly, it would appear somewhat counter-intuitive to have the
broader mix of dust temperatures where there are fewer dust grains (the dust
surface density is considerably lower in the outer disk), particularly with respect to the central regions. This agrees with \cite{Malinen}, showing that the observed $\beta$ is closer to the true $\beta$ in regions with lower dust densities. A smaller temperature variation in the outer than the inner parts is already visible in Fig.~\ref{fig:sed} by comparing $T_c$ and $T_w$ in each of those regions. \\\
-Secondly, several flux cut-offs were used for the single-component model (Fig.~\ref{fig:radial}) and irrespective of the cutoff a clear
gradient in $\beta$ is observed.  Selecting pixels with strong fluxes, we select outer disk
regions with a level of star formation similar to what is observed in the inner disk and
indeed the dust is fairly warm.  However, these pixels have a low $\beta$. Similarly, using the two-component model, the two hot giant HII regions in the outer parts, IC\,131 and IC\,133, appear as weak sources in the $\beta$ map   (Fig.~\ref{fig:beta}), unlike the HII regions in the inner disk.\\  
-Thirdly, not only the single-component but also the two-component MBB model, which is designed to disentangle the temperature mixing at first order of approximation, leads to a radial decrease in $\beta$. Comparing $\beta$ from the two methods, it is deduced that the single-component method indeed leads to a smaller $\beta$, particularly in the center.  However, the difference in $\beta$ is not larger in the outer than in the inner disk  (see Fig.~\ref{fig:beta_dif}, showing the difference in $\beta$ obtained from the two methods vs. radius).}

\section{Summary}
Using the Herschel SPIRE/PACS and Spitzer MIPS maps at $\lambda \geq$\,70\,$\mu$m, we investigate the physical properties of dust using single and two-component modified black body models and study the variation of the dust emissivity index $\beta$ across M~33. Although the single and two-component models use modified black-bodies, the procedures are very different and both have been designed to avoid the $\beta-T$ degeneracy.  The single-component model assumes that $\beta$ does not vary rapidly from pixel to pixel at the scale observed in M33 (160\,pc) and, by testing many
values of $\beta$, determines the value leading to the lowest residuals.  The
two-component models solve exactly for the parameters ($T$, $\beta$, surface density)
for a combination of warm and cold dust for data covering 70\,$\mu$m to 500\,$\mu$m whereas
the single component models use the data from 160\,$\mu$m to 500\,$\mu$m.

Both analyses find that ($i$) $\beta$ is higher in actively star forming regions,
($ii$) $\beta$ is higher in the inner disk and decreases sharply beyond a
galactocentric distance of 4 kpc, and of course ($iii$) that the dust temperature
decreases with radius.  {\it Both $\beta$ and $T$ decrease with radius in M33,
contrary to reported anti-correlations.} 

In addition, the two component model shows that within the inner disk, $\beta$ is
higher along the spiral arms.  The warm dust follows the H$\alpha$ emission,
indicating that the separation into warm and cold dust is appropriate.  
Proper behavior of the two-component model has also been verified by extensive Monte-Carlo simulations. 

In an attempt to identify the reason for the variation of $\beta$, we find that $\beta$ is correlated both with H$\alpha$ intensity and with CO(2-1) line strength but not significantly with the HI column density. {  As CO and H$\alpha$ emission decrease radially}, we are not able to distinguish their effects on $\beta$. The radial decrease in $\beta$ appears {\it not} to be due to an increasingly broad mixture of dust temperatures. 

%{  It is also not due to a decrease in the strength of the interstellar radiation field or temperature as there is no positive correlation between $T_c$ and $\beta$ and as not only the weaker but also the FIR-bright regions in the outer disk have a low $\beta$.}

%

\acknowledgement We thank Eva Schinnerer for support and helpful discussions. We also thank {  the anonymous referee}, Brent Groves, Hendrik Linz, Svitlana Zhukovska, and Amy Stutz {  for discussions and useful comments}. FST acknowledges the support by the DFG via the grant TA 801/1-1. 

\bibliography{s.bib}   
     
\end{document}